\newif\ifpreprint
 
\preprintfalse 
  
\ifpreprint
\documentclass[preprint,superscriptaddress]{revtex4-2}
\else
\documentclass[prx,reprint,superscriptaddress]{revtex4-2}
\fi

\usepackage{newunicodechar}

\usepackage[T1]{fontenc} 

\usepackage{amsmath}
\usepackage{amssymb}

\usepackage{xcolor}
\usepackage[version=4]{mhchem} 
\usepackage{color,soul}

\usepackage{xspace}
\usepackage{braket}
\usepackage{ifthen}
\usepackage{comment}
\usepackage{booktabs}
\usepackage{tablefootnote} 
\usepackage{threeparttable}
\usepackage{graphicx}
\usepackage{dcolumn}

\usepackage{url}
\usepackage{svg}

\usepackage{multirow}
\usepackage{xspace}

\newcommand*{\abinitio}{{\it ab initio}\xspace}

\newcommand*{\kcal}{kcal mol$^{-1}$\xspace}
\newcommand*{\sunit}{$E_{\rm h}^{-2}$\xspace}
\newcommand*{\Eh}{$E_{\rm h}$\xspace}

\newcommand{\name}[0]{QDSRG\xspace}
\newcommand{\appA}[0]{1-\name}
\newcommand{\appB}[0]{2-\name}
\newcommand{\stgap}[0]{\Delta E_\mathrm{ST}}

\newcommand{\rdc}[1]{\boldsymbol\lambda_{#1}}
\newcommand{\rdm}[1]{\boldsymbol\gamma_{#1}}

\newcommand{\tens}[3]{{#1}_{#2}^{#3}}

\newcommand{\aphystei}[2]{\bra{#1}\!\!\ket{#2}}

\newcommand{\density}[2]{\gamma_{#2}^{#1}}

\newcommand{\no}[1]{ \{ {#1} \}}

\mathchardef\mhyphen="2D

\usepackage[colorlinks = true,
            linkcolor = blue,
            urlcolor  = black,
            citecolor = blue,
            anchorcolor = black]{hyperref}

\definecolor{goodorange}{RGB}{225,125,0}
\definecolor{goodgreen}{RGB}{5,130,5}
\definecolor{goodred}{RGB}{220,50,25}
\definecolor{goodblue}{RGB}{30,144,255}
\definecolor{OliveGreen}{RGB}{5,100,5}

\newcommand{\note}[2]{
\ifthenelse{\equal{#1}{F}}{
\colorbox{goodred}{\textcolor{white}{\footnotesize \fontfamily{phv}\selectfont #1}}
    \textcolor{goodred}{{\footnotesize \fontfamily{phv}\selectfont #2}}\xspace
}{}
\ifthenelse{\equal{#1}{R}}{
\colorbox{goodblue}{\textcolor{white}{\footnotesize \fontfamily{phv}\selectfont #1}}
    \textcolor{goodblue}{{\footnotesize \fontfamily{phv}\selectfont #2}}\xspace
}{}
\ifthenelse{\equal{#1}{Y}}{
\colorbox{goodgreen}{\textcolor{white}{\footnotesize \fontfamily{phv}\selectfont #1}}
    \textcolor{goodgreen}{{\footnotesize \fontfamily{phv}\selectfont #2}}\xspace
}{}
}

\date{\today}

\begin{document}

\title{Leveraging small scale quantum computers with unitarily downfolded Hamiltonians}

\author{Renke Huang}
\address{Department of Chemistry and Cherry Emerson Center for Scientific Computation, Emory University, Atlanta, GA, 30322}

\author{Chenyang Li}
\address{Key Laboratory of Theoretical and Computational Photochemistry, Ministry of Education, College of Chemistry, Beijing Normal University, Beijing 100875, China}

\author{Francesco A. Evangelista}
\email{francesco.evangelista@emory.edu}
\address{Department of Chemistry and Cherry Emerson Center for Scientific Computation, Emory University, Atlanta, GA, 30322}

\date{\today}

\begin{abstract}
In this work, we propose a quantum unitary downfolding formalism based on the driven similarity renormalization group (QDSRG) that may be combined with quantum algorithms for both noisy and fault-tolerant hardware.
The \name is a classical polynomially-scaling downfolding method that avoids the evaluation of costly three- and higher-body reduced density matrices while retaining the accuracy of classical multireference many-body theories.
We calibrate and test the \name on several challenging chemical problems and propose a strategy for avoiding classical exponential-scaling steps in the QDSRG scheme.
We report \name computations of two chemical systems using the variational quantum eigensolver on IBM quantum devices: i) the dissociation curve of \ce{H2} using a quintuple-$\zeta$ basis and ii) the bicyclobutane isomerization reaction to \textit{trans}-butadiene, demonstrating the reduction of problems that require several hundred qubits to a single qubit.
Our work shows that the \name is a viable approach to leverage near-term quantum devices for the accurate estimation of molecular properties.
\end{abstract}

\maketitle

\section{Introduction}

Molecule and materials that display strong electron correlation are hard to simulate with classical computers due to the exponential growth of the many-body Hilbert space \cite{Laughlin:2000br}.
Quantum computers \cite{Feynman:1982gn,manin1980computable}  are particularly well-suited to simulate many-body systems, as they can efficiently represent and perform operations on many-particle wave functions. These features make them a promising solution to the electron correlation problem \cite{Abrams:1997ha,Abrams:1999ur}, which, in its most general form, still defies classical algorithms.
However, the accurate modeling of realistic many-electron systems requires the use of large computational bases and, therefore, it is unlikely to be accessible to small-scale quantum computers with up to only a few hundred qubits.
One of the most promising strategies to maximize the impact of near-term quantum devices is to pair a quantum algorithm with classical approaches that perform a reduction of the number of qubits required in a quantum computation \cite{Bauer:2016fc}.

Several strategies have been proposed to minimize quantum resources by combining quantum computations with traditional quantum chemistry approaches.
Takeshita \textit{el al.} \cite{Takeshita:2020dh}  applied quantum algorithms in combination with the active-space approximation and orbital optimization.
This idea was demonstrated experimentally for model molecules with up to ten orbitals using just two qubits \cite{Urbanek:2020fh}.
Boyn \textit{el al.} \cite{Boyn:2021kp} obtained active-space 2-RDMs from quantum computations and post-processed them with two classical correlation methods, the anti-Hermitian contracted Schrödinger equation (ACSE) theory \cite{Mazziotti:2006iw, Smart:2021hd, Smart:2022ik} and multiconﬁgurational pair-density functional theory (MC-PDFT) \cite{LiManni:2014kg}.
Fujii \textit{el al.} combined the divide-and-conquer technique with quantum computations to solve the ground state of a 64-qubit two-dimensional Heisenberg model with 20-qubit simulations, \cite{fujii2022deep} and later extended it to obtain excited states of periodic hydrogen chain \cite{mizuta2021deep}.
Ryabinkin \textit{et al.} \cite{Ryabinkin_2021} used low-order perturbative corrections to reduce the quantum resources required in their iQCC-VQE algorithm.
Tammaro \textit{et al.} \cite{tammaro2022n} investigated the use of $N$-electron valence perturbation theory (NEVPT) \cite{Angeli:2001vf, Angeli:2006by} in combination with the variational quantum eigensolver (VQE) \cite{Peruzzo:2014kc, Yung:2014iv, McClean:2015bs} and quantum subspace expansion \cite{McClean:2017ct, Colless:2018hp} algorithms.


Huggins \textit{et al.} recently proposed a hybrid algorithm that combines quantum shadow tomography with the auxiliary-field quantum Monte Carlo method, achieving the largest-to-date chemical simulation on hardware of a 16-qubit system \cite{Huggins:2022iw}.
Other works have focused on reducing the number of qubits and gates required by variational quantum algorithms using embedding techniques.
For example, density-matrix embedding theory \cite{Knizia:2012do, wouters2016practical} has been applied to simulate a \ce{H10} ring on an ion-trap quantum computer by decomposing this 20-qubit system into ten two-qubit problems \cite{Kawashima:2021hv}.
Huang \textit{el al.} have used quantum defect embedding theory to simulate spin defects on quantum computers \cite{Huang:2022bu}.
Combing explicitly correlated methods with quantum algorithms is another strategy explored in several works \cite{Motta:2020ic, mcardle2020improving, Schleich:2022fa,  sokolov2022orders,2018projectq, kumar2022accurate} to achieve higher accuracy without increasing quantum resources.
Motta \textit{et al.} \cite{Motta:2020ic} used a canonical transcorrelated F12 (CT-F12) Hamiltonian \cite{Yanai:2012cj} in conjunction with the variational quantum eigensolver method, whereas McArdle \textit{et al.} \cite{mcardle2020improving} used Boys and Handy's transcorrelated approach, which produces a non-Hermitian Hamiltonian containing up to three-body terms.
A recent contribution also considers an \textit{a posteriori} perturbative correction based on the explicitly correlated $\mathrm{[2]_{R12}}$ approach \cite{Schleich:2022fa}.

Quantum chemistry effective Hamiltonian theories \cite{White:2002uj,Yanai:2006gi,Yanai:2010kf,Spiegelmann1984,Lyakh:2012cn,Kohn:2013cp,Evangelista:2018bt}
offer another way to downfold correlation effects into a small active-space quantum computation.
In such approaches one partitions the orbital space into two sets, active and inactive, and a transformation is applied to the Hamiltonian to eliminate terms that couple these two spaces.
The resulting Hamiltonian then accounts for electron correlation effects in the inactive orbitals via effective many-body interactions.
In principle, it is straightforward to adapt this strategy to a quantum-classical hybrid setting whereby a highly entangled quantum state involving only the active orbitals is solved for on a quantum computer and the remaining weak correlation effects are recovered with a polynomially scaling classical algorithm.
However, there are several major potential issues with effective Hamiltonian methods as formulated in a classical setting.
A particularly severe limitation of some approaches is the need to measure three- and four-body reduced density matrices (RDMs), introducing a prohibitively large prefactor in quantum computations that scales as the sixth to the eight power of the number of active orbitals.
A second important issues is the impact of noise in the measured RDMs on the energy (and other properties) and the numerical stability of methods that require the solution of nonlinear equations.
Lastly, it is often necessary to go beyond low-order perturbative treatments to achieve accurate energetics.
Several quantum downfolding methods have been proposed.
The double unitary coupled-cluster (DUCC) approach \cite{Bauman:2019dx, Metcalf:2020fe, Nicholas:2021cy} 
is a downfolding procedure based on a mean-field reference state.
Le and Tran \cite{Le2022} have employed their one-body second-order Møller-Plesset perturbation theory (OBMP2) to create an effective Hamiltonian with modified one-body interactions for the VQE method.

In this work, we present a quantum downfolding approach based on the driven similarity renormalization group (DSRG) \cite{evangelista2014driven, li2015multireference, li2016towards, li2019multireference} that addresses the challenges highlighted above.
The DSRG is an integral reformulation of numerical flow-renormalization group methods \cite{Wegner:1994kh,Glazek:1994uu,Tsukiyama:2012dw,Hergert:2017bk}.
Our quantum formulation of the DSRG (\name) is compatible with any quantum algorithm capable of producing low-rank RDMs (up to partial or full second-order) and augments it with an accurate, numerically-robust, and nonperturbative treatment of weak (dynamical) correlation.
We benchmark the performance of the \name scheme in computing the dissociation curve of the nitrogen molecule and the adiabatic singlet-triplet splittings of the para-benzyne diradical. 
In addition to exact simulations, we demonstrate the usefulness of this strategy in the presence of realistic noise by combining the \name with VQE experiments on IBM quantum computers. 
We compute the dissociation curve of the hydrogen molecule with a nearly-complete quintuple-$\zeta$ basis, as well as model the bicyclobutane isomerization pathways to \textit{trans}-butadiene, which, to the best of our knowledge, is the first example of modeling an organic chemistry reaction on near-term quantum devices.

\section{Theory}
\label{theory}

\subsection{Unitary Hamiltonian downfolding via the DSRG}
\label{dsrg}

The DSRG method \cite{evangelista2014driven, li2015multireference, li2016towards, li2019multireference, Li:2021eb}  starts from a reference correlated state $\Psi_0$ and performs a unitary transformation of the Hamiltonian, $H$, that brings it to a block-diagonal form
\begin{equation}
H \mapsto \bar{H} = e^{-A} H e^{A},
\end{equation}
where the operator $A$ is anti-Hermitian.
The goal of this transformation is to remove the second-quantized components of $\bar{H}$ that couple $\Psi_0$ to excited configurations, which we refer to as the nondiagonal components of $\bar{H}$ (denoted as $\bar{H}^\mathrm{N}$).

One of the challenges associated with eliminating these couplings (i.e., solving for $\bar{H}^\mathrm{N} = 0$) is the emergence of numerical instabilities, which are related to excitations with small energy denominators.
To avoid this issue, the DSRG achieves only a partial block-diagonalization of $H$ by solving a set of regularized equations
\begin{equation}
\label{eq:dsrg}
\bar{H}^{\mathrm{N}}= R(s).
\end{equation}
In this equation, $R(s)$ is an operator that depends on the flow parameter $s \in [0, \infty)$, and its purpose is to suppress excited configurations associated with an energy denominator smaller than the energy cutoff $\Lambda = 1/\sqrt{s}$.
Hence, solving Eq.~\eqref{eq:dsrg} imparts a dependence on $s$ onto the $A$ operator and the transformed Hamiltonian.

The DSRG operator $A(s)$ is expressed in terms of a $s$-dependent coupled cluster particle-hole excitation operator \cite{Watts:1989ve, Musial:2008kn, Musial:2008ea, Crawford:2000ub} as $A = T - T^\dagger$ with $T = T_1 + T_2 + \ldots$ where each $k$-body component is
\begin{equation}
	T_k = \frac{1}{(k !)^{2}} \sum_{i j \cdots}^{\mathrm{hole}} \sum_{a b \cdots}^{\mathrm{particle}} t_{a b \cdots}^{i j \cdots}(s)\{\hat{a}_{i j \cdots}^{a b \cdots}\},
\end{equation}
where we write the normal-ordered creation and annihilation operators in a compact form $\{\hat{a}_{i j \cdots}^{a b \cdots}\}=\{\hat{a}^{a} \hat{a}^{b} \cdots \hat{a}_{j} \hat{a}_{i}\}$ \cite{Mukherjee:1997tk, Kutzelnigg:1997ut}, 
and the cluster amplitudes ($t_{a b \cdots}^{i j \cdots}$) are tensors antisymmetric with respect to the individual permutation of upper and lower indices. 
The hole space contains the occupied and partially occupied orbitals, while the particle space contains the partially occupied and unoccupied orbitals of the reference $\Psi_0$.
One of the simplest nonperturbative truncation schemes is the linearized DSRG with one- and two-body operators [LDSRG(2)] \cite{li2016towards} where:
1) $T$ is truncated as $T \approx T_1 + T_2$, and 2) every commutator in the Baker--Campbell--Hausdorff expansion of $\bar{H}$ contains only one- and two-body operators (indicated with subscript ``1,2'')
\begin{equation}
\bar{H} \approx H+\sum_{k=1}^{\infty} \frac{1}{k !} \underbrace{ [ \cdots [ [H, A]_{1,2}, A]_{1,2}, \ldots ]_{1,2}}_{k \text { nested commutators }}.
\end{equation}
The resulting DSRG transformed Hamiltonian contains up to two-body interactions.
Once Eq.~\eqref{eq:dsrg} is solved, the  energy may computed as the expectation value of $\bar{H}$
\begin{equation}
\label{eq:unrelaxed_energy}
E = \braket{\Psi_{0} | \bar{H} | \Psi_{0}}.
\end{equation}
Alternatively, one may solve the eigenvalue problem
\begin{equation}
\label{eq:relaxed_energy}
\bar{H} \ket{\tilde{\Psi}_0} = \tilde{E} \ket{\tilde{\Psi}_0},
\end{equation}
and obtain a \textit{relaxed} reference state $\tilde{\Psi}_0$ and its corresponding energy $\tilde{E}$.
It is often the case that multireference quantum chemistry methods, like for the example, the CASPT2 \cite{andersson1992second} or NEVPT2 \cite{Angeli:2001vf, Angeli:2006by} methods, only evaluate the energy as an expectation value via equations analogous to Eq.~\eqref{eq:unrelaxed_energy}.
In this case one talks of a ``diagonalize-then-perturb'' approach and the resulting formalism only provides an energy correction rather than a properly downfolded Hamiltonian.

Solving the DSRG equations [Eq.~\eqref{eq:dsrg}] requires the reduced density cumulants (RDCs) of the reference state (which we also simply refer to as ``cumulants'') \cite{Mazziotti_1998,Kutzelnigg_Mukherjee_1999}.
A generic $k$-body reduced density cumulant ($\rdc{k}$) is the connected part of the corresponding $k$-body RDM ($\boldsymbol\gamma_k$), defined as
$\gamma^{pq\cdots}_{rs\cdots} = 
\braket{\Psi_{0} | a_p^\dagger a_q^\dagger \cdots a_s a_r | \Psi_{0}}$,
where the product $a_p^\dagger a_q^\dagger \cdots a_s a_r$ contains $k$ creation and $k$ annihilation operators.
The RDCs of the reference state encode all the information required to include correlation effects outside of the active orbitals.
Therefore, any computational method capable of generating $\Psi_0$ and its RDMs can be interfaced with the DSRG downfolding procedure.
It is convenient to express the DSRG equations in terms of cumulants as any truncated scheme preserves the size extensivity of the energy.
Reduced density cumulants enter in the LDSRG(2) equations in the following way.
Evaluation of the operator $A$ requires $\rdc{1}$ and $\rdc{2}$, while evaluation of the energy additionally requires $\rdc{3}$ which is challenging to measure on near-term devices.
In the next section, we analyze a modified DSRG approach amenable for a hybrid quantum-classical scheme.

\subsection{Hybrid quantum-classical DSRG downfolding}
\label{outline}

\begin{figure*}[htbp]
   \includegraphics[width=6.5in]{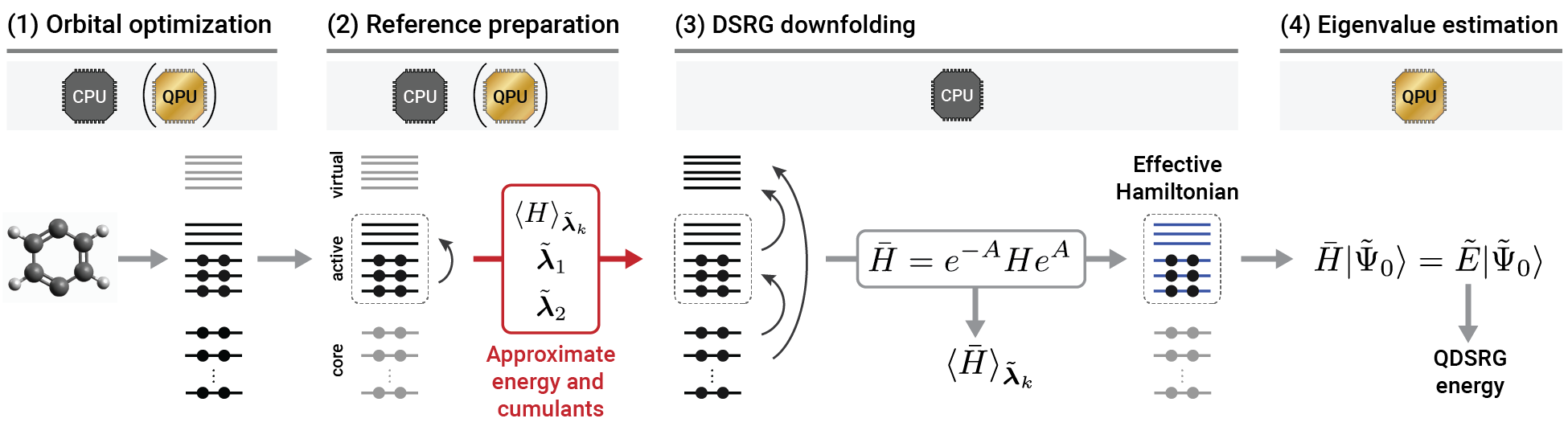} 
   \caption{
   	\linespread{1}\small
   	The \name scheme for performing hybrid classical-quantum computations on strongly correlated molecules. The computation begins with an a classical or hybrid orbital optimization (1), followed by the preparation of correlated reference state defined in the subset of active molecular orbitals (2). This step yields the reduced density cumulants ($\tilde{\boldsymbol \lambda}_k$) and the approximate energy ($\braket{H}_{\tilde{\boldsymbol\lambda}_k}$), which are passed to a classical DSRG algorithm (3) to produce the effective Hamiltonian ($\bar{H}$) and its expectation value with respect to the approximate cumulants ($\braket{\bar{H}}_{\tilde{\boldsymbol\lambda}_k}$). (4) In the last step, an eigenvalue of the DSRG effective Hamiltonian is found via a quantum algorithm.}
   \label{fig:vqe-dsrg}
\end{figure*}

An outline of how the DSRG downfolding scheme may be adapted to a hybrid quantum-classical scheme is illustrated in Figure~\ref{fig:vqe-dsrg}.
We can break down the procedure into four steps:
\begin{enumerate}
\item \textit{Orbital optimization}. The first step is an optimization of the molecular orbitals for a target electronic state.
To ensure that this scheme is applicable to large active spaces, it is important that the cost of the orbital optimization scales as a low-order polynomial of the systems size.
This step could employ a mean-field approximation (e.g. Hartree--Fock) or optimize the orbitals of a correlated state, as in the complete-active-space self-consistent-field (CASSCF) method.
In the latter case, a quantum computation may be used to optimize the correlated state.

\item \textit{Reference preparation}. In the second step, we propose to employ an approximate reference state that might be obtained from a classical or a  quantum computation.
In this step, the spinorbital basis is partitioned into three subsets: core (doubly occupied), active (partially occupied), and virtual (empty).
The quantum computation involves only the active orbitals and uses a modified one-electron operator that accounts for the interaction with the occupied core orbitals.
With these restrictions, the quantum computation requires at most 2 $N_\mathrm{A}$ qubits, where $N_\mathrm{A}$ is the number of active orbitals.
As part of the quantum computation, the low-rank reduced density cumulants ($\tilde{\boldsymbol \lambda}_k$) and the approximate  energy ($\braket{H}_{\tilde{\boldsymbol\lambda}_k}$) of the reference state are evaluated.

\item \textit{DSRG downfolding}. The third step consists of a classical DSRG computation using the approximate reduced density cumulants ($\tilde{\boldsymbol\lambda}_k$) from step 2. This step produces the anti-Hermitian operator $A$ and the DSRG transformed Hamiltonian.
We also obtain the expectation value of $\bar{H}$ with respect to the approximate RDCs, ($\braket{\bar{H}}_{\tilde{\boldsymbol\lambda}_k}$); however, this quantity is generally a poor approximation to the exact energy.
This step has polynomial scaling in the number of active and total orbitals.

\item \textit{Eigenvalue estimation}.
In the last step of this procedure, the DSRG downfolded Hamiltonian is used in a quantum computation  that estimates one of its eigenvalues ($\tilde{E}$).
\end{enumerate}
For generality we separate steps 1 and 2 of the \name scheme; however, if the orbital optimization step minimizes the energy of a correlated state generated via a quantum computation (e.g., like in CASSCF), then these steps can be combined into a single one.

The most crucial differences between the conventional DSRG formulation and the \name scheme are the use of approximate density cumulants in the DSRG downfolding procedure and the eigenvalue estimation (step 4).
Here we consider approximations of the cumulants that are consistent with the measurement of at most a quartic-scaling number of elements.
The simplest approximation (\appA), retains only the diagonal elements of the one-body density cumulant, that is,
\begin{equation}
\tilde{\lambda}^{u}_{v}  =
\begin{cases}
\lambda^{u}_{u}, & \text{if } u = v, \\
0, & \text{otherwise}.
\end{cases}
\end{equation}
Since $\rdc{1} = \rdm{1}$, this scheme requires only the diagonal parts of the one-body density matrix $\gamma^{u}_{u} = \braket{\Psi_0 | a^\dagger_u a_u | \Psi_0}$.
Therefore, if $\Psi_0$ is generated via a quantum computation, this would require performing $N_\mathrm{A} + 1$ experiments to measure the energy ($\braket{H}$) and the diagonal elements of $\rdm{1}$.

The next approximation (\appB) requires access to a quadratic number of elements of the RDCs, and consists in taking the full one-body density matrix ($\tilde{\boldsymbol \lambda}_1 = \boldsymbol \lambda_1$) plus the diagonal components of the two-body reduced density cumulant
\begin{equation}
\tilde{\lambda}^{uv}_{xy}  =
\begin{cases}
\lambda^{uv}_{uv}, & \text{if } u = x \text{ and } v = y, \\
\lambda^{uv}_{vu} = - \lambda^{uv}_{uv}, & \text{if } u = y \text{ and } v = x, \\
0, & \text{otherwise}.
\end{cases}
\end{equation}

These cumulant approximations may be justified using a perturbative argument.
At first order in perturbation theory (assuming a one-body diagonal zeroth-order operator, $H_0 = \sum_p \epsilon_p \no{a^\dagger_p a_p}$) one may show that the amplitudes corresponding to single and double excitations are given by \cite{li2015multireference}
\begin{align}
\label{eq:t_1_amps}
\tens{t}{a}{i,(1)} &= \big[ \tens{f}{a}{i} + \sum\limits_{ux}^{\rm A} \tens{\Delta}{u}{x} \tens{t}{ax}{iu,(1)} \density{x}{u} \big] \frac{1 - e^{-s(\tens{\Delta}{a}{i})^2}}{\tens{\Delta}{a}{i}}, \\
\label{eq:t_2_amps}
\tens{t}{ab}{ij,(1)} &= \aphystei{ab}{ij} \frac{ 1 - e^{-s(\tens{\Delta}{ab}{ij})^2} }{\tens{\Delta}{ab}{ij}},
\end{align}
where the quantities $\tens{f}{a}{i}$ and $\aphystei{ab}{ij}$ are elements of an effective one-body operator and antisymmetrized two-electron integrals, respectively, while the denominators are defined as $\tens{\Delta}{a}{i} = \epsilon_i  - \epsilon_a$ and $\tens{\Delta}{ab}{ij} = \epsilon_i  + \epsilon_j  - \epsilon_a- \epsilon_b$, with $\epsilon_i = \tens{f}{i}{i}$ \cite{li2015multireference}.
These equations show that at first order, the double excitation component of $A$ is independent of the reference cumulants, and the single excitation component of $A$ depends only on the off-diagonal elements of the active-active block of  $\rdm{1}$ [since when $x = u$ in Eq.~\eqref{eq:t_1_amps} we have that $\density{u}{u} \tens{\Delta}{u}{u} =  \density{u}{u} (\epsilon_u  - \epsilon_u) = 0$].
Therefore, the \appB approximation is already consistent with a first-order approximation to the DSRG operator $A$, while \appA neglects the off-diagonal terms of $\density{x}{u}$ that enter into Eq.~\eqref{eq:t_1_amps}.
The three-body cumulant $\rdc{3}$ is neglected in both the \appA and \appB, and it may be shown to enter in the energy at second-order in perturbation theory.
The impact of neglecting $\rdc{3}$ in the DSRG was analyzed in previous studies \cite{li2015multireference,Wang.2021.10.1021/acs.jctc.1c00980,He.2022.10.1021/acs.jctc.1c01099}.

To enable the pipeline of \name computations, we implement functionalities that export integrals and read/write the reference density matrices from external files in \texttt{Forte} \cite{Evangelista2021Forte}, an open-source plugin for the \abinitio quantum chemistry package \texttt{Psi4} \cite{Smith:2020ci}.
We obtain the \name effective Hamiltonian from \texttt{Forte} using a spin-free implementation \cite{Li:2021eb}.

\section{ Calibration}

\subsection{\label{sec:calibration} Noiseless simulations}
\begin{table*}[ht!]
\renewcommand{\arraystretch}{1.}
\footnotesize
\caption{\label{tab: h2_benchmark} 
	\linespread{1}\small
	Energy error for the hydrogen molecule (in m\Eh) computed with the LDSRG(2) and QLDSRG(2) methods. All computations use a cc-pVTZ basis \cite{cc-pvdz-dk-H-Ne} and the flow parameter value $s$ = 0.5 \sunit. Energy errors are computed with respect to full configuration interaction (FCI) energies reported in the last row of the table. For the LDSRG(2), the expectation value of the energy is designated with ``$\braket{\bar{H}}$'', while the lowest eigenvalue is indicated with ``eig. $\bar{H}$''. For the \name methods we report two sets of data. The ones labeled ``$\rdm{3} = 0$'' use an approximate three-body cumulant reconstructed from $\tilde{\boldsymbol\lambda}_1$ and $\tilde{\boldsymbol\lambda}_2$.
	Results for line M employ CCSD natural orbitals and the one-body relaxed CCSD reduced density matrix as input to the QDSRG procedure.
}

\begin{ruledtabular}
\scriptsize
\begin{tabular}{@{} clcccccc @{}}
       Case &  Method &  Orbital Type&  Active orbitals & \multicolumn{4}{c}{$r_\text{HH}$ (\AA)} \\
      \cline{5-8}
    & &  &  & 0.75 & 1.50 & 2.25 & 3.00\\
   \hline
A & LDSRG(2) ($\braket{\bar{H}}$) & RHF &  $\{1 \sigma_g,1 \sigma_u\}$ & 0.306  & 7.337 & 15.454 & 17.428 \\
B & LDSRG(2) ($\braket{\bar{H}}$) & CASSCF(2,2) & $\{1 \sigma_g,1 \sigma_u\}$ & 0.791 & 1.007 & 0.307 & 0.041 \\
C & LDSRG(2) ($\braket{\bar{H}}$) & CASSCF(2,2) & $\{1 \sigma_g,1 \sigma_u,2 \sigma_g,2 \sigma_u\}$ & 0.064 & 0.280 & 0.080 & 0.012 \\
D & LDSRG(2) (eig. $\bar{H}$) & CASSCF(2,2) & $\{1 \sigma_g,1 \sigma_u\}$ & 0.331 & 0.070 & 0.242 & 0.041 \\
E & 2-QLDSRG(2) ($\rdm{3} = 0$) & CASSCF(2,2) & $\{1 \sigma_g,1 \sigma_u\}$ & 0.327 & 0.055 & 0.235 & 0.039 \\
F & 1-QLDSRG(2) ($\rdm{3} = 0$) & CASSCF(2,2) & $\{1 \sigma_g,1 \sigma_u\}$ & 2.455 & 6.422 & 4.591 & 1.203 \\
G & 1-QLDSRG(2) ($\rdm{3} = 0$) & CASSCF(2,2)  & $\{1 \sigma_g,1 \sigma_u,2 \sigma_g,2 \sigma_u\}$ & 0.601 & 2.193 & 1.383 & 0.564 \\
H & 1-QLDSRG(2) ($\rdm{3} = 0$) & CASSCF(2,2) & $\{1 \sigma_g,1 \sigma_u,2 \sigma_g,2 \sigma_u,3 \sigma_g,3 \sigma_u\}$ & 0.324 & 0.511 & 0.071 & -0.051 \\
I & 2-QLDSRG(2) & CASSCF(2,2) & $\{1 \sigma_g,1 \sigma_u\}$ & 1.065 & 2.981 & 2.152 & 0.354 \\
J & 1-QLDSRG(2) & CASSCF(2,2) & $\{1 \sigma_g,1 \sigma_u\}$ & 1.061 & 2.732 & 1.890 & 0.327 \\
K & 1-QLDSRG(2) & CASSCF(2,2)  & $\{1 \sigma_g,1 \sigma_u,2 \sigma_g,2 \sigma_u\}$ & 0.016 & 1.392 & 0.935 & 0.070 \\
L & 1-QLDSRG(2) & CASSCF(2,2) & $\{1 \sigma_g,1 \sigma_u,2 \sigma_g,2 \sigma_u,3 \sigma_g,3 \sigma_u\}$ & -0.104 & 0.293 & 0.130 & 0.008 \\
M & 1-QLDSRG(2) & CCSD NOs & $\{1 \sigma_g,1 \sigma_u\}$ & 1.059	& 2.927 & 2.095 & 0.351 \\
\hline
&    FCI & RHF & All MOs & -1.172301 & -1.066168 & -1.010114 & -1.000726\\
\end{tabular}
\end{ruledtabular}
\end{table*}

To investigate the accuracy of the \name procedure for quantum computing, we examine a numerical example.
We consider the \ce{H2} molecule at four geometries using a triple-$\zeta$ basis.
As the \ce{H2} molecule is stretched, the $1 \sigma_g$ and $1 \sigma_u$ orbitals become near-degenerate,  and the reference state must be taken of the form $\ket{\Psi_0} = c_g \ket{(1 \sigma_g)^2} + c_u \ket{(1 \sigma_u)^2}$ to guarantee a continuous and qualitatively correct solution for all bond lengths.
Table~\ref{tab: h2_benchmark} reports the energy error with respect to a full configuration interaction (FCI) computation for the LDSRG(2) and the QLDSRG(2) methods. We report both the average energy $\braket{\bar{H}}$ and the eigenvalue of $\bar{H}$ using active spaces of various size and different orbitals choices.
In analyzing these results we focus on the largest error and use the labels A--H to refer to a specific row of Table~\ref{tab: h2_benchmark}.

The importance of optimizing the orbitals is reflected in the significant difference in the accuracy of the LDSRG(2) $\braket{\bar{H}}$ when the orbitals $1 \sigma_g$ and $1 \sigma_u$ come from a  restricted Hartree--Fock (RHF) or CASSCF orbitals (using only the $1 \sigma_g$ and $1 \sigma_u$ MOs), whereby the latter optimize both the orbitals and coefficients of the determinant that define $\Psi_0$.
The LDSRG(2) error with  RHF orbitals (A) is as large as 17.4 m\Eh $\approx$ 0.47 eV (at $r_\text{H-H}$ = 3 \AA{}), whereas CASSCF orbitals (B) give an error of ca. 1 m\Eh $\approx$ 0.03 eV, and this error can be further reduced to less than 0.3 m\Eh (C) if the reference state is augmented with determinants formed out of a larger active space that includes the 
$2 \sigma_g$ and $2 \sigma_u$ orbitals.
Diagonalization of $\bar{H}$ in an active space containing the $1 \sigma_g$ and $1 \sigma_u$ MOs (D) yields a maximum error similar to the one of case (C).

For the \name methods, we report two sets of results.
We first examine the ones denoted with ``$\rdm{3} = 0$'', which use a three-body cumulant reconstructed from $\tilde{\boldsymbol\lambda}_1$ and $\tilde{\boldsymbol\lambda}_2$ \cite{Colmenero_C_Valdemoro_1993,Mazziotti_1998,Kutzelnigg_Mukherjee_1999,DePrince_Mazziotti_2007} and, therefore, differ slightly from the approximations defined in the previous section. However, this definition is consistent with the fact that a reference containing two electrons always yields a zero three-body RDM.
We note that the \appB approach (E) leads to small errors (max 0.4 m\Eh) that are similar to those of case D, where the energy comes from diagonalization of the LDSRG(2) $\bar{H}$.
The more drastic approximation (\appA) gives a large maximum error (6.4 m\Eh, F). 
In this case it is possible to improve the accuracy by expanding the active space with a single or double set of $\sigma_g/\sigma_u$ orbitals, reducing the energy error to 2.2 and 0.5 m\Eh (G, H), respectively.
Interestingly, imposing $\rdm{3} = 0$ seems to benefit the approximate \name schemes only when using two active orbitals (compare E,F with I,J).
The results obtained from imposing $\tilde{\boldsymbol\lambda}_3 = 0$ with four and six active orbitals (K,L) are more accurate than the corresponding ones from imposing $\rdm{3} = 0$ (G,H).

As mentioned earlier, a practical realization of the \name scheme requires either two quantum computations (one to generate the orbitals and approximate cumulants plus a final diagonalization step) or it may use orbitals and cumulants from a polynomially-scaling classical method as a starting point.
Here we demonstrate how this second option may be realized in practice using natural orbitals from coupled cluster theory \cite{Abrams_Sherrill_2004}.
In the results labeled ``M'', we use coupled cluster theory with singles and doubles (CCSD) to compute an approximate density matrix $\boldsymbol\gamma^\mathrm{CCSD}_1$ that spans the entire orbital space.
The orbitals are then rotated to the natural basis (defined as the basis in which $\boldsymbol\gamma^\mathrm{CCSD}_1$ is diagonal). 
The active space occupation numbers are then scaled so that their sum equals to the number of electrons in the active orbitals (2) and these are used to reconstruct an approximate diagonal $\tilde{\boldsymbol\lambda}_1$.
A comparison of the \appA data using CCSD NOs (M) and CASSCF(2,2) orbitals (J), shows that these two procedures give energies that are within 0.2 m\Eh. 
We reexamine the use of CCSD NOs as a way to reduce the cost of \name computations in Sec.~\ref{results}.

In Appendix \ref{compare_ducc}, we provide a comparison of the \name scheme with the DUCC downfolding approach for the \ce{H2} molecule and the beryllium atom using data from Ref.~\cite{Bauman:2019dx}.
Both methods employ an exponential unitary transformation of the Hamiltonian, but differ in several ways.
For example, whereas, DUCC is formulated in a single-reference setting, the  \name method derives the $A$ operator from a correlated state.
This and other differences, have important consequences on the accuracy of these two methods, with our comparison showing that the \name leads to smaller errors (up to an order of magnitude smaller), especially in computations with fewer active orbitals.

\subsection{\label{sec:noise} Sensitivity to noise}

\begin{figure}[htbp]
   \centering
    \includegraphics[width=3.375in]{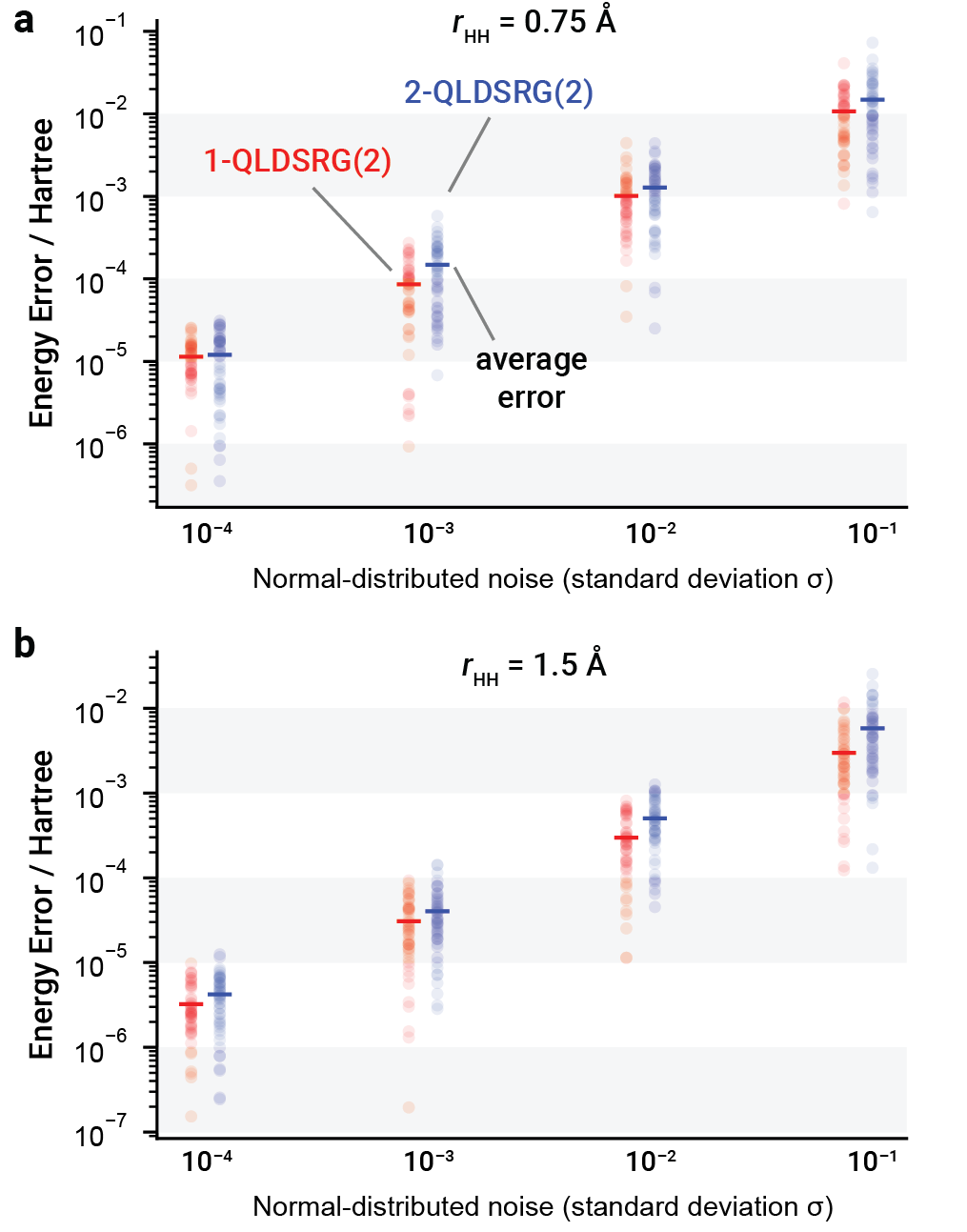}   
   \caption{\label{fig_h2_noise} 	
   		\linespread{1}\small
   	Energy error for the hydrogen molecule (in m\Eh) computed with the QLDSRG(2) with various amounts of stochastic noise ($\sigma$) added to the RDMs. For a given value of $\sigma$,  semi-opaque circles aligned vertically show the distribution of errors from 50 computations, while horizontal bars represent the average error.
    All computations use a cc-pVTZ basis \cite{cc-pvdz-dk-H-Ne} and the flow parameter value $s$ = 0.5 \sunit. Energy errors are computed with respect to noiseless values.
   Data obtained imposing $\rdm{3} = 0$ in the reconstruction of the approximate three-body cumulant.}
\end{figure}

We conclude our initial assessment of the \name approach by analyzing the sensitivity to stochastic errors introduced by quantum devices.
As shown in Figure~\ref{fig:vqe-dsrg}, step 2 of the \name procedure allows for the approximate cumulants to be obtained from a quantum computation.
In this case, there will be a compounding of errors due to the fact that the measured densities (later converted into cumulants) will be subject to finite  measurement errors and gate and measurement noise.

To study the effect of noise on the measured RDMs, we performed \name computations on the \ce{H2} molecule at bond distances of 0.75 and 1.5 \AA{}.
Following Ref.~\cite{Romero:2019hk}, we model noise by augmenting the cumulants with stochastic error sampled from a Gaussian distribution with standard deviation $\sigma$ and zero mean [$\mathcal{N}(0,\sigma^2)$],
\begin{equation}
\gamma^{uv\cdots, \text{measured}}_{xy \cdots} 
= \gamma^{uv\cdots}_{xy \cdots}  + \mathcal{N}(0,\sigma^2).
\end{equation}
This is a simple noise model that can mimic finite measurement errors, but cannot account for correlated noise among qubits and decoherence.
Noise is added to the unique elements of the RDMs to avoid breaking antisymmetry with respect to permutation of the upper/lower indices (e.g., $\gamma^{uv}_{xy} = -\gamma^{vu}_{xy} = - \gamma^{uv}_{yx} = \gamma^{vu}_{yx}$); however, we do not enforce fermionic $N$-representability conditions \cite{Percus64_1756, Erdahl78_697, Percus04_2095, Mazziotti12_263002} onto the resulting RDMs, which likely leads to overestimating the resulting energy errors.
Several works discuss how to utilize the $N$-representability constraints to accelerate and improve hybrid quantum algorithms mainly via reducing the measurement scaling \cite{rubin2018application, GoogleAIQuantumandCollaborators:2020bs}, which might be combined with the \name approach to improve its accuracy.

Figure~\ref{fig_h2_noise} shows the energy error computed with respect to noiseless results for the 1- and 2-QLDSRG(2) schemes (enforcing $\rdm{3} = 0$).
At both geometries we observe that the 1-QLDSRG(2) is less sensitive to noise, and that the average energy error increases linearly with $\sigma$.
Interestingly, the average error is slightly higher at the shorter bond distance (0.75 \AA{}) than at the elongated one (1.5 \AA{}).
In both cases, a value of $\sigma = 0.01$ seems sufficient to recover the energy with an error less than 1 kcal/mol ($\approx$ 1.6 m\Eh).
These results can then inform an analysis of the quantum resources necessary to measure the RDMs with an accuracy sufficient for a hybrid quantum-classical procedure based on the \name.

In summary, the preliminary results reported in Sec.~\ref{sec:calibration} and \ref{sec:noise} show that even a very drastic approximation of the cumulants that enter the DSRG \textit{combined with} diagonalization of the resulting transformed Hamiltonian can yield energies with small absolute energy errors, even under the presence of noise.
We expand this analysis to molecules with more complex electronic structures and larger basis sets in Sec.~\ref{results}. There, we also report the results of experiments on NISQ devices that show the potential usefulness of \name in leveraging near-term quantum computers.

\section{Results and Discussion}
\label{results}

In this section, we report two types of \name results: 
the noiseless exact computations in Secs. \ref{N2} and \ref{PBenzyne}, 
and device computations (Sec. \ref{Hardware}) where we combine the \name with variational quantum computations performed on IBM hardware.

\subsection{Dissociation curve of the nitrogen molecule}
\label{N2}
\begin{figure}[htbp]
   \centering
    \includegraphics[width=3.375in]{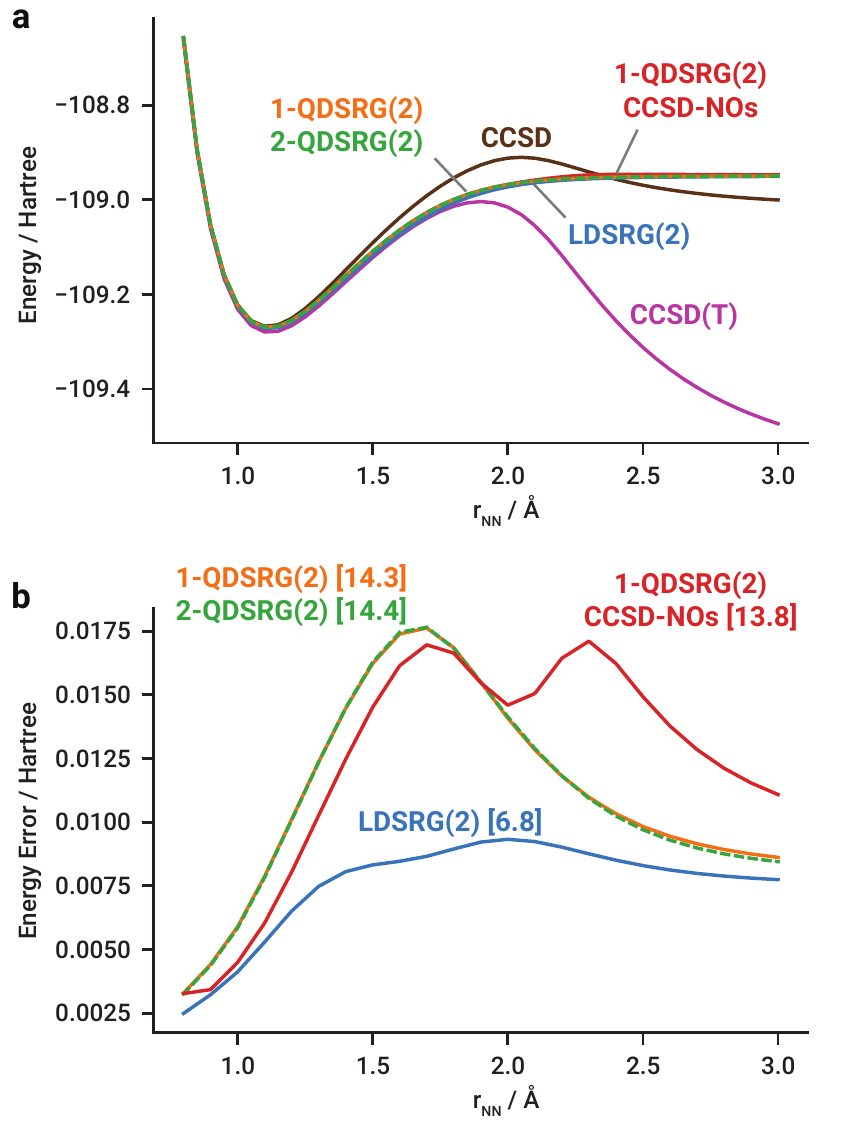}   
   \caption{\label{fig_n2} 
   		\linespread{1}\small
   	Dissociation curve for the nitrogen molecule computed with the LDSRG(2) and QLDSRG(2).
   (a) Total energy and (b) energy error with respect to FCI in units of \Eh.
   Nonparallelism error (in m\Eh) for each method are reported in square brackets.
    All computations use a reference state containing six 2p N atomic orbitals, a cc-pVDZ basis \cite{cc-pvdz-dk-H-Ne}, and the flow parameter value $s$ = 0.5 \sunit.
    The 1-QDSRG(2)-CCSD data employ natural orbitals and $\rdm{1}$ from CCSD as the input to the QDSRG computation. All other results employed CASSCF(6,6) orbitals.
    }
\end{figure}

As the first benchmark of the \name scheme we compute the potential energy curve for the ground singlet state of \ce{N2} using an active orbital containing six orbitals (build from combinations of the six 2p N orbitals).
Figure~\ref{fig_n2}a shows the potential energy curve for the LDSRG(2), the two approximate variants of the \name, CCSD \cite{shavitt2009many}, and CCSD(T) \cite{Raghavachari:1989vx}.
In the DSRG computations, we employ CASSCF(6,6) orbitals and use the corresponding state as a reference, while the CCSD and CCSD(T) results employ restricted Hartree--Fock references.
The DSRG methods produce curves that are nearly indistinguishable, except in the recoupling region (1.6--2.0 \AA{}) where the \name energy is slightly higher than the LDSRG(2) one.
In comparison, the CCSD and CCSD(T) curves, although accurate in the equilibrium region, deviate significantly from the DSRG one for large N--N distances.
In the bottom panel of Figure~\ref{fig_n2} we report the energy error with respect to FCI.
Here we notice that the 1- and 2-QDSRG(2) lead to errors as large as 17.5 m\Eh, while the LDSRG(2) is more accurate, with the maximum deviation from FCI being always less than 10 m\Eh.
As observed for \ce{H2}, the 1-QLDSRG(2) potential energy curve based on the CCSD-NOs reference is still accurate and display maximum errors with respect to FCI smaller than those obtained using CASSCF orbitals and cumulants from exact diagonalization.


\subsection{Singlet-triplet gaps of \textit{para}-benzyne}
\label{PBenzyne}
\begin{table*}[ht!]
\caption{\label{tab: pbenzyne} 
		\linespread{1}\small
	Adiabatic singlet-triplet splittings ($\stgap = E_\mathrm{T} - E_\mathrm{S}$) in \kcal of \textit{para}-benzyne computed with the LDSRG(2) and QLDSRG(2). All computations use the cc-pVDZ basis \cite{cc-pvdz-dk-H-Ne} and the flow parameter value $s$ = 1.0 \sunit. 
	All computational results include a zero-point vibrational energy (ZPVE) correction equal to $+0.30$ \kcal. Geometries and the ZPVE correction are taken from Ref.~\cite{Evangelista:2012fo}.
	We use CASSCF(2,2) optimized orbitals for all computations of the singlet state, and ROHF orbitals for the triplet state.
	For the LDSRG(2), the expectation value of the energy is designated with ''$\braket{\bar{H}}$'', while the lowest eigenvalue is indicated with ''eig. $\bar{H}$''. For the \name methods we report two sets of data. The ones labeled ``$\rdm{3} = 0$'' use an approximate three-body cumulant reconstructed from approximate $\tilde{\boldsymbol\lambda}_1$ and $\tilde{\boldsymbol\lambda}_2$.  	}
\begin{ruledtabular}
\begin{tabular}{@{} lccc @{}}
Method 	&	$\stgap$ / \kcal	& \mbox{ $E_\mathrm{T}$ /\Eh}  & \mbox{$E_\mathrm{S}$ /\Eh} 	\\
\hline
\multicolumn{4}{c}{Active orbitals: $\{\sigma_g, \sigma_u\}$ }  \\							
CASSCF(2,2) 	&	0.27	&	-229.416126	&	-229.416074	\\
LDSRG(2) ($\braket{\bar{H}}$)   	&	2.72	&	-230.194973	&	-230.198828	\\
LDSRG(2) (eig. $\bar{H}$)    	&	3.42	&	-230.194973	&	-230.199944	\\
2-QLDSRG(2) ($\rdm{3} = 0$) 	&	3.15	&	-230.194762	&	-230.199308	\\
1-QLDSRG(2) ($\rdm{3} = 0$) 	&	3.15	&	-230.191807	&	-230.196354	\\
2-QLDSRG(2)  	&	2.69	&	-230.194762	&	-230.198567	\\
1-QLDSRG(2) 	&	3.30	&	-230.194411	&	-230.199197	\\ [3pt]
\multicolumn{4}{c}{Active orbitals: $\{\sigma_g, \sigma_u, 6 \times \pi \}$ } 	\\						
LDSRG(2) (eig. $\bar{H}$) 	&	2.61	&	-230.196267	&	-230.199943	\\
2-QLDSRG(2)	&	2.85	&	-230.196262	&	-230.200325	\\
1-QLDSRG(2) 	&	4.16	&	-230.196225	&	-230.202375	\\
\hline
Experiment \cite{Leopold:1986vo}   & $3.8 \pm 0.4$   &  & \\

\end{tabular}
\end{ruledtabular}
\end{table*}

In our next example, we apply the \name scheme to a medium-sized molecule.
We compute the adiabatic singlet-triplet splitting ($\Delta E_\mathrm{ST} = E_\mathrm{T} - E_\mathrm{S}$) of \textit{para}-benzyne.
The singlet ground state of this molecule exhibits pronounced diradical character and is dominated by two closed-shell determinants.
\textit{para}-benzyne and its isomers, have been studied extensively both in experiments \cite{Leopold:1986vo} and in theory \cite{Evangelista:2007hz, Evangelista:2012fo, Hanauer:2012gf, Li:2008tk, Cramer:1997uh, Lindh:1999um, Slipchenko:2002un, li2007s1, wang2008extended}.

Here we compute the singlet-triplet splitting using CASSCF(2,2) orbitals for the singlet state and ROHF orbitals for the triplet state.
The experimental splitting is taken from the ultraviolet photoelectron spectroscopy results of Ref.~\cite{Leopold:1986vo}.
We utilize the singlet and triplet geometries from Ref.~\cite{Evangelista:2012fo}, which were optimized at the Mk-MRCCSD/cc-pVTZ level of theory.
All computations use the cc-pVDZ basis set \cite{cc-pvdz-dk-H-Ne}, and the value of the DSRG flow parameter is set to 1.0 \sunit, based on previous studies \cite{Li:2016hb, Wang.2021.10.1021/acs.jctc.1c00980, Wang:2019kf}.
We freeze the six 1s-like orbitals on carbon atoms in the DSRG correlation treatment.

Table~\ref{tab: pbenzyne} reports the singlet-triplet splitting obtained by the LDSRG(2) and the QLDSRG(2) methods. All splittings are shifted by 0.30 \kcal to account for zero-point vibrational energy (ZPVE) corrections \cite{Evangelista:2012fo}.
Labels for methods are consistent with those in Table~\ref{tab: h2_benchmark}.
For the active space consisting of two $\sigma$ orbitals, compared to the LDSRG(2) average energy $\braket{\bar{H}}$ (2.72 \kcal), diagonalizing the LDSRG(2) $\bar{H}$ (3.42 \kcal) improves the singlet-triplet gap by 0.7 \kcal.
For the \name methods, interestingly, the more drastic approximation 1-QLDSRG(2) (3.30 \kcal) gives better prediction than the 2-QLDSRG(2) (2.69 \kcal), which might be attributed to error cancellation.
Imposing $\rdm{3} = 0$ seems to have different effects on two approximations: 
the error worsens for the 1-QLDSRG(2) by 0.15 \kcal while improves for the 2-QLDSRG(2) by 0.46 \kcal.
Unlike the \ce{H2} cases reported in Table~\ref{tab: h2_benchmark}, for this problem, we do not see consistent improvements of results when enlarging the active space by adding six $\pi$ orbitals from the set of CASSCF(2,2) orbitals.

\subsection{Hardware implementation}
\label{Hardware}

In this section, we combine the 2-QLDSRG(2) method with the VQE \cite{Peruzzo:2014kc, Yung:2014iv, McClean:2015bs} on the IBM quantum computers to demonstrate the ability of this hybrid scheme to compute the total energies under realistic noise from near-term quantum devices.
We use the \texttt{Qiskit} \cite{Qiskit} package to construct circuits and execute them on hardware.

Ideally we would measure both density matrices and the \name energy (steps 2 and 4 in Figure~\ref{fig:vqe-dsrg}) from a quantum computation. 
Due to the high level of noise from near-term devices and the fact that density matrices are more sensitive to noise than the energy, we employ a quantum computer only to estimate the eigenvalue of the 2-QLDSRG(2) effective Hamiltonian $\bar{H}$.
We use the VQE algorithm to optimize a trial wave function and measure its energy.
To reduce the quantum resources (the number of qubits, the circuit depth, etc) and minimize errors, 
we explore a symmetry-preserving one-qubit ansatz (see Appendix \ref{ansatz} for details).

For each experiment on the device, we carry out the maximum number of measurements allowed, which differs by device.
To ameliorate measurement errors, we utilize readout-error-mitigation tools in the \texttt{Qiskit-Ignis} module to construct a calibration matrix and apply its inverse to the raw measurement counts of each experiment.

Our first example is a computation of the dissociation curve of \ce{H2}, which is a representative benchmark system for quantum computing. 
Figure~\ref{fig:vqe-dsrg} shows the dissociation curve and the energy error for the \ce{H2} molecule in the cc-pV5Z basis \cite{cc-pvdz-dk-H-Ne} (110 orbitals) obtained by the one-qubit 2-QLDSRG(2) computations on the \texttt{ibmq\_lagos} quantum computer.
A direct second-quantized quantum computation would require 220 qubits (ignoring qubit tapering or other symmetry adaptation techniques).
We also report the 2-QLDSRG(2) energy errors and the standard deviations of the device results in Table~\ref{tab:h2_hardware}.
The effectiveness of the \name downfolding method can be seen from the small errors of the 2-QLDSRG(2) energies, which differ from noiseless simulations at most by 0.5 m\Eh for all geometries.
The 2-QLDSRG(2) energies from the device have unsigned average errors lower than 1 m\Eh for over half of the geometries, with a maximum error of 2.0 m\Eh. 
Empirically, it is important to collect the measurement statistics of $10^5$ shots to obtain a reliable estimate of the average energy from the device.
From this example, we see that errors from hardware (finite measurements, decoherence, etc.) are more significant than errors from the \name downfolding.

\begin{figure}[htbp]
   \centering
    \includegraphics[width=3.375in]{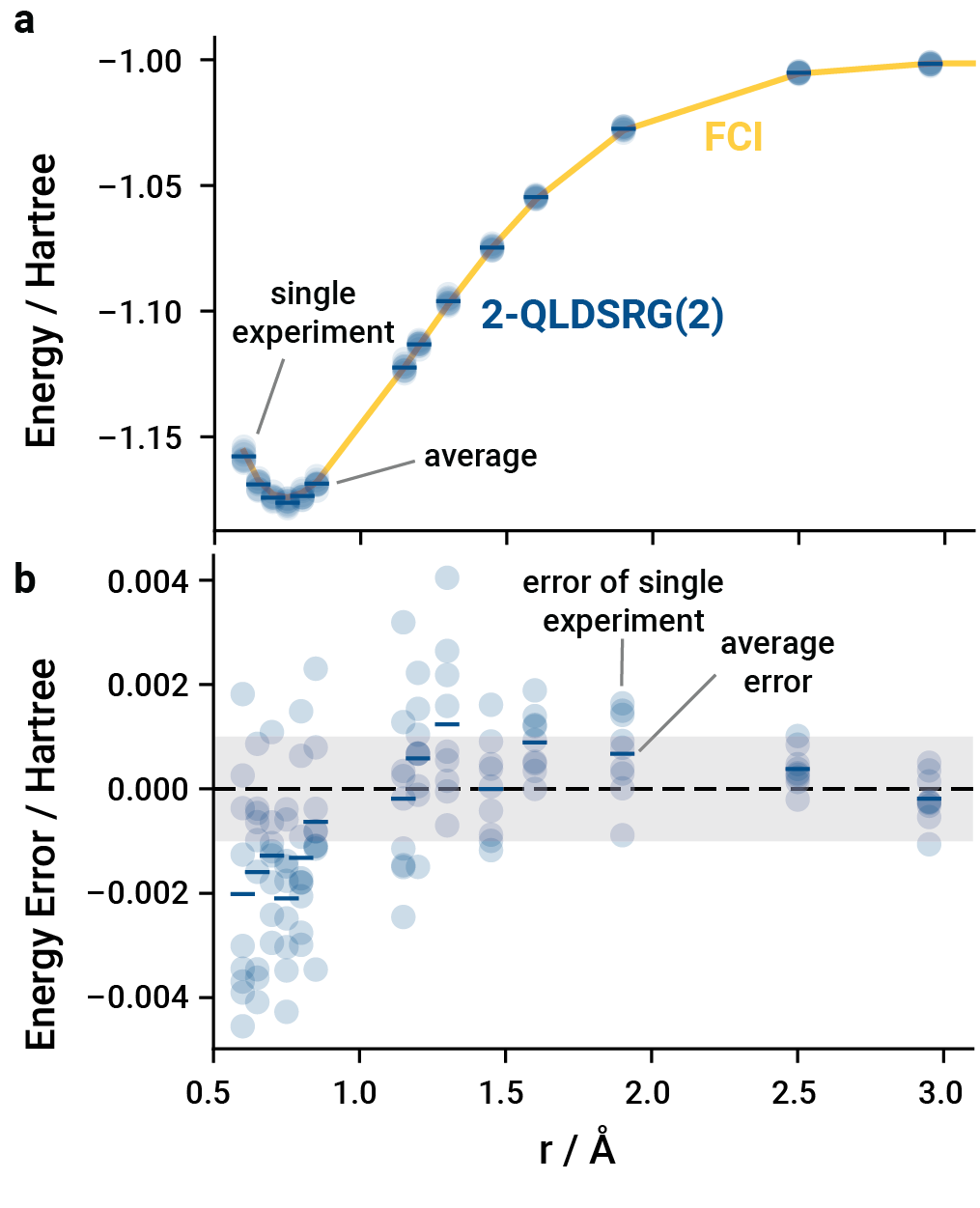}   
   \caption{\label{fig_h2_hardware}  	\linespread{1}\small
   	Dissociation curve (a) and energy error (b) for the hydrogen molecule computed with the 2-QLDSRG(2) using one qubit on the \texttt{ibm\_lagos} device.
   	Energy errors are with respect to FCI energies.
   	For each geometry, semi-opaque blue circles aligned vertically show the distribution of energies (energy errors) from 9 experments, with each experiemt consisting of 32000 measurements, while horizontal bars in blue denote average energies or average energy errors.
  All computations use the cc-pV5Z basis \cite{cc-pvdz-dk-H-Ne} (110 basis functions), CASSCF(2,2) orbitals, and the flow parameter value $s$ = 0.5 \sunit.
  The grey-shaded area indicates unsigned energy errors below 1 m\Eh.
  The unsigned energy errors and the standard deviations are reported in Table~\ref{tab:h2_hardware}.
 }
\end{figure}
\begin{table}
	\renewcommand{\arraystretch}{0.9}
	\footnotesize
	\caption{\label{tab:h2_hardware}   \linespread{1}\small
	The errors of 2-QLDSRG(2) energies in m\Eh (with respect to FCI energies) along the \ce{H2} dissociation curve (Figure~\ref{fig_h2_hardware}). For results from the \texttt{ibm\_lagos} device, we show the \textbf{unsigned} average energy errors and standard deviations (in m\Eh). Unsigned energy errors blow 1 m\Eh are highlighted in bold font. 
	}
\scriptsize
\begin{tabular}{@{} lccc@{}}
\hline
\hline
	$r/$\AA{}	&	$\Delta E_\mathrm{noiseless}$	&	$\Delta E_\mathrm{device}$	&	Standard Deviation	\\
\hline
	0.6	&	0.53	&	2.02	&	2.08	\\
	0.65	&	0.54	&	1.59	&	1.63	\\
	0.7	&	0.54	&	1.28	&	1.14	\\
	0.75	&	0.55	&	2.10	&	1.24	\\
	0.8	&	0.54	&	1.32	&	1.41	\\
	0.85	&	0.54	&	\textbf{0.63}	&	1.47	\\
	1.15	&	0.40	&	\textbf{0.19}	&	1.61	\\
	1.2	&	0.37	&	\textbf{0.59}	&	1.00	\\
	1.3	&	0.32	&	1.24	&	1.42	\\
	1.45	&	0.28	&	\textbf{0.00}	&	0.89	\\
	1.6	&	0.29	&	\textbf{0.89}	&	0.56	\\
	1.9	&	0.36	&	\textbf{0.68}	&	0.77	\\
	2.5	&	0.20	&	\textbf{0.38}	&	0.35	\\
	2.95	&	0.07	&	\textbf{0.19}	&	0.45	\\
	6.0	&	0.00	&	\textbf{0.31}	&	0.15	\\
\hline
\hline
\end{tabular}
\end{table}

\begin{figure*}[htbp]
   \centering
    \includegraphics[width=4.5in]{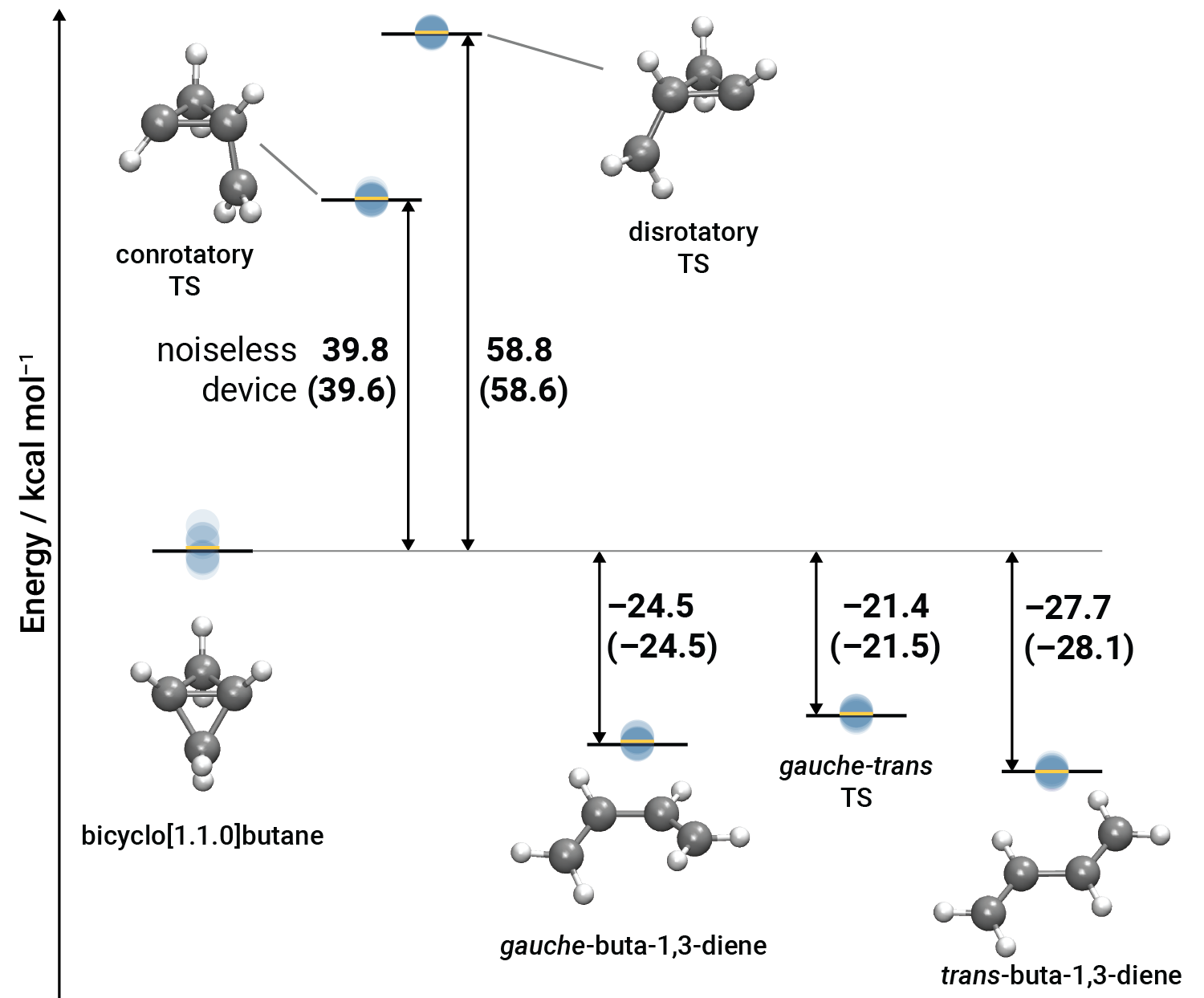}   
   \caption{\label{fig_iso_hardware}    \linespread{1}\small
   	Conrotatory and disrotatory pathways describing the isomerization of bicyclo[1.1.0]butane (bicyclobutane) to \textit{trans}-buta-1,3-diene (\textit{trans}-butadiene). Enthalpies in \kcal are relative to the reactant for relevant stationary points computed with the 2-QLDSRG(2) method using one qubit on the device \texttt{ibmq\_manila}. 
   	Black horizontal bars represent the 2-QLDSRG(2) results obtained with noiseless simulations.
   	The device results are shown in the parentheses.   	
   	Semi-opaque blue circles aligned vertically show the distribution of the relative enthalpies from 8 experiments (20000 measurements per experiment), while yellow horizontal bars denote the average relative enthalpies.
   	We use a cc-pVTZ basis \cite{cc-pvdz-dk-H-Ne} (204 basis functions) and CASSCF(2,2) natural orbitals, and the flow parameter value $s$ = 1.0 \sunit for all six stationary points.
    }
\end{figure*}
\begin{table*}[ht!]
\setlength{\tabcolsep}{6pt}
	\renewcommand{\arraystretch}{0.9}
	\footnotesize
	\caption{\label{tab:iso_hardware} 	\linespread{1}\small
	   LDSRG(2) and QLDSRG(2) (exact and device) computations of the relative enthalpies (in \kcal) with respect to the bicyclobutane reactant of the conrotatory transition state and the disrotatory transition state, the \textit{gauche}-butadiene intermediate, the transition state connecting \textit{gauche}-butadiene and \textit{trans}-butadiene, and the \textit{trans}-butadiene product. 
		We use the cc-pVTZ basis \cite{cc-pvdz-dk-H-Ne} (204 basis functions) and CASSCF(2,2) natural orbitals, and the flow parameter value $s$ = 1.0 \sunit.
		All results include zero-point vibrational energy (ZPVE) corrections taken from Ref. \cite{kinal2007computational}.
		OMR3-DMC and CC(t;3) results are also in the cc-pVTZ basis.
		}
	\begin{threeparttable}
		\begin{tabular}{@{} lccccc@{}}
			\hline
			\hline
			   & conrotatory TS   &disrotatory TS     &\textit{gauche}-butadiene   & \textit{gauche}-\textit{trans} TS     &\textit{trans}-butadiene   \\
			\hline
			LDSRG(2) ($\braket{\bar{H}}$) 	&	41.3	&	58.5	&	-24.5	&	-21.5	&	-27.7	\\
			LDSRG(2) (eig. $\bar{H}$)	&	39.6	&	58.6	&	-24.5	&	-21.4	&	-27.8	\\
			\multicolumn{6}{c}{2-QLDSRG(2)} \\											
			Exact	&	41.5	&	58.9	&	-24.3	&	-21.3	&	-27.5	\\
			\texttt{ibmq\_manila}\tnote{a}	&	42.6	&	57.8	&	-23.8	&	-22.2	&	-27.4	\\
			\texttt{ibm\_lagos}\tnote{b}	&	42.7	&	59.6	&	-23.3	&	-20.7	&	-26.4	\\
            \texttt{ibmq\_jakarta}\tnote{c}	&	40.5	&	57.5	&	-25.6	&	-22.4	&	-28.4	\\
			\multicolumn{6}{c}{2-QLDSRG(2) ($\rdm{3} = 0$)} \\											
			Exact	&	39.8	&	58.8	&	-24.5	&	-21.4	&	-27.7	\\
			\texttt{ibmq\_manila}\tnote{a}	&	\textbf{39.6}	&	\textbf{58.6}	&	\textbf{-24.5}	&	\textbf{-21.5}	&	\textbf{-28.1}	\\
			\texttt{ibm\_lagos}\tnote{b}	&	39.0	&	57.7	&	-25.2	&	-22.2	&	-28.6	\\
			\texttt{ibmq\_jakarta}\tnote{c}	&	39.1	&	58.4	&	-25.2	&	-21.7	&	-28.0	\\
			[3pt]
			
			OMR3-DMC\tnote{d} & 40.4(5) & 58.6(5) & -25.2(5) &-22.2(5) & -27.9(5) \\
			CC(t;3)\tnote{e} & 40.2 & 60.1 & -25.3 & -22.6 & -28.3 \\  
			
			Experiment &$40.6 \pm 2.5$\tnote{f} & & & & $-25.9 \pm 0.4$\tnote{g}  \\
			\hline
			\hline
		\end{tabular}
		\begin{tablenotes}
			\linespread{1}\small 
			\item [a] The average is over 8 experiments (20000 shots per experiment).
			\item [b] The average is over 10 experiments (32000 shots per experiment).
			\item [c] The average is over 10 experiments (20000 shots per experiment).
			\item [d]  From Ref.~\cite{berner2010isomerization}.
			\item [e]  From Ref.~\cite{shen2012combining}.
			\item [f] From Ref.~\cite{srinivasan1965thermal}.
			\item [g] The reaction enthalpy at 298 K is based on enthalpies of the formation of bicy-clo[1.1.0]butane and buta-1,3-diene reported in Ref.~\cite{wiberg1968heats}.
		\end{tablenotes}
	\end{threeparttable}
\end{table*}


For our second set of hardware experiments, we consider a larger and more chemically-relevant problem, the pericyclic rearrangement reaction of bicyclo[1.1.0]butane (bicyclobutane) to \textit{trans}-buta-1,3-diene (\textit{trans}-butadiene). 
This isomerization process goes through a concerted conrotatory movement of the methylene groups with an activation barrier (enthalpy) of $40.6 \pm 2.5$ \kcal \cite{srinivasan1965thermal}, suggested by early experimental studies \cite{srinivasan1965thermal, blanchard1966bicyclo, frey1965thermal, wiberg1966formation, closs1968steric}.
This mechanism has been investigated in many computational studies using high-level electronic structure methods, including 
MRMBPT \cite{kinal2007computational, mazziotti2008energy, nguyen1995isomerization},
MRCI \cite{nguyen1995isomerization}, 
variants of single-reference coupled-cluster methods including 
CR-CC(2,3) \cite{kinal2007computational, lutz2008extrapolating} and CC(t;3) \cite{shen2012combining}, 
the diffusion quantum Monte Carlo \cite{berner2010isomerization}
and the ACSE method \cite{mazziotti2008energy, boyn2022elucidating}.

These theoretical studies also investigate the unfavored concerted disrotatory pathway \cite{kinal2007computational, lutz2008extrapolating, mazziotti2008energy, berner2010isomerization, nguyen1995isomerization}, 
characterized by a transition state that is estimated to be 15--25 \kcal higher in energy than the conrotatory one. 
Both transition states display significant biradical character \cite{kinal2007computational} and their ground state wave functions have large contributions from multiple determinants, requiring a multireference treatment. 
This makes the system suitable to treat with the \name method.
Previous studies also confirmed that for both concerted pathways, the reaction will continue to reach an intermediate \textit{gauche}-buta-1,3-diene intermediate (\textit{gauche}-butadiene), and then proceeds through a low-energy rotational barrier to get to the \textit{trans}-butadiene product.

We compute the reaction enthalpies along the full concerted conrotatory and disrotatory pathways from bicyclobutane to \textit{trans}-butadiene. 
Cartesian coordinates of the structures of all six stationary points optimized at the CASSCF(10,10)/cc-pVDZ level of theory are taken from Ref.~\cite{kinal2007computational}.
Zero-point vibrational energies (ZPVE) obtained at the same level of theory are used to convert total electronic energies to enthalpies.
For all six stationary points, we use CASSCF(2,2) natural orbitals, which yield a two-configuration reference that can be mapped to a one-qubit ansatz (see Appendix \ref{ansatz} for details).
All computations use the cc-pVTZ basis \cite{cc-pvdz-dk-H-Ne} (204 orbitals)
and we freeze four 1s-like orbitals on carbon atoms in the DSRG correlation treatment.

Figure~\ref{fig_iso_hardware} shows the concerted conrotatory and disrotatory pathways of the bicyclobutane $\rightarrow$ \textit{trans}-butadiene reaction and Table~\ref{tab:iso_hardware} reports the relative enthalpies (with respect to the bicyclobutane reactant) from the LDSRG(2) and the 2-QLDSRG(2) methods obtained via noiseless simulations and via VQE on three quantum devices. We show the best device results in Figure~\ref{fig_iso_hardware}.

Compared to the experimental value \cite{srinivasan1965thermal},
both LDSRG(2) and the 2-QLDSRG(2) methods give relative enthalpies of the conrotatory transition state that are in good agreement, while the relative enthalpies of \textit{trans}-butadiene predicted by the two methods are slightly underestimated \cite{wiberg1968heats}.
For the other three stationary points, experimental data are not available; therefore, we compare our results with data from two high-level approaches, the optimized multireference diffusion quantum Monte Carlo (OMR3-DMC) \cite{berner2010isomerization} and an active-space coupled-cluster method with corrected triple excitations termed CC(t;3) \cite{shen2012combining} using the same cc-pVTZ basis. 
The relative enthalpy of the disrotatory transition state predicted by the LDSRG(2) and 2-QLDSRG(2) methods agrees with the OMR3-DMC result (58.6 \kcal) \cite{berner2010isomerization} and are slightly lower than the CC(t;3) value (60.1 \kcal) \cite{shen2012combining}.
For both \textit{gauche}-butadiene and the transition state connecting \textit{gauche}-butadiene and \textit{trans}-butadiene, the LDSRG(2) and the 2-QLDSRG(2) results are about 1 \kcal lower than the OMR3-DMC and CC(t;3) values.      
The 2-QLDSRG(2) results from three devices are in good agreement with the result from noiseless simulations, with most devices yielding values within 1 \kcal from the exact result. The best device results (data in bold font in Table~\ref{tab:iso_hardware}) give unsigned errors less than 0.5 \kcal for all six stationary points. 

The results for the bicyclobutane $\rightarrow$ \textit{trans}-butadiene reaction demonstrate that the \name method can effectively downfold the dynamical correlation for a large basis with 204 orbitals, reducing the number of qubits from several hundreds to just one.

\section{Conclusion}
\label{conclusion}

In this work, we introduced a practical unitary downfolding method that enables accurate molecular computations on near-term quantum computers.
The \name is agnostic to the type of quantum algorithm (e.g., variational, phase estimation) and can be used with both noisy near-term computers and future fault-tolerant hardware.
Therefore, we expect that the \name will be a useful method to leverage small quantum computers in applications to large molecules and large basis sets.

The \name is based on the driven similarity renormalization group (DSRG) \cite{evangelista2014driven}, a classical numerically-robust and polynomial-scaling approach to block-diagonalize many-body Hamiltonians.
In this work, we propose a ``diagonalize-then-dress-then-diagonalize'' strategy that combines truncation of the reduced density cumulants provided to the DSRG with diagonalization of the resulting similarity-transformed Hamiltonian.
This downfolding procedure may be justified by a perturbative analysis of the DSRG equations and leads to two practical computational schemes: in the 1-\name we retain only the diagonal part of the one-body RDM, whereas in the 2-\name we retain the full one-body RDM and the diagonal part of the two-body cumulants.
These two schemes require the estimation of a number of reduced density matrix elements that is at most linear or quadratic in the number of active orbitals ($N_\mathrm{A}$), substantially reducing the demands of conventional multireference theories, which require $N_\mathrm{A}^6$ to $N_\mathrm{A}^8$ RDM elements.

Our calibration of the \name shows that the use of orbitals optimized for a reference correlated state is crucial to compute accurate energies.
The \name results show that the 2- approximation is able to accurately predict energies along the bond-breaking coordinate in a minimal active space. The 1- approximation leads to larger errors, but these can be suppressed by increasing the active space size.
To simulate the effect of noise, we examined \name computations starting with inaccurate RDMs and found that milliHartree accuracy can be retained when the standard deviation of the RDMs errors is as large as $10^{-3}$--$10^{-2}$.

In our computations on the more challenging \ce{N2} and \textit{p}-benzyne molecules, we were also able to accurately predict potential energy curves and singlet-triplet gaps using the \name.
In the case of \ce{N2}, we demonstrate how the first two steps of the \name procedure (orbital optimization and reference preparation) could be approximated with the classical polynomial-scaling CCSD method, using the corresponding one-body reduced density matrix.
Finally, we demonstrate the \name procedure in combination with the VQE algorithm on the IBM quantum devices.
We extend computations of the \ce{H2} dissociation curve with a nearly-complete quintuple-$\zeta$ basis, corresponding to a full computation with 220 qubits.
In this example, we find that hardware errors still remain the most significant source of error in comparison to the \name  downfolding error.
We also apply the \name method to model the reaction pathways of the bicyclobutane $\rightarrow$ \textit{trans}-butadiene isomerization process using a basis of 204 orbitals.
We are able to obtain high-quality device results that reach sub-\kcal accuracy with respect to the exact \name and two high-level classical electronic structure approaches.

The extension of the \name with explicit correlation methods and to electronically excited states are two interesting extension worth exploring.
We expect that with the availability of more accurate and larger number of qubits, the \name will provide a systematic path to perform accurate quantum chemistry computations on chemically relevant systems.
%
%


\section*{Acknowledgements}
This work was supported by the U.S. Department of Energy under Award No.  DE-SC0019374.
The authors thank Nicholas Stair and Nan He for their contribution to the definition and implementation of the integral exchange format in \textsc{Forte}, and Jonathon Misiewicz and Tom O'Brien for helpful conversations.

\appendix

\section{Symmetry-preserving ansatz for two-configuration wave functions}
\label{ansatz}

We exploit spin, particle number and spatial symmetries to construct a hardware-efficient ansatz for two-configuration wave functions.
Consider a two electron wave function in a basis of two molecular orbitals $\psi_1$, $\psi_2$.
The singlet ground state in the most general case includes three configurations (bars over the number denote $\beta$ spin orbitals),
\begin{equation}
\begin{split}
	\ket{\Phi_1} &= \ket{ \psi_1 \psi_{\bar{1}}  },  \\
	\ket{\Phi_2} &= \ket{ \psi_2 \psi_{\bar{2}}  },  \\
	\ket{\Phi_3} &=  \frac{1}{\sqrt{2}} ( \ket{ \psi_1 \psi_{\bar{2}}  }  -  \ket{  \psi_{\bar{1}} \psi_2 } ). \\
\end{split}
\end{equation}
We can remove the contribution of the open-shell configuration $\ket{\Phi_3}$ from the normalized ground-state wave function $\ket{\Psi_0} = C_1^\prime \ket{\Phi_1} + C_2^\prime \ket{\Phi_2} + C_3^\prime \ket{\Phi_3}$ without changing the energy via an orbital rotation
\begin{equation}
	\label{orb_rot}
	\begin{split}
		\ket{\psi_1^\prime} &=  \cos \theta \ket{\psi_1}  +  \sin \theta \ket{\psi_2},  \\
		\ket{\psi_2^\prime} &=  \sin \theta \ket{\psi_1}  -  \cos \theta \ket{\psi_2},  \\
	\end{split}
\end{equation}
where $\tan 2\theta = \frac{\sqrt{2}  C_3^\prime }{C_1^\prime - C_2^\prime}$ \cite{Allen:1986tf, Allen:1987uu, Evangelista:2007hz}.
We refer interested readers to Ref.~\cite{Evangelista:2007hz} (Sec IV.) for detailed discussions on this basis transformation.

For \ce{H2}, the ground-state wave function expanded in the CASSCF(2,2) orbitals can be accurately described by two closed-shell configurations. 
While for two transition states in the bicyclobutane $\rightarrow$ \textit{trans}-butadiene reaction, the open-shell contribution to the wave function cannot be neglected due to strong biradical character (especially for the disrotatory TS). Therefore, we transform to the CASSCF(2,2) natural orbital basis, which is mathematically equivalent to enforcing Eq.~\ref{orb_rot}. 
The ground states of other four stationary points in the pathways are generally well described by a single determinant in the CASSCF(2,2) basis; however, for consistency, we employ CASSCF(2,2) natural orbitals for all computations.

The resulting two-configuration wave function
$\ket{\Psi_0} = C_1  \ket{ \psi_1^\prime  \psi_{\bar{1}}^\prime  } + C_2  \ket{ \psi_2^\prime  \psi_{\bar{2}}^\prime }$ can be mapped to a one-qubit space
\begin{equation}
\begin{split}
     \ket{ \psi_1^\prime  \psi_{\bar{1}}^\prime  } &\rightarrow \ket{0},  \\
     \ket{ \psi_2^\prime  \psi_{\bar{2}}^\prime } &\rightarrow \ket{1}, \\
\end{split}
\end{equation}
which leads to the one-qubit wave function ansatz
\begin{equation}
    \ket{\Psi} = C_1\ket{0} + C_2\ket{1}.
\end{equation}
This state can be prepared by applying to the $\ket{0}$ state a single Y-rotation gate $\hat{R}_{y}(t)=e^{-\frac{i t Y}{2}}$ with one variational parameter $t$, giving $C_1= \cos{\frac{t}{2}}$, $C_2 = \sin{\frac{t}{2}}$. The Hamiltonian in the one-qubit basis is represented as
$
    H =h_{00}|0\rangle\langle 0|+h_{11}|1\rangle\langle1| +h_{10} (| 0\rangle\langle 1|+ | 1\rangle\langle 0|)
$
where $h_{00}$, $h_{11}$, $h_{10}$ are calculated from the one- and two-electron integrals.
The Hamiltonian can be decomposed into a weighted sum of single-qubit Pauli operators
\begin{equation}
   H = c_0 + c_z Z+ c_x X,
\end{equation}
with coefficients given by $c_0 = (h_{00}+h_{11})/2$, $c_z = (h_{00}-h_{11})/2$, $c_x = h_{10}$.

The expectation value of this one-qubit Hamiltonian with respect to $ \ket{\Psi}$ has the definite tomography \cite{parrish2019jacobi, Nakanishi:2020in} given by
\begin{equation}
	\braket{H}_t=a+b \cos t+c \sin t,
\end{equation}
The coefficients $a, b, c$ can be found using a three-point Fourier quadrature \cite{parrish2019jacobi} that requires measuring expectation values for three parameters $t_0,\ t_1, \ t_2$.
The corresponding linear equation to solve is:
\begin{equation*}
	\left( \begin{array}{ccc}
		1 & \langle Z\rangle_{t_{0}} & \langle X\rangle_{t_{0}} \\
		1 & \langle Z\rangle_{t_{1}} & \langle X\rangle_{t_{1}} \\
		1 & \langle Z\rangle_{t_2} & \langle X\rangle_{t_{2}} 
	\end{array} \right)
	\left(\begin{array}{c}
		a \\
		b \\
		c
	\end{array}\right)
	=\left(\begin{array}{c}
		\braket{H}_{t_{0}} \\
		\braket{H}_{t_{1}} \\
		\braket{H}_{t_{2}}
	\end{array}\right) .
\end{equation*}
In this work, we use the following three-point Fourier quadrature: 
\begin{equation}
	t_0, t_0 - \pi/3, t_0 + \pi/3 ,
\end{equation}
where $t_0$ is arbitrary. For convenience, we use the analytic solution for the optimal angle
\begin{equation}
	t_0 = \mathrm{arctan2} \left(\frac{c_{x}}{c_{z}}\right).
\end{equation}

For the reference preparation (step 2 in Figure~\ref{fig:vqe-dsrg}), we run the VQE algorithm to obtain the 1- and 2-RDMs.
These quantities can be measured from the state tomography of the optimal wave function. Table~\ref{tab:1q_rdm} summarizes the expressions for the 1- and 2-RDM in terms of analytical expressions of the variational parameter $t$ and quantities from direct measurements. 

Note that for both the 1- and 2-QDSRG methods, we only need to measure the Pauli $Z$ operator to compute the full 1-RDM and the approximate 2-RDM, while the Pauli $X$ operator only contributes to the non-diagonal components of the two-body reduced density cumulant.  
Finally, we use the VQE algorithm to estimate the eigenvalue of the DSRG effective Hamiltonian (step 4 in Figure~\ref{fig:vqe-dsrg}).

\begin{table}
\renewcommand{\arraystretch}{0.9}
  \caption{\label{tab:1q_rdm} 	
  \linespread{1}\small
   	Relations between elements of the fermionic 1-RDM ($\gamma^p_q=\braket{ a_p^\dagger a_q}$) and 2-RDM ($ \gamma^{p q}_{r s}=\braket{a_{p}^{\dagger} a_{q}^{\dagger} a_{s} a_{r}}$) and the measured quantites for the one-qubit ansatz. Only nonzero elements are shown; others are zero due to symmetries.
  	$|C_1|^2 = \braket{\Psi | 0} \braket{0|\Psi}$, $|C_2|^2 =  \braket{\Psi | 1} \braket{1|\Psi}$ are obtained from projective measurements of the optimized state in the computational basis;
  	$\braket{X}$ is the expectation value of the Pauli $X$ operator.
  }
  \begin{tabular}{lcc}
	\hline
  	RDM Element & Analytic & Measurement \\
  	\hline
  	$\gamma_1^1, \gamma_{\overline{1}}^{\overline{1}}$,\quad $\gamma_{\overline{1} 1}^{1 \overline{1}}, \gamma_{1 \overline{1}}^{\overline{1} 1}$ & $\cos^2{\frac{t}{2}}$      & $|C_1|^2$ \\
  	$\gamma_2^2$, $\gamma_{\overline{2}}^{\overline{2}}$,\quad $\gamma_{\overline{2} 2}^{2 \overline{2}}, \gamma_{2 \overline{2}}^{\overline{2} 2}$ & $\sin^2{\frac{t}{2}}$    & $|C_2|^2$ \\
  	$\gamma_{\overline{1} 1}^{\overline{1} 1}, \gamma_{1 \overline{1}}^{1 \overline{1}}$ & $ - \cos^2{\frac{t}{2}}$      & $ - |C_1|^2$ \\
	$\gamma_{\overline{2} 2}^{\overline{2} 2}, \gamma_{2 \overline{2}}^{2 \overline{2}}$ & $-\sin^2{\frac{t}{2}}$      & $-|C_2|^2$ \\
	$\gamma^{1 \overline{1}}_{\overline{2} 2}, \gamma^{\overline{1} 1}_{2 \overline{2}}$ $\gamma^{2 \overline{2}}_{\overline{1} 1}, \gamma^{\overline{2} 2}_{1 \overline{1}}$ & $\cos{\frac{t}{2}} \sin{\frac{t}{2}}$  & $\braket{X}/2$ \\
	$\gamma^{\overline{1} 1}_{\overline{2} 2}, \gamma^{1 \overline{1}}_{2 \overline{2}}$
  	$\gamma^{\overline{2} 2}_{\overline{1} 1}, \gamma^{2 \overline{2}}_{1 \overline{1}}$
  	& $-\cos{\frac{t}{2}} \sin{\frac{t}{2}}$  & $-\braket{X}/2$ \\
	\hline
  \end{tabular}
\end{table}

\section{Comparisons with the double unitary coupled cluster approach}
\label{compare_ducc}
\begin{table*}[ht!]
	\renewcommand{\arraystretch}{0.9}
	\footnotesize
	\caption{\label{tab:h2_ducc} 	\linespread{1}\small
		Comparison of the energy errors (in m\Eh) of the \name and the DUCC for the hydrogen molecule at four bond lengths in the cc-pVTZ basis using DSRG flow parameter $s$ = 0.5 \sunit. 
		The size of the active space is denoted in the parenthesis. 
		The energy errors are with respect to the full-space (30-orbital) FCI computations (absolute energies are shown in the first row).
		The DUCC data are taken from Ref.~\cite{Bauman:2019dx}.
	}
	\begin{ruledtabular}
		\scriptsize
		\begin{tabular}{@{} lccccc@{}}
			Method	&	Orbital Type		($N_\mathrm{act}$)		&	0.8~a.u.	&	1.4008~a.u.	&	4.00~a.u.		&	10.00~a.u.		\\
			\hline
			FCI (full space) 	&	RHF	(	30	)	&	-1.015729	&	-1.172455	&	-1.014872	&	-0.999623	\\
			FCI (active space) 	&	RHF	(	4	)	&	32.729	&	25.755	&	7.872	&	2.523	\\
			DUCC	&	RHF	(	4	)	&	7.129	&	4.555	&	-0.328	&	-1.977	\\
			LDSRG(2) (diag. $\bar{H}$)    	&	RHF	(	4	)	&	-0.130	&	-0.234	&	1.920	&	0.122	\\
			1-QLDSRG(2)	&	RHF	(	4	)	&	-0.094	&	-0.098	&	4.047	&	0.524	\\
			2-QLDSRG(2)	&	RHF	(	4	)	&	-0.087	&	-0.084	&	3.794	&	0.126	\\
			1-QLDSRG(2)	&	 CASSCF(2,2)      	(	4	)	&	0.455	&	0.169	&	1.498	&	0.006	\\
			2-QLDSRG(2)	&	 CASSCF(2,2)      	(	4	)	&	0.460	&	0.179	&	1.662	&	0.006	\\
		\end{tabular}
	\end{ruledtabular}
\end{table*}

\begin{table*}[ht!]
	\renewcommand{\arraystretch}{0.9}
	\footnotesize
	\caption{\label{tab:be_ducc}  	\linespread{1}\small  
		Comparison of the energy errors (in m\Eh) of the \name and the DUCC for the beryllium atom with different basis sets and active spaces. The DSRG flow parameter $s$ = 2.0 \sunit. All computations use RHF orbitals.
		The energy errors are with respect to the full-space FCI computations (absolute energies are shown in the last column), which use 14, 30, 55 orbitals for cc-pVDZ, cc-pVTZ, and cc-pVQZ basis sets, respectively.
		The DUCC data are taken from Ref.~\cite{Bauman:2019dx}.
	}
	\begin{ruledtabular}
		\scriptsize
		\begin{tabular}{@{} lcccc@{}}																
			Method	&	5 orbitals	&	6 orbitals	&	9 orbitals	&	All orbitals	\\
			\hline
			\multicolumn{5}{c}{cc-pVDZ}  \\									
			FCI	&	22.242	&	20.575	&	0.493	&	-14.617409	\\
			DUCC	&	19.009	&	18.109	&	1.809	&		\\
			1-QLDSRG(2)	&	3.408	&	1.998	&	3.067	&		\\
			2-QLDSRG(2)	&	3.430	&	2.074	&	3.065	&		\\
			\multicolumn{5}{c}{cc-pVTZ}  \\									
			FCI	&	34.881	&	33.621	&	7.024	&	-14.623810	\\
			DUCC	&	26.710	&	24.410	&	5.010	&		\\
			1-QLDSRG(2)	&	4.199	&	3.202	&	1.216	&		\\
			2-QLDSRG(2)	&	4.200	&	3.192	&	2.751	&		\\
			\multicolumn{5}{c}{cc-pVQZ}  \\									
			FCI	&	54.950	&	54.366	&	26.798	&	-14.640147	\\
			DUCC	&	30.547	&	27.547	&	7.347	&		\\
			1-QLDSRG(2)	&	4.927	&	4.166	&	0.879	&		\\
			2-QLDSRG(2)	&	4.923	&	4.158	&	2.909	&		\\
			
		\end{tabular}
	\end{ruledtabular}
\end{table*}
Here we compare our \name method with the DUCC downfolding technique for the hydrogen molecule and the beryllium atom.
In Table~\ref{tab:h2_ducc}, we report energies of \ce{H2} at four geometries obtained by diagonalizing the bare, the \name-downfolded, and the DUCC downfolded Hamiltonians in a four-orbital active space, together with the energy errors with respect to the full-space FCI results which use 30 orbitals. We also compute \name energies using different types of orbitals. 
The DUCC data are taken from Ref.~\cite{Bauman:2019dx}.
We observe that for three geometries, all \name computations consistently give less significant energy errors than the DUCC results.
For instance, the largest error for the DUCC using RHF orbitals is 7.13 m\Eh, while the \name shows smaller errors (at most 4.1 m\Eh).
The use of CASSCF orbitals further reduces the maximum \name errors to at most 1.7 m\Eh.

Tab.~\ref{tab:be_ducc} shows the comparison of the beryllium atom results using active spaces of different sizes and three basis sets.
Here we see that the DUCC method introduce errors in the range of 18--30.5 m\Eh when 5 or 6 active orbitals are used, and that this error reduced to smaller values (1.8--7.3 m\Eh) when using 9 active orbitals, 
while \name results consistently show much smaller energy errors (0.9--4.9 m\Eh) for all active spaces and basis sets.
Notably, the \name downfolding is most effective for the large cc-pVQZ basis, which significantly reduces the energy errors of the active-space FCI results by 50, 50.2, 26 m\Eh for 5-, 6- and 9-orbital active spaces, while the DUCC method merely gives a reduction of 24, 27, 19 m\Eh.

\newpage
\bibliography{bibs/vqe-dsrg.bib,bibs/refs.bib,bibs/others.bib}{}

\begin{thebibliography}{112}%
\makeatletter
\providecommand \@ifxundefined [1]{%
 \@ifx{#1\undefined}
}%
\providecommand \@ifnum [1]{%
 \ifnum #1\expandafter \@firstoftwo
 \else \expandafter \@secondoftwo
 \fi
}%
\providecommand \@ifx [1]{%
 \ifx #1\expandafter \@firstoftwo
 \else \expandafter \@secondoftwo
 \fi
}%
\providecommand \natexlab [1]{#1}%
\providecommand \enquote  [1]{``#1''}%
\providecommand \bibnamefont  [1]{#1}%
\providecommand \bibfnamefont [1]{#1}%
\providecommand \citenamefont [1]{#1}%
\providecommand \href@noop [0]{\@secondoftwo}%
\providecommand \href [0]{\begingroup \@sanitize@url \@href}%
\providecommand \@href[1]{\@@startlink{#1}\@@href}%
\providecommand \@@href[1]{\endgroup#1\@@endlink}%
\providecommand \@sanitize@url [0]{\catcode `\\12\catcode `\$12\catcode
  `\&12\catcode `\#12\catcode `\^12\catcode `\_12\catcode `\%12\relax}%
\providecommand \@@startlink[1]{}%
\providecommand \@@endlink[0]{}%
\providecommand \url  [0]{\begingroup\@sanitize@url \@url }%
\providecommand \@url [1]{\endgroup\@href {#1}{\urlprefix }}%
\providecommand \urlprefix  [0]{URL }%
\providecommand \Eprint [0]{\href }%
\providecommand \doibase [0]{https://doi.org/}%
\providecommand \selectlanguage [0]{\@gobble}%
\providecommand \bibinfo  [0]{\@secondoftwo}%
\providecommand \bibfield  [0]{\@secondoftwo}%
\providecommand \translation [1]{[#1]}%
\providecommand \BibitemOpen [0]{}%
\providecommand \bibitemStop [0]{}%
\providecommand \bibitemNoStop [0]{.\EOS\space}%
\providecommand \EOS [0]{\spacefactor3000\relax}%
\providecommand \BibitemShut  [1]{\csname bibitem#1\endcsname}%
\let\auto@bib@innerbib\@empty
\bibitem [{\citenamefont {Laughlin}\ and\ \citenamefont
  {Pines}(2000)}]{Laughlin:2000br}%
  \BibitemOpen
  \bibfield  {author} {\bibinfo {author} {\bibfnamefont {R.~B.}\ \bibnamefont
  {Laughlin}}\ and\ \bibinfo {author} {\bibfnamefont {D.}~\bibnamefont
  {Pines}},\ }\bibfield  {title} {\bibinfo {title} {{The theory of
  everything.}},\ }\href@noop {} {\bibfield  {journal} {\bibinfo  {journal}
  {Proc. Natl. Acad. Sci. U.S.A.}\ }\textbf {\bibinfo {volume} {97}},\ \bibinfo
  {pages} {28} (\bibinfo {year} {2000})}\BibitemShut {NoStop}%
\bibitem [{\citenamefont {Feynman}(1982)}]{Feynman:1982gn}%
  \BibitemOpen
  \bibfield  {author} {\bibinfo {author} {\bibfnamefont {R.~P.}\ \bibnamefont
  {Feynman}},\ }\bibfield  {title} {\bibinfo {title} {{Simulating Physics with
  Computers}},\ }\href@noop {} {\bibfield  {journal} {\bibinfo  {journal} {Int.
  J. Theor. Phys.}\ }\textbf {\bibinfo {volume} {21}},\ \bibinfo {pages} {467}
  (\bibinfo {year} {1982})}\BibitemShut {NoStop}%
\bibitem [{\citenamefont {Manin}(1980)}]{manin1980computable}%
  \BibitemOpen
  \bibfield  {author} {\bibinfo {author} {\bibfnamefont {Y.}~\bibnamefont
  {Manin}},\ }\bibfield  {title} {\bibinfo {title} {Computable and
  uncomputable},\ }\href@noop {} {\bibfield  {journal} {\bibinfo  {journal}
  {Sovetskoye Radio, Moscow}\ }\textbf {\bibinfo {volume} {128}} (\bibinfo
  {year} {1980})}\BibitemShut {NoStop}%
\bibitem [{\citenamefont {Abrams}\ and\ \citenamefont
  {Lloyd}(1997)}]{Abrams:1997ha}%
  \BibitemOpen
  \bibfield  {author} {\bibinfo {author} {\bibfnamefont {D.~S.}\ \bibnamefont
  {Abrams}}\ and\ \bibinfo {author} {\bibfnamefont {S.}~\bibnamefont {Lloyd}},\
  }\bibfield  {title} {\bibinfo {title} {{Simulation of Many-Body Fermi Systems
  on a Universal Quantum Computer}},\ }\href@noop {} {\bibfield  {journal}
  {\bibinfo  {journal} {Phys. Rev. Lett.}\ }\textbf {\bibinfo {volume} {79}},\
  \bibinfo {pages} {2586} (\bibinfo {year} {1997})}\BibitemShut {NoStop}%
\bibitem [{\citenamefont {Abrams}\ and\ \citenamefont
  {Lloyd}(1999)}]{Abrams:1999ur}%
  \BibitemOpen
  \bibfield  {author} {\bibinfo {author} {\bibfnamefont {D.}~\bibnamefont
  {Abrams}}\ and\ \bibinfo {author} {\bibfnamefont {S.}~\bibnamefont {Lloyd}},\
  }\bibfield  {title} {\bibinfo {title} {{Quantum algorithm providing
  exponential speed increase for finding eigenvalues and eigenvectors}},\
  }\href@noop {} {\bibfield  {journal} {\bibinfo  {journal} {Phys. Rev. Lett.}\
  }\textbf {\bibinfo {volume} {83}},\ \bibinfo {pages} {5162} (\bibinfo {year}
  {1999})}\BibitemShut {NoStop}%
\bibitem [{\citenamefont {Bauer}\ \emph {et~al.}(2016)\citenamefont {Bauer},
  \citenamefont {Wecker}, \citenamefont {Millis}, \citenamefont {Hastings},\
  and\ \citenamefont {Troyer}}]{Bauer:2016fc}%
  \BibitemOpen
  \bibfield  {author} {\bibinfo {author} {\bibfnamefont {B.}~\bibnamefont
  {Bauer}}, \bibinfo {author} {\bibfnamefont {D.}~\bibnamefont {Wecker}},
  \bibinfo {author} {\bibfnamefont {A.~J.}\ \bibnamefont {Millis}}, \bibinfo
  {author} {\bibfnamefont {M.~B.}\ \bibnamefont {Hastings}},\ and\ \bibinfo
  {author} {\bibfnamefont {M.}~\bibnamefont {Troyer}},\ }\bibfield  {title}
  {\bibinfo {title} {{Hybrid Quantum-Classical Approach to Correlated
  Materials}},\ }\href@noop {} {\bibfield  {journal} {\bibinfo  {journal}
  {Phys. Rev. X}\ }\textbf {\bibinfo {volume} {6}},\ \bibinfo {pages} {031045}
  (\bibinfo {year} {2016})}\BibitemShut {NoStop}%
\bibitem [{\citenamefont {Takeshita}\ \emph {et~al.}(2020)\citenamefont
  {Takeshita}, \citenamefont {Rubin}, \citenamefont {Jiang}, \citenamefont
  {Lee}, \citenamefont {Babbush},\ and\ \citenamefont
  {McClean}}]{Takeshita:2020dh}%
  \BibitemOpen
  \bibfield  {author} {\bibinfo {author} {\bibfnamefont {T.}~\bibnamefont
  {Takeshita}}, \bibinfo {author} {\bibfnamefont {N.~C.}\ \bibnamefont
  {Rubin}}, \bibinfo {author} {\bibfnamefont {Z.}~\bibnamefont {Jiang}},
  \bibinfo {author} {\bibfnamefont {E.}~\bibnamefont {Lee}}, \bibinfo {author}
  {\bibfnamefont {R.}~\bibnamefont {Babbush}},\ and\ \bibinfo {author}
  {\bibfnamefont {J.~R.}\ \bibnamefont {McClean}},\ }\bibfield  {title}
  {\bibinfo {title} {{Increasing the Representation Accuracy of Quantum
  Simulations of Chemistry without Extra Quantum Resources}},\ }\href@noop {}
  {\bibfield  {journal} {\bibinfo  {journal} {Phys. Rev. X}\ }\textbf {\bibinfo
  {volume} {10}},\ \bibinfo {pages} {011004} (\bibinfo {year}
  {2020})}\BibitemShut {NoStop}%
\bibitem [{\citenamefont {Urbanek}\ \emph {et~al.}(2020)\citenamefont
  {Urbanek}, \citenamefont {Camps}, \citenamefont {Van~Beeumen},\ and\
  \citenamefont {de~Jong}}]{Urbanek:2020fh}%
  \BibitemOpen
  \bibfield  {author} {\bibinfo {author} {\bibfnamefont {M.}~\bibnamefont
  {Urbanek}}, \bibinfo {author} {\bibfnamefont {D.}~\bibnamefont {Camps}},
  \bibinfo {author} {\bibfnamefont {R.}~\bibnamefont {Van~Beeumen}},\ and\
  \bibinfo {author} {\bibfnamefont {W.~A.}\ \bibnamefont {de~Jong}},\
  }\bibfield  {title} {\bibinfo {title} {{Chemistry on Quantum Computers with
  Virtual Quantum Subspace Expansion}},\ }\href@noop {} {\bibfield  {journal}
  {\bibinfo  {journal} {J. Chem. Theory Comput.}\ }\textbf {\bibinfo {volume}
  {16}},\ \bibinfo {pages} {5425} (\bibinfo {year} {2020})}\BibitemShut
  {NoStop}%
\bibitem [{\citenamefont {Boyn}\ \emph {et~al.}(2021)\citenamefont {Boyn},
  \citenamefont {Lykhin}, \citenamefont {Smart}, \citenamefont {Gagliardi},\
  and\ \citenamefont {Mazziotti}}]{Boyn:2021kp}%
  \BibitemOpen
  \bibfield  {author} {\bibinfo {author} {\bibfnamefont {J.-N.}\ \bibnamefont
  {Boyn}}, \bibinfo {author} {\bibfnamefont {A.~O.}\ \bibnamefont {Lykhin}},
  \bibinfo {author} {\bibfnamefont {S.~E.}\ \bibnamefont {Smart}}, \bibinfo
  {author} {\bibfnamefont {L.}~\bibnamefont {Gagliardi}},\ and\ \bibinfo
  {author} {\bibfnamefont {D.~A.}\ \bibnamefont {Mazziotti}},\ }\bibfield
  {title} {\bibinfo {title} {{Quantum-classical hybrid algorithm for the
  simulation of all-electron correlation}},\ }\href@noop {} {\bibfield
  {journal} {\bibinfo  {journal} {J. Chem. Phys.}\ }\textbf {\bibinfo {volume}
  {155}},\ \bibinfo {pages} {244106} (\bibinfo {year} {2021})}\BibitemShut
  {NoStop}%
\bibitem [{\citenamefont {Mazziotti}(2006)}]{Mazziotti:2006iw}%
  \BibitemOpen
  \bibfield  {author} {\bibinfo {author} {\bibfnamefont {D.~A.}\ \bibnamefont
  {Mazziotti}},\ }\bibfield  {title} {\bibinfo {title} {{Anti-Hermitian
  Contracted Schr{\"o}dinger Equation: Direct Determination of the Two-Electron
  Reduced Density Matrices of Many-Electron Molecules}},\ }\href@noop {}
  {\bibfield  {journal} {\bibinfo  {journal} {Phys. Rev. Lett.}\ }\textbf
  {\bibinfo {volume} {97}},\ \bibinfo {pages} {143002} (\bibinfo {year}
  {2006})}\BibitemShut {NoStop}%
\bibitem [{\citenamefont {Smart}\ and\ \citenamefont
  {Mazziotti}(2021)}]{Smart:2021hd}%
  \BibitemOpen
  \bibfield  {author} {\bibinfo {author} {\bibfnamefont {S.~E.}\ \bibnamefont
  {Smart}}\ and\ \bibinfo {author} {\bibfnamefont {D.~A.}\ \bibnamefont
  {Mazziotti}},\ }\bibfield  {title} {\bibinfo {title} {{Quantum Solver of
  Contracted Eigenvalue Equations for Scalable Molecular Simulations on Quantum
  Computing Devices}},\ }\href@noop {} {\bibfield  {journal} {\bibinfo
  {journal} {Phys. Rev. Lett.}\ }\textbf {\bibinfo {volume} {126}},\ \bibinfo
  {pages} {070504} (\bibinfo {year} {2021})}\BibitemShut {NoStop}%
\bibitem [{\citenamefont {Smart}\ \emph {et~al.}(2022)\citenamefont {Smart},
  \citenamefont {Boyn},\ and\ \citenamefont {Mazziotti}}]{Smart:2022ik}%
  \BibitemOpen
  \bibfield  {author} {\bibinfo {author} {\bibfnamefont {S.~E.}\ \bibnamefont
  {Smart}}, \bibinfo {author} {\bibfnamefont {J.-N.}\ \bibnamefont {Boyn}},\
  and\ \bibinfo {author} {\bibfnamefont {D.~A.}\ \bibnamefont {Mazziotti}},\
  }\bibfield  {title} {\bibinfo {title} {{Resolving correlated states of
  benzyne with an error-mitigated contracted quantum eigensolver}},\
  }\href@noop {} {\bibfield  {journal} {\bibinfo  {journal} {Phys. Rev. A}\
  }\textbf {\bibinfo {volume} {105}},\ \bibinfo {pages} {022405} (\bibinfo
  {year} {2022})}\BibitemShut {NoStop}%
\bibitem [{\citenamefont {Li~Manni}\ \emph {et~al.}(2014)\citenamefont
  {Li~Manni}, \citenamefont {Carlson}, \citenamefont {Luo}, \citenamefont {Ma},
  \citenamefont {Olsen}, \citenamefont {Truhlar},\ and\ \citenamefont
  {Gagliardi}}]{LiManni:2014kg}%
  \BibitemOpen
  \bibfield  {author} {\bibinfo {author} {\bibfnamefont {G.}~\bibnamefont
  {Li~Manni}}, \bibinfo {author} {\bibfnamefont {R.~K.}\ \bibnamefont
  {Carlson}}, \bibinfo {author} {\bibfnamefont {S.}~\bibnamefont {Luo}},
  \bibinfo {author} {\bibfnamefont {D.}~\bibnamefont {Ma}}, \bibinfo {author}
  {\bibfnamefont {J.}~\bibnamefont {Olsen}}, \bibinfo {author} {\bibfnamefont
  {D.~G.}\ \bibnamefont {Truhlar}},\ and\ \bibinfo {author} {\bibfnamefont
  {L.}~\bibnamefont {Gagliardi}},\ }\bibfield  {title} {\bibinfo {title}
  {{Multiconfiguration Pair-Density Functional Theory}},\ }\href@noop {}
  {\bibfield  {journal} {\bibinfo  {journal} {J. Chem. Theory Comput.}\
  }\textbf {\bibinfo {volume} {10}},\ \bibinfo {pages} {3669} (\bibinfo {year}
  {2014})}\BibitemShut {NoStop}%
\bibitem [{\citenamefont {Fujii}\ \emph {et~al.}(2022)\citenamefont {Fujii},
  \citenamefont {Mizuta}, \citenamefont {Ueda}, \citenamefont {Mitarai},
  \citenamefont {Mizukami},\ and\ \citenamefont {Nakagawa}}]{fujii2022deep}%
  \BibitemOpen
  \bibfield  {author} {\bibinfo {author} {\bibfnamefont {K.}~\bibnamefont
  {Fujii}}, \bibinfo {author} {\bibfnamefont {K.}~\bibnamefont {Mizuta}},
  \bibinfo {author} {\bibfnamefont {H.}~\bibnamefont {Ueda}}, \bibinfo {author}
  {\bibfnamefont {K.}~\bibnamefont {Mitarai}}, \bibinfo {author} {\bibfnamefont
  {W.}~\bibnamefont {Mizukami}},\ and\ \bibinfo {author} {\bibfnamefont
  {Y.~O.}\ \bibnamefont {Nakagawa}},\ }\bibfield  {title} {\bibinfo {title}
  {Deep variational quantum eigensolver: a divide-and-conquer method for
  solving a larger problem with smaller size quantum computers},\ }\href@noop
  {} {\bibfield  {journal} {\bibinfo  {journal} {PRX Quantum}\ }\textbf
  {\bibinfo {volume} {3}},\ \bibinfo {pages} {010346} (\bibinfo {year}
  {2022})}\BibitemShut {NoStop}%
\bibitem [{\citenamefont {Mizuta}\ \emph {et~al.}(2021)\citenamefont {Mizuta},
  \citenamefont {Fujii}, \citenamefont {Fujii}, \citenamefont {Ichikawa},
  \citenamefont {Imamura}, \citenamefont {Okuno},\ and\ \citenamefont
  {Nakagawa}}]{mizuta2021deep}%
  \BibitemOpen
  \bibfield  {author} {\bibinfo {author} {\bibfnamefont {K.}~\bibnamefont
  {Mizuta}}, \bibinfo {author} {\bibfnamefont {M.}~\bibnamefont {Fujii}},
  \bibinfo {author} {\bibfnamefont {S.}~\bibnamefont {Fujii}}, \bibinfo
  {author} {\bibfnamefont {K.}~\bibnamefont {Ichikawa}}, \bibinfo {author}
  {\bibfnamefont {Y.}~\bibnamefont {Imamura}}, \bibinfo {author} {\bibfnamefont
  {Y.}~\bibnamefont {Okuno}},\ and\ \bibinfo {author} {\bibfnamefont {Y.~O.}\
  \bibnamefont {Nakagawa}},\ }\bibfield  {title} {\bibinfo {title} {Deep
  variational quantum eigensolver for excited states and its application to
  quantum chemistry calculation of periodic materials},\ }\href@noop {}
  {\bibfield  {journal} {\bibinfo  {journal} {Phys. Rev. Res.}\ }\textbf
  {\bibinfo {volume} {3}},\ \bibinfo {pages} {043121} (\bibinfo {year}
  {2021})}\BibitemShut {NoStop}%
\bibitem [{\citenamefont {Ryabinkin}\ \emph {et~al.}(2021)\citenamefont
  {Ryabinkin}, \citenamefont {Izmaylov},\ and\ \citenamefont
  {Genin}}]{Ryabinkin_2021}%
  \BibitemOpen
  \bibfield  {author} {\bibinfo {author} {\bibfnamefont {I.~G.}\ \bibnamefont
  {Ryabinkin}}, \bibinfo {author} {\bibfnamefont {A.~F.}\ \bibnamefont
  {Izmaylov}},\ and\ \bibinfo {author} {\bibfnamefont {S.~N.}\ \bibnamefont
  {Genin}},\ }\bibfield  {title} {\bibinfo {title} {A posteriori corrections to
  the iterative qubit coupled cluster method to minimize the use of quantum
  resources in large-scale calculations},\ }\href
  {https://doi.org/10.1088/2058-9565/abda8e} {\bibfield  {journal} {\bibinfo
  {journal} {Quantum Sci. Technol.}\ }\textbf {\bibinfo {volume} {6}},\
  \bibinfo {pages} {024012} (\bibinfo {year} {2021})}\BibitemShut {NoStop}%
\bibitem [{\citenamefont {Tammaro}\ \emph {et~al.}(2022)\citenamefont
  {Tammaro}, \citenamefont {Gall}, \citenamefont {Rice},\ and\ \citenamefont
  {Motta}}]{tammaro2022n}%
  \BibitemOpen
  \bibfield  {author} {\bibinfo {author} {\bibfnamefont {A.}~\bibnamefont
  {Tammaro}}, \bibinfo {author} {\bibfnamefont {D.~E.}\ \bibnamefont {Gall}},
  \bibinfo {author} {\bibfnamefont {J.~E.}\ \bibnamefont {Rice}},\ and\
  \bibinfo {author} {\bibfnamefont {M.}~\bibnamefont {Motta}},\ }\bibfield
  {title} {\bibinfo {title} {N-electron valence perturbation theory with
  reference wavefunctions from quantum computing: application to the relative
  stability of hydroxide anion and hydroxyl radical},\ }\href@noop {}
  {\bibfield  {journal} {\bibinfo  {journal} {arXiv preprint arXiv:2202.13002}\
  } (\bibinfo {year} {2022})}\BibitemShut {NoStop}%
\bibitem [{\citenamefont {Angeli}\ \emph {et~al.}(2001)\citenamefont {Angeli},
  \citenamefont {Cimiraglia}, \citenamefont {Evangelisti}, \citenamefont
  {Leininger},\ and\ \citenamefont {Malrieu}}]{Angeli:2001vf}%
  \BibitemOpen
  \bibfield  {author} {\bibinfo {author} {\bibfnamefont {C.}~\bibnamefont
  {Angeli}}, \bibinfo {author} {\bibfnamefont {R.}~\bibnamefont {Cimiraglia}},
  \bibinfo {author} {\bibfnamefont {S.}~\bibnamefont {Evangelisti}}, \bibinfo
  {author} {\bibfnamefont {T.}~\bibnamefont {Leininger}},\ and\ \bibinfo
  {author} {\bibfnamefont {J.~P.}\ \bibnamefont {Malrieu}},\ }\bibfield
  {title} {\bibinfo {title} {{Introduction of n-electron valence states for
  multireference perturbation theory}},\ }\href@noop {} {\bibfield  {journal}
  {\bibinfo  {journal} {J. Chem. Phys.}\ }\textbf {\bibinfo {volume} {114}},\
  \bibinfo {pages} {10252} (\bibinfo {year} {2001})}\BibitemShut {NoStop}%
\bibitem [{\citenamefont {Angeli}\ \emph {et~al.}(2007)\citenamefont {Angeli},
  \citenamefont {Pastore},\ and\ \citenamefont {Cimiraglia}}]{Angeli:2006by}%
  \BibitemOpen
  \bibfield  {author} {\bibinfo {author} {\bibfnamefont {C.}~\bibnamefont
  {Angeli}}, \bibinfo {author} {\bibfnamefont {M.}~\bibnamefont {Pastore}},\
  and\ \bibinfo {author} {\bibfnamefont {R.}~\bibnamefont {Cimiraglia}},\
  }\bibfield  {title} {\bibinfo {title} {{New perspectives in multireference
  perturbation theory: the n-electron valence state approach}},\ }\href@noop {}
  {\bibfield  {journal} {\bibinfo  {journal} {Theor. Chem. Acc.}\ }\textbf
  {\bibinfo {volume} {117}},\ \bibinfo {pages} {743} (\bibinfo {year}
  {2007})}\BibitemShut {NoStop}%
\bibitem [{\citenamefont {Peruzzo}\ \emph {et~al.}(2014)\citenamefont
  {Peruzzo}, \citenamefont {McClean}, \citenamefont {Shadbolt}, \citenamefont
  {Yung}, \citenamefont {Zhou}, \citenamefont {Love}, \citenamefont {Brien},\
  and\ \citenamefont {Aspuru-Guzik}}]{Peruzzo:2014kc}%
  \BibitemOpen
  \bibfield  {author} {\bibinfo {author} {\bibfnamefont {A.}~\bibnamefont
  {Peruzzo}}, \bibinfo {author} {\bibfnamefont {J.}~\bibnamefont {McClean}},
  \bibinfo {author} {\bibfnamefont {P.}~\bibnamefont {Shadbolt}}, \bibinfo
  {author} {\bibfnamefont {M.-H.}\ \bibnamefont {Yung}}, \bibinfo {author}
  {\bibfnamefont {X.-Q.}\ \bibnamefont {Zhou}}, \bibinfo {author}
  {\bibfnamefont {P.~J.}\ \bibnamefont {Love}}, \bibinfo {author}
  {\bibfnamefont {J.~L. O.~r.}\ \bibnamefont {Brien}},\ and\ \bibinfo {author}
  {\bibfnamefont {A.}~\bibnamefont {Aspuru-Guzik}},\ }\bibfield  {title}
  {\bibinfo {title} {{A variational eigenvalue solver on a photonic quantum
  processor}},\ }\href@noop {} {\bibfield  {journal} {\bibinfo  {journal} {Nat.
  Commun.}\ }\textbf {\bibinfo {volume} {5}},\ \bibinfo {pages} {4213}
  (\bibinfo {year} {2014})}\BibitemShut {NoStop}%
\bibitem [{\citenamefont {Yung}\ \emph {et~al.}(2014)\citenamefont {Yung},
  \citenamefont {Casanova}, \citenamefont {Lamata}, \citenamefont {Solano},
  \citenamefont {Aspuru-Guzik}, \citenamefont {McClean},\ and\ \citenamefont
  {Mezzacapo}}]{Yung:2014iv}%
  \BibitemOpen
  \bibfield  {author} {\bibinfo {author} {\bibfnamefont {M.-H.}\ \bibnamefont
  {Yung}}, \bibinfo {author} {\bibfnamefont {J.}~\bibnamefont {Casanova}},
  \bibinfo {author} {\bibfnamefont {L.}~\bibnamefont {Lamata}}, \bibinfo
  {author} {\bibfnamefont {E.}~\bibnamefont {Solano}}, \bibinfo {author}
  {\bibfnamefont {A.}~\bibnamefont {Aspuru-Guzik}}, \bibinfo {author}
  {\bibfnamefont {J.~R.}\ \bibnamefont {McClean}},\ and\ \bibinfo {author}
  {\bibfnamefont {A.}~\bibnamefont {Mezzacapo}},\ }\bibfield  {title} {\bibinfo
  {title} {{From transistor to trapped-ion computers for quantum chemistry}},\
  }\href@noop {} {\bibfield  {journal} {\bibinfo  {journal} {Sci. Rep.}\
  }\textbf {\bibinfo {volume} {4}},\ \bibinfo {pages} {3589} (\bibinfo {year}
  {2014})}\BibitemShut {NoStop}%
\bibitem [{\citenamefont {McClean}\ \emph {et~al.}(2016)\citenamefont
  {McClean}, \citenamefont {Romero}, \citenamefont {Babbush},\ and\
  \citenamefont {Aspuru-Guzik}}]{McClean:2015bs}%
  \BibitemOpen
  \bibfield  {author} {\bibinfo {author} {\bibfnamefont {J.~R.}\ \bibnamefont
  {McClean}}, \bibinfo {author} {\bibfnamefont {J.}~\bibnamefont {Romero}},
  \bibinfo {author} {\bibfnamefont {R.}~\bibnamefont {Babbush}},\ and\ \bibinfo
  {author} {\bibfnamefont {A.}~\bibnamefont {Aspuru-Guzik}},\ }\bibfield
  {title} {\bibinfo {title} {{The theory of variational hybrid
  quantum-classical algorithms}},\ }\href@noop {} {\bibfield  {journal}
  {\bibinfo  {journal} {New J. Phys.}\ }\textbf {\bibinfo {volume} {18}},\
  \bibinfo {pages} {023023} (\bibinfo {year} {2016})}\BibitemShut {NoStop}%
\bibitem [{\citenamefont {McClean}\ \emph {et~al.}(2017)\citenamefont
  {McClean}, \citenamefont {Kimchi-Schwartz}, \citenamefont {Carter},\ and\
  \citenamefont {de~Jong}}]{McClean:2017ct}%
  \BibitemOpen
  \bibfield  {author} {\bibinfo {author} {\bibfnamefont {J.~R.}\ \bibnamefont
  {McClean}}, \bibinfo {author} {\bibfnamefont {M.~E.}\ \bibnamefont
  {Kimchi-Schwartz}}, \bibinfo {author} {\bibfnamefont {J.}~\bibnamefont
  {Carter}},\ and\ \bibinfo {author} {\bibfnamefont {W.~A.}\ \bibnamefont
  {de~Jong}},\ }\bibfield  {title} {\bibinfo {title} {{Hybrid quantum-classical
  hierarchy for mitigation of decoherence and determination of excited states
  }},\ }\href@noop {} {\bibfield  {journal} {\bibinfo  {journal} {Phys. Rev.
  A}\ }\textbf {\bibinfo {volume} {95}},\ \bibinfo {pages} {042308} (\bibinfo
  {year} {2017})}\BibitemShut {NoStop}%
\bibitem [{\citenamefont {Colless}\ \emph {et~al.}(2018)\citenamefont
  {Colless}, \citenamefont {Ramasesh}, \citenamefont {Dahlen}, \citenamefont
  {Blok}, \citenamefont {Kimchi-Schwartz}, \citenamefont {McClean},
  \citenamefont {Carter}, \citenamefont {de~Jong},\ and\ \citenamefont
  {Siddiqi}}]{Colless:2018hp}%
  \BibitemOpen
  \bibfield  {author} {\bibinfo {author} {\bibfnamefont {J.~I.}\ \bibnamefont
  {Colless}}, \bibinfo {author} {\bibfnamefont {V.~V.}\ \bibnamefont
  {Ramasesh}}, \bibinfo {author} {\bibfnamefont {D.}~\bibnamefont {Dahlen}},
  \bibinfo {author} {\bibfnamefont {M.~S.}\ \bibnamefont {Blok}}, \bibinfo
  {author} {\bibfnamefont {M.~E.}\ \bibnamefont {Kimchi-Schwartz}}, \bibinfo
  {author} {\bibfnamefont {J.~R.}\ \bibnamefont {McClean}}, \bibinfo {author}
  {\bibfnamefont {J.}~\bibnamefont {Carter}}, \bibinfo {author} {\bibfnamefont
  {W.~A.}\ \bibnamefont {de~Jong}},\ and\ \bibinfo {author} {\bibfnamefont
  {I.}~\bibnamefont {Siddiqi}},\ }\bibfield  {title} {\bibinfo {title}
  {{Computation of Molecular Spectra on a Quantum Processor with an
  Error-Resilient Algorithm}},\ }\href@noop {} {\bibfield  {journal} {\bibinfo
  {journal} {Phys. Rev. X}\ }\textbf {\bibinfo {volume} {8}},\ \bibinfo {pages}
  {011021} (\bibinfo {year} {2018})}\BibitemShut {NoStop}%
\bibitem [{\citenamefont {Huggins}\ \emph {et~al.}(2022)\citenamefont
  {Huggins}, \citenamefont {O'Gorman}, \citenamefont {Rubin}, \citenamefont
  {Reichman}, \citenamefont {Babbush},\ and\ \citenamefont
  {Lee}}]{Huggins:2022iw}%
  \BibitemOpen
  \bibfield  {author} {\bibinfo {author} {\bibfnamefont {W.~J.}\ \bibnamefont
  {Huggins}}, \bibinfo {author} {\bibfnamefont {B.~A.}\ \bibnamefont
  {O'Gorman}}, \bibinfo {author} {\bibfnamefont {N.~C.}\ \bibnamefont {Rubin}},
  \bibinfo {author} {\bibfnamefont {D.~R.}\ \bibnamefont {Reichman}}, \bibinfo
  {author} {\bibfnamefont {R.}~\bibnamefont {Babbush}},\ and\ \bibinfo {author}
  {\bibfnamefont {J.}~\bibnamefont {Lee}},\ }\bibfield  {title} {\bibinfo
  {title} {{Unbiasing fermionic quantum Monte Carlo with a quantum
  computer.}},\ }\href@noop {} {\bibfield  {journal} {\bibinfo  {journal}
  {Nature}\ }\textbf {\bibinfo {volume} {603}},\ \bibinfo {pages} {416}
  (\bibinfo {year} {2022})}\BibitemShut {NoStop}%
\bibitem [{\citenamefont {Knizia}\ and\ \citenamefont
  {Chan}(2012)}]{Knizia:2012do}%
  \BibitemOpen
  \bibfield  {author} {\bibinfo {author} {\bibfnamefont {G.}~\bibnamefont
  {Knizia}}\ and\ \bibinfo {author} {\bibfnamefont {G.~K.-L.}\ \bibnamefont
  {Chan}},\ }\bibfield  {title} {\bibinfo {title} {{Density Matrix Embedding: A
  Simple Alternative to Dynamical Mean-Field Theory}},\ }\href@noop {}
  {\bibfield  {journal} {\bibinfo  {journal} {Phys. Rev. Lett.}\ }\textbf
  {\bibinfo {volume} {109}},\ \bibinfo {pages} {186404} (\bibinfo {year}
  {2012})}\BibitemShut {NoStop}%
\bibitem [{\citenamefont {Wouters}\ \emph {et~al.}(2016)\citenamefont
  {Wouters}, \citenamefont {Jim{\'e}nez-Hoyos}, \citenamefont {Sun},\ and\
  \citenamefont {Chan}}]{wouters2016practical}%
  \BibitemOpen
  \bibfield  {author} {\bibinfo {author} {\bibfnamefont {S.}~\bibnamefont
  {Wouters}}, \bibinfo {author} {\bibfnamefont {C.~A.}\ \bibnamefont
  {Jim{\'e}nez-Hoyos}}, \bibinfo {author} {\bibfnamefont {Q.}~\bibnamefont
  {Sun}},\ and\ \bibinfo {author} {\bibfnamefont {G.~K.-L.}\ \bibnamefont
  {Chan}},\ }\bibfield  {title} {\bibinfo {title} {A practical guide to density
  matrix embedding theory in quantum chemistry},\ }\href@noop {} {\bibfield
  {journal} {\bibinfo  {journal} {J. Chem. Theory Comput.}\ }\textbf {\bibinfo
  {volume} {12}},\ \bibinfo {pages} {2706} (\bibinfo {year}
  {2016})}\BibitemShut {NoStop}%
\bibitem [{\citenamefont {Kawashima}\ \emph {et~al.}(2021)\citenamefont
  {Kawashima}, \citenamefont {Lloyd}, \citenamefont {Coons}, \citenamefont
  {Nam}, \citenamefont {Matsuura}, \citenamefont {Garza}, \citenamefont
  {Johri}, \citenamefont {Huntington}, \citenamefont {Senicourt}, \citenamefont
  {Maksymov}, \citenamefont {Nguyen}, \citenamefont {Kim}, \citenamefont
  {Alidoust}, \citenamefont {Zaribafiyan},\ and\ \citenamefont
  {Yamazaki}}]{Kawashima:2021hv}%
  \BibitemOpen
  \bibfield  {author} {\bibinfo {author} {\bibfnamefont {Y.}~\bibnamefont
  {Kawashima}}, \bibinfo {author} {\bibfnamefont {E.}~\bibnamefont {Lloyd}},
  \bibinfo {author} {\bibfnamefont {M.~P.}\ \bibnamefont {Coons}}, \bibinfo
  {author} {\bibfnamefont {Y.}~\bibnamefont {Nam}}, \bibinfo {author}
  {\bibfnamefont {S.}~\bibnamefont {Matsuura}}, \bibinfo {author}
  {\bibfnamefont {A.~J.}\ \bibnamefont {Garza}}, \bibinfo {author}
  {\bibfnamefont {S.}~\bibnamefont {Johri}}, \bibinfo {author} {\bibfnamefont
  {L.}~\bibnamefont {Huntington}}, \bibinfo {author} {\bibfnamefont
  {V.}~\bibnamefont {Senicourt}}, \bibinfo {author} {\bibfnamefont {A.~O.}\
  \bibnamefont {Maksymov}}, \bibinfo {author} {\bibfnamefont {J.~H.~V.}\
  \bibnamefont {Nguyen}}, \bibinfo {author} {\bibfnamefont {J.}~\bibnamefont
  {Kim}}, \bibinfo {author} {\bibfnamefont {N.}~\bibnamefont {Alidoust}},
  \bibinfo {author} {\bibfnamefont {A.}~\bibnamefont {Zaribafiyan}},\ and\
  \bibinfo {author} {\bibfnamefont {T.}~\bibnamefont {Yamazaki}},\ }\bibfield
  {title} {\bibinfo {title} {{Optimizing electronic structure simulations on a
  trapped-ion quantum computer using problem decomposition}},\ }\href@noop {}
  {\bibfield  {journal} {\bibinfo  {journal} {Commun Phys}\ }\textbf {\bibinfo
  {volume} {4}},\ \bibinfo {pages} {245} (\bibinfo {year} {2021})}\BibitemShut
  {NoStop}%
\bibitem [{\citenamefont {Huang}\ \emph {et~al.}(2022)\citenamefont {Huang},
  \citenamefont {Govoni},\ and\ \citenamefont {Galli}}]{Huang:2022bu}%
  \BibitemOpen
  \bibfield  {author} {\bibinfo {author} {\bibfnamefont {B.}~\bibnamefont
  {Huang}}, \bibinfo {author} {\bibfnamefont {M.}~\bibnamefont {Govoni}},\ and\
  \bibinfo {author} {\bibfnamefont {G.}~\bibnamefont {Galli}},\ }\bibfield
  {title} {\bibinfo {title} {{Simulating the electronic structure of spin
  defects on quantum computers}},\ }\href@noop {} {\bibfield  {journal}
  {\bibinfo  {journal} {PRX Quantum}\ }\textbf {\bibinfo {volume} {3}},\
  \bibinfo {pages} {010339} (\bibinfo {year} {2022})}\BibitemShut {NoStop}%
\bibitem [{\citenamefont {Motta}\ \emph {et~al.}(2020)\citenamefont {Motta},
  \citenamefont {Gujarati}, \citenamefont {Rice}, \citenamefont {Kumar},
  \citenamefont {Masteran}, \citenamefont {Latone}, \citenamefont {Lee},
  \citenamefont {Valeev},\ and\ \citenamefont {Takeshita}}]{Motta:2020ic}%
  \BibitemOpen
  \bibfield  {author} {\bibinfo {author} {\bibfnamefont {M.}~\bibnamefont
  {Motta}}, \bibinfo {author} {\bibfnamefont {T.~P.}\ \bibnamefont {Gujarati}},
  \bibinfo {author} {\bibfnamefont {J.~E.}\ \bibnamefont {Rice}}, \bibinfo
  {author} {\bibfnamefont {A.}~\bibnamefont {Kumar}}, \bibinfo {author}
  {\bibfnamefont {C.}~\bibnamefont {Masteran}}, \bibinfo {author}
  {\bibfnamefont {J.~A.}\ \bibnamefont {Latone}}, \bibinfo {author}
  {\bibfnamefont {E.}~\bibnamefont {Lee}}, \bibinfo {author} {\bibfnamefont
  {E.~F.}\ \bibnamefont {Valeev}},\ and\ \bibinfo {author} {\bibfnamefont
  {T.~Y.}\ \bibnamefont {Takeshita}},\ }\bibfield  {title} {\bibinfo {title}
  {{Quantum simulation of electronic structure with a transcorrelated
  Hamiltonian: improved accuracy with a smaller footprint on the quantum
  computer}},\ }\href@noop {} {\bibfield  {journal} {\bibinfo  {journal} {Phys.
  Chem. Chem. Phys.}\ }\textbf {\bibinfo {volume} {22}},\ \bibinfo {pages}
  {24270} (\bibinfo {year} {2020})}\BibitemShut {NoStop}%
\bibitem [{\citenamefont {McArdle}\ and\ \citenamefont
  {Tew}(2020)}]{mcardle2020improving}%
  \BibitemOpen
  \bibfield  {author} {\bibinfo {author} {\bibfnamefont {S.}~\bibnamefont
  {McArdle}}\ and\ \bibinfo {author} {\bibfnamefont {D.~P.}\ \bibnamefont
  {Tew}},\ }\bibfield  {title} {\bibinfo {title} {Improving the accuracy of
  quantum computational chemistry using the transcorrelated method},\
  }\href@noop {} {\bibfield  {journal} {\bibinfo  {journal} {arXiv preprint
  arXiv:2006.11181}\ } (\bibinfo {year} {2020})}\BibitemShut {NoStop}%
\bibitem [{\citenamefont {Schleich}\ \emph {et~al.}(2022)\citenamefont
  {Schleich}, \citenamefont {Kottmann},\ and\ \citenamefont
  {Aspuru-Guzik}}]{Schleich:2022fa}%
  \BibitemOpen
  \bibfield  {author} {\bibinfo {author} {\bibfnamefont {P.}~\bibnamefont
  {Schleich}}, \bibinfo {author} {\bibfnamefont {J.~S.}\ \bibnamefont
  {Kottmann}},\ and\ \bibinfo {author} {\bibfnamefont {A.}~\bibnamefont
  {Aspuru-Guzik}},\ }\bibfield  {title} {\bibinfo {title} {{Improving the
  accuracy of the variational quantum eigensolver for molecular systems by the
  explicitly-correlated perturbative [2] R12 - correction}},\ }\href@noop {}
  {\bibfield  {journal} {\bibinfo  {journal} {Phys. Chem. Chem. Phys.}\
  }\textbf {\bibinfo {volume} {24}},\ \bibinfo {pages} {13550} (\bibinfo {year}
  {2022})}\BibitemShut {NoStop}%
\bibitem [{\citenamefont {Sokolov}\ \emph {et~al.}(2022)\citenamefont
  {Sokolov}, \citenamefont {Dobrautz}, \citenamefont {Luo}, \citenamefont
  {Alavi},\ and\ \citenamefont {Tavernelli}}]{sokolov2022orders}%
  \BibitemOpen
  \bibfield  {author} {\bibinfo {author} {\bibfnamefont {I.~O.}\ \bibnamefont
  {Sokolov}}, \bibinfo {author} {\bibfnamefont {W.}~\bibnamefont {Dobrautz}},
  \bibinfo {author} {\bibfnamefont {H.}~\bibnamefont {Luo}}, \bibinfo {author}
  {\bibfnamefont {A.}~\bibnamefont {Alavi}},\ and\ \bibinfo {author}
  {\bibfnamefont {I.}~\bibnamefont {Tavernelli}},\ }\bibfield  {title}
  {\bibinfo {title} {Orders of magnitude reduction in the computational
  overhead for quantum many-body problems on quantum computers via an exact
  transcorrelated method},\ }\href@noop {} {\bibfield  {journal} {\bibinfo
  {journal} {arXiv preprint arXiv:2201.03049}\ } (\bibinfo {year}
  {2022})}\BibitemShut {NoStop}%
\bibitem [{\citenamefont {Steiger}\ \emph {et~al.}(2018)\citenamefont
  {Steiger}, \citenamefont {H{\"a}ner},\ and\ \citenamefont
  {Troyer}}]{2018projectq}%
  \BibitemOpen
  \bibfield  {author} {\bibinfo {author} {\bibfnamefont {D.~S.}\ \bibnamefont
  {Steiger}}, \bibinfo {author} {\bibfnamefont {T.}~\bibnamefont {H{\"a}ner}},\
  and\ \bibinfo {author} {\bibfnamefont {M.}~\bibnamefont {Troyer}},\
  }\bibfield  {title} {\bibinfo {title} {{ProjectQ: An Open Source Software
  Framework for Quantum Computing}},\ }\href@noop {} {\bibfield  {journal}
  {\bibinfo  {journal} {Quantum}\ }\textbf {\bibinfo {volume} {2}},\ \bibinfo
  {pages} {49} (\bibinfo {year} {2018})}\BibitemShut {NoStop}%
\bibitem [{\citenamefont {Kumar}\ \emph {et~al.}(2022)\citenamefont {Kumar},
  \citenamefont {Asthana}, \citenamefont {Masteran}, \citenamefont {Valeev},
  \citenamefont {Zhang}, \citenamefont {Cincio}, \citenamefont {Tretiak},\ and\
  \citenamefont {Dub}}]{kumar2022accurate}%
  \BibitemOpen
  \bibfield  {author} {\bibinfo {author} {\bibfnamefont {A.}~\bibnamefont
  {Kumar}}, \bibinfo {author} {\bibfnamefont {A.}~\bibnamefont {Asthana}},
  \bibinfo {author} {\bibfnamefont {C.}~\bibnamefont {Masteran}}, \bibinfo
  {author} {\bibfnamefont {E.~F.}\ \bibnamefont {Valeev}}, \bibinfo {author}
  {\bibfnamefont {Y.}~\bibnamefont {Zhang}}, \bibinfo {author} {\bibfnamefont
  {L.}~\bibnamefont {Cincio}}, \bibinfo {author} {\bibfnamefont
  {S.}~\bibnamefont {Tretiak}},\ and\ \bibinfo {author} {\bibfnamefont {P.~A.}\
  \bibnamefont {Dub}},\ }\bibfield  {title} {\bibinfo {title} {Accurate quantum
  simulation of molecular ground and excited states with a transcorrelated
  hamiltonian},\ }\href@noop {} {\bibfield  {journal} {\bibinfo  {journal}
  {arXiv preprint arXiv:2201.09852}\ } (\bibinfo {year} {2022})}\BibitemShut
  {NoStop}%
\bibitem [{\citenamefont {Yanai}\ and\ \citenamefont
  {Shiozaki}(2012)}]{Yanai:2012cj}%
  \BibitemOpen
  \bibfield  {author} {\bibinfo {author} {\bibfnamefont {T.}~\bibnamefont
  {Yanai}}\ and\ \bibinfo {author} {\bibfnamefont {T.}~\bibnamefont
  {Shiozaki}},\ }\bibfield  {title} {\bibinfo {title} {{Canonical
  transcorrelated theory with projected Slater-type geminals}},\ }\href@noop {}
  {\bibfield  {journal} {\bibinfo  {journal} {J. Chem. Phys.}\ }\textbf
  {\bibinfo {volume} {136}},\ \bibinfo {pages} {084107} (\bibinfo {year}
  {2012})}\BibitemShut {NoStop}%
\bibitem [{\citenamefont {White}(2002)}]{White:2002uj}%
  \BibitemOpen
  \bibfield  {author} {\bibinfo {author} {\bibfnamefont {S.~R.}\ \bibnamefont
  {White}},\ }\bibfield  {title} {\bibinfo {title} {{Numerical canonical
  transformation approach to quantum many-body problems}},\ }\href@noop {}
  {\bibfield  {journal} {\bibinfo  {journal} {J. Chem. Phys.}\ }\textbf
  {\bibinfo {volume} {117}},\ \bibinfo {pages} {7472} (\bibinfo {year}
  {2002})}\BibitemShut {NoStop}%
\bibitem [{\citenamefont {Yanai}\ and\ \citenamefont
  {Chan}(2006)}]{Yanai:2006gi}%
  \BibitemOpen
  \bibfield  {author} {\bibinfo {author} {\bibfnamefont {T.}~\bibnamefont
  {Yanai}}\ and\ \bibinfo {author} {\bibfnamefont {G.~K.-L.}\ \bibnamefont
  {Chan}},\ }\bibfield  {title} {\bibinfo {title} {{Canonical transformation
  theory for multireference problems}},\ }\href@noop {} {\bibfield  {journal}
  {\bibinfo  {journal} {J. Chem. Phys.}\ }\textbf {\bibinfo {volume} {124}},\
  \bibinfo {pages} {194106} (\bibinfo {year} {2006})}\BibitemShut {NoStop}%
\bibitem [{\citenamefont {Yanai}\ \emph {et~al.}(2010)\citenamefont {Yanai},
  \citenamefont {Kurashige}, \citenamefont {Neuscamman},\ and\ \citenamefont
  {Chan}}]{Yanai:2010kf}%
  \BibitemOpen
  \bibfield  {author} {\bibinfo {author} {\bibfnamefont {T.}~\bibnamefont
  {Yanai}}, \bibinfo {author} {\bibfnamefont {Y.}~\bibnamefont {Kurashige}},
  \bibinfo {author} {\bibfnamefont {E.}~\bibnamefont {Neuscamman}},\ and\
  \bibinfo {author} {\bibfnamefont {G.~K.-L.}\ \bibnamefont {Chan}},\
  }\bibfield  {title} {\bibinfo {title} {{Multireference quantum chemistry
  through a joint density matrix renormalization group and canonical
  transformation theory}},\ }\href@noop {} {\bibfield  {journal} {\bibinfo
  {journal} {J. Chem. Phys.}\ }\textbf {\bibinfo {volume} {132}},\ \bibinfo
  {pages} {024105} (\bibinfo {year} {2010})}\BibitemShut {NoStop}%
\bibitem [{\citenamefont {Spiegelmann}\ and\ \citenamefont
  {Malrieu}(1984)}]{Spiegelmann1984}%
  \BibitemOpen
  \bibfield  {author} {\bibinfo {author} {\bibfnamefont {F.}~\bibnamefont
  {Spiegelmann}}\ and\ \bibinfo {author} {\bibfnamefont {J.~P.}\ \bibnamefont
  {Malrieu}},\ }\bibfield  {title} {\bibinfo {title} {{The use of effective
  Hamiltonians for the treatment of avoided crossings. I. Adiabatic potential
  curves}},\ }\href {https://doi.org/10.1088/0022-3700/17/7/012} {\bibfield
  {journal} {\bibinfo  {journal} {J. Phys. B At. Mol. Phys.}\ }\textbf
  {\bibinfo {volume} {17}},\ \bibinfo {pages} {1235} (\bibinfo {year}
  {1984})}\BibitemShut {NoStop}%
\bibitem [{\citenamefont {Lyakh}\ \emph {et~al.}(2012)\citenamefont {Lyakh},
  \citenamefont {Musia{\l}}, \citenamefont {Lotrich},\ and\ \citenamefont
  {Bartlett}}]{Lyakh:2012cn}%
  \BibitemOpen
  \bibfield  {author} {\bibinfo {author} {\bibfnamefont {D.~I.}\ \bibnamefont
  {Lyakh}}, \bibinfo {author} {\bibfnamefont {M.}~\bibnamefont {Musia{\l}}},
  \bibinfo {author} {\bibfnamefont {V.~F.}\ \bibnamefont {Lotrich}},\ and\
  \bibinfo {author} {\bibfnamefont {R.~J.}\ \bibnamefont {Bartlett}},\
  }\bibfield  {title} {\bibinfo {title} {{Multireference Nature of Chemistry:
  The Coupled-Cluster View}},\ }\href@noop {} {\bibfield  {journal} {\bibinfo
  {journal} {Chem. Rev.}\ }\textbf {\bibinfo {volume} {112}},\ \bibinfo {pages}
  {182} (\bibinfo {year} {2012})}\BibitemShut {NoStop}%
\bibitem [{\citenamefont {K{\"o}hn}\ \emph {et~al.}(2013)\citenamefont
  {K{\"o}hn}, \citenamefont {Hanauer}, \citenamefont {M{\"u}ck}, \citenamefont
  {Jagau},\ and\ \citenamefont {Gauss}}]{Kohn:2013cp}%
  \BibitemOpen
  \bibfield  {author} {\bibinfo {author} {\bibfnamefont {A.}~\bibnamefont
  {K{\"o}hn}}, \bibinfo {author} {\bibfnamefont {M.}~\bibnamefont {Hanauer}},
  \bibinfo {author} {\bibfnamefont {L.~A.}\ \bibnamefont {M{\"u}ck}}, \bibinfo
  {author} {\bibfnamefont {T.-C.}\ \bibnamefont {Jagau}},\ and\ \bibinfo
  {author} {\bibfnamefont {J.}~\bibnamefont {Gauss}},\ }\bibfield  {title}
  {\bibinfo {title} {{State-specific multireference coupled-cluster theory}},\
  }\href@noop {} {\bibfield  {journal} {\bibinfo  {journal} {Wiley Interdiscip.
  Rev.: Comput. Mol. Sci.}\ }\textbf {\bibinfo {volume} {3}},\ \bibinfo {pages}
  {176} (\bibinfo {year} {2013})}\BibitemShut {NoStop}%
\bibitem [{\citenamefont {Evangelista}(2018)}]{Evangelista:2018bt}%
  \BibitemOpen
  \bibfield  {author} {\bibinfo {author} {\bibfnamefont {F.~A.}\ \bibnamefont
  {Evangelista}},\ }\bibfield  {title} {\bibinfo {title} {{Perspective:
  Multireference coupled cluster theories of dynamical electron correlation}},\
  }\href@noop {} {\bibfield  {journal} {\bibinfo  {journal} {J. Chem. Phys.}\
  }\textbf {\bibinfo {volume} {149}},\ \bibinfo {pages} {030901} (\bibinfo
  {year} {2018})}\BibitemShut {NoStop}%
\bibitem [{\citenamefont {Bauman}\ \emph {et~al.}(2019)\citenamefont {Bauman},
  \citenamefont {Bylaska}, \citenamefont {Krishnamoorthy}, \citenamefont {Low},
  \citenamefont {Wiebe}, \citenamefont {Granade}, \citenamefont {Roetteler},
  \citenamefont {Troyer},\ and\ \citenamefont {Kowalski}}]{Bauman:2019dx}%
  \BibitemOpen
  \bibfield  {author} {\bibinfo {author} {\bibfnamefont {N.~P.}\ \bibnamefont
  {Bauman}}, \bibinfo {author} {\bibfnamefont {E.~J.}\ \bibnamefont {Bylaska}},
  \bibinfo {author} {\bibfnamefont {S.}~\bibnamefont {Krishnamoorthy}},
  \bibinfo {author} {\bibfnamefont {G.~H.}\ \bibnamefont {Low}}, \bibinfo
  {author} {\bibfnamefont {N.}~\bibnamefont {Wiebe}}, \bibinfo {author}
  {\bibfnamefont {C.~E.}\ \bibnamefont {Granade}}, \bibinfo {author}
  {\bibfnamefont {M.}~\bibnamefont {Roetteler}}, \bibinfo {author}
  {\bibfnamefont {M.}~\bibnamefont {Troyer}},\ and\ \bibinfo {author}
  {\bibfnamefont {K.}~\bibnamefont {Kowalski}},\ }\bibfield  {title} {\bibinfo
  {title} {{Downfolding of many-body Hamiltonians using active-space models:
  Extension of the sub-system embedding sub-algebras approach to unitary
  coupled cluster formalisms}},\ }\href@noop {} {\bibfield  {journal} {\bibinfo
   {journal} {J. Chem. Phys.}\ }\textbf {\bibinfo {volume} {151}},\ \bibinfo
  {pages} {014107} (\bibinfo {year} {2019})}\BibitemShut {NoStop}%
\bibitem [{\citenamefont {Metcalf}\ \emph {et~al.}(2020)\citenamefont
  {Metcalf}, \citenamefont {Bauman}, \citenamefont {Kowalski},\ and\
  \citenamefont {de~Jong}}]{Metcalf:2020fe}%
  \BibitemOpen
  \bibfield  {author} {\bibinfo {author} {\bibfnamefont {M.}~\bibnamefont
  {Metcalf}}, \bibinfo {author} {\bibfnamefont {N.~P.}\ \bibnamefont {Bauman}},
  \bibinfo {author} {\bibfnamefont {K.}~\bibnamefont {Kowalski}},\ and\
  \bibinfo {author} {\bibfnamefont {W.~A.}\ \bibnamefont {de~Jong}},\
  }\bibfield  {title} {\bibinfo {title} {{Resource-Efficient Chemistry on
  Quantum Computers with the Variational Quantum Eigensolver and the Double
  Unitary Coupled-Cluster Approach.}},\ }\href@noop {} {\bibfield  {journal}
  {\bibinfo  {journal} {J. Chem. Theory Comput.}\ }\textbf {\bibinfo {volume}
  {16}},\ \bibinfo {pages} {6165} (\bibinfo {year} {2020})}\BibitemShut
  {NoStop}%
\bibitem [{\citenamefont {Bauman}\ \emph {et~al.}(2021)\citenamefont {Bauman},
  \citenamefont {Chladek}, \citenamefont {Veis}, \citenamefont {Pittner},\ and\
  \citenamefont {Kowalski}}]{Nicholas:2021cy}%
  \BibitemOpen
  \bibfield  {author} {\bibinfo {author} {\bibfnamefont {N.~P.}\ \bibnamefont
  {Bauman}}, \bibinfo {author} {\bibfnamefont {J.}~\bibnamefont {Chladek}},
  \bibinfo {author} {\bibfnamefont {L.}~\bibnamefont {Veis}}, \bibinfo {author}
  {\bibfnamefont {J.}~\bibnamefont {Pittner}},\ and\ \bibinfo {author}
  {\bibfnamefont {K.}~\bibnamefont {Kowalski}},\ }\bibfield  {title} {\bibinfo
  {title} {{Variational quantum eigensolver for approximate diagonalization of
  downfolded Hamiltonians using generalized unitary coupled cluster ansatz}},\
  }\href@noop {} {\bibfield  {journal} {\bibinfo  {journal} {Quantum Sci.
  Technol.}\ }\textbf {\bibinfo {volume} {6}},\ \bibinfo {pages} {034008}
  (\bibinfo {year} {2021})}\BibitemShut {NoStop}%
\bibitem [{\citenamefont {Le}\ and\ \citenamefont {Tran}(2022)}]{Le2022}%
  \BibitemOpen
  \bibfield  {author} {\bibinfo {author} {\bibfnamefont {N.~T.}\ \bibnamefont
  {Le}}\ and\ \bibinfo {author} {\bibfnamefont {L.~N.}\ \bibnamefont {Tran}},\
  }\href@noop {} {\bibinfo {title} {Correlated reference-assisted variational
  quantum eigensolver}},\ \bibinfo {howpublished} {e-print arXiv:2205.03539
  [cond-mat.str-el]} (\bibinfo {year} {2022})\BibitemShut {NoStop}%
\bibitem [{\citenamefont {Evangelista}(2014)}]{evangelista2014driven}%
  \BibitemOpen
  \bibfield  {author} {\bibinfo {author} {\bibfnamefont {F.~A.}\ \bibnamefont
  {Evangelista}},\ }\bibfield  {title} {\bibinfo {title} {A driven similarity
  renormalization group approach to quantum many-body problems},\ }\href@noop
  {} {\bibfield  {journal} {\bibinfo  {journal} {J. Chem. Phys.}\ }\textbf
  {\bibinfo {volume} {141}},\ \bibinfo {pages} {054109} (\bibinfo {year}
  {2014})}\BibitemShut {NoStop}%
\bibitem [{\citenamefont {Li}\ and\ \citenamefont
  {Evangelista}(2015)}]{li2015multireference}%
  \BibitemOpen
  \bibfield  {author} {\bibinfo {author} {\bibfnamefont {C.}~\bibnamefont
  {Li}}\ and\ \bibinfo {author} {\bibfnamefont {F.~A.}\ \bibnamefont
  {Evangelista}},\ }\bibfield  {title} {\bibinfo {title} {Multireference driven
  similarity renormalization group: a second-order perturbative analysis},\
  }\href@noop {} {\bibfield  {journal} {\bibinfo  {journal} {J. Chem. Theory
  Comput.}\ }\textbf {\bibinfo {volume} {11}},\ \bibinfo {pages} {2097}
  (\bibinfo {year} {2015})}\BibitemShut {NoStop}%
\bibitem [{\citenamefont {Li}\ and\ \citenamefont
  {Evangelista}(2016{\natexlab{a}})}]{li2016towards}%
  \BibitemOpen
  \bibfield  {author} {\bibinfo {author} {\bibfnamefont {C.}~\bibnamefont
  {Li}}\ and\ \bibinfo {author} {\bibfnamefont {F.~A.}\ \bibnamefont
  {Evangelista}},\ }\bibfield  {title} {\bibinfo {title} {Towards numerically
  robust multireference theories: The driven similarity renormalization group
  truncated to one-and two-body operators},\ }\href@noop {} {\bibfield
  {journal} {\bibinfo  {journal} {J. Chem. Phys.}\ }\textbf {\bibinfo {volume}
  {144}},\ \bibinfo {pages} {164114} (\bibinfo {year}
  {2016}{\natexlab{a}})}\BibitemShut {NoStop}%
\bibitem [{\citenamefont {Li}\ and\ \citenamefont
  {Evangelista}(2019)}]{li2019multireference}%
  \BibitemOpen
  \bibfield  {author} {\bibinfo {author} {\bibfnamefont {C.}~\bibnamefont
  {Li}}\ and\ \bibinfo {author} {\bibfnamefont {F.~A.}\ \bibnamefont
  {Evangelista}},\ }\bibfield  {title} {\bibinfo {title} {Multireference
  theories of electron correlation based on the driven similarity
  renormalization group},\ }\href@noop {} {\bibfield  {journal} {\bibinfo
  {journal} {Annu. Rev. Phys. Chem.}\ }\textbf {\bibinfo {volume} {70}},\
  \bibinfo {pages} {245} (\bibinfo {year} {2019})}\BibitemShut {NoStop}%
\bibitem [{\citenamefont {Wegner}(1994)}]{Wegner:1994kh}%
  \BibitemOpen
  \bibfield  {author} {\bibinfo {author} {\bibfnamefont {F.}~\bibnamefont
  {Wegner}},\ }\bibfield  {title} {\bibinfo {title} {{Flow-equations for
  Hamiltonians}},\ }\href@noop {} {\bibfield  {journal} {\bibinfo  {journal}
  {Ann. Phys.}\ }\textbf {\bibinfo {volume} {506}},\ \bibinfo {pages} {77}
  (\bibinfo {year} {1994})}\BibitemShut {NoStop}%
\bibitem [{\citenamefont {G{\l}azek}\ and\ \citenamefont
  {Wilson}(1994)}]{Glazek:1994uu}%
  \BibitemOpen
  \bibfield  {author} {\bibinfo {author} {\bibfnamefont {S.~D.}\ \bibnamefont
  {G{\l}azek}}\ and\ \bibinfo {author} {\bibfnamefont {K.~G.}\ \bibnamefont
  {Wilson}},\ }\bibfield  {title} {\bibinfo {title} {{Perturbative
  renormalization group for Hamiltonians}},\ }\href@noop {} {\bibfield
  {journal} {\bibinfo  {journal} {Phys. Rev. D}\ }\textbf {\bibinfo {volume}
  {49}},\ \bibinfo {pages} {4214} (\bibinfo {year} {1994})}\BibitemShut
  {NoStop}%
\bibitem [{\citenamefont {Tsukiyama}\ \emph {et~al.}(2012)\citenamefont
  {Tsukiyama}, \citenamefont {Bogner},\ and\ \citenamefont
  {Schwenk}}]{Tsukiyama:2012dw}%
  \BibitemOpen
  \bibfield  {author} {\bibinfo {author} {\bibfnamefont {K.}~\bibnamefont
  {Tsukiyama}}, \bibinfo {author} {\bibfnamefont {S.~K.}\ \bibnamefont
  {Bogner}},\ and\ \bibinfo {author} {\bibfnamefont {A.}~\bibnamefont
  {Schwenk}},\ }\bibfield  {title} {\bibinfo {title} {{In-medium similarity
  renormalization group for open-shell nuclei}},\ }\href@noop {} {\bibfield
  {journal} {\bibinfo  {journal} {Phys. Rev. C}\ }\textbf {\bibinfo {volume}
  {85}},\ \bibinfo {pages} {061304} (\bibinfo {year} {2012})}\BibitemShut
  {NoStop}%
\bibitem [{\citenamefont {Hergert}(2017)}]{Hergert:2017bk}%
  \BibitemOpen
  \bibfield  {author} {\bibinfo {author} {\bibfnamefont {H.}~\bibnamefont
  {Hergert}},\ }\bibfield  {title} {\bibinfo {title} {{In-medium similarity
  renormalization group for closed and open-shell nuclei}},\ }\href@noop {}
  {\bibfield  {journal} {\bibinfo  {journal} {Physica Scripta}\ }\textbf
  {\bibinfo {volume} {92}},\ \bibinfo {pages} {023002} (\bibinfo {year}
  {2017})}\BibitemShut {NoStop}%
\bibitem [{\citenamefont {Li}\ and\ \citenamefont
  {Evangelista}(2021)}]{Li:2021eb}%
  \BibitemOpen
  \bibfield  {author} {\bibinfo {author} {\bibfnamefont {C.}~\bibnamefont
  {Li}}\ and\ \bibinfo {author} {\bibfnamefont {F.~A.}\ \bibnamefont
  {Evangelista}},\ }\bibfield  {title} {\bibinfo {title} {{Spin-free
  formulation of the multireference driven similarity renormalization group: A
  benchmark study of first-row diatomic molecules and spin-crossover
  energetics}},\ }\href@noop {} {\bibfield  {journal} {\bibinfo  {journal} {J.
  Chem. Phys.}\ }\textbf {\bibinfo {volume} {155}},\ \bibinfo {pages} {114111}
  (\bibinfo {year} {2021})}\BibitemShut {NoStop}%
\bibitem [{\citenamefont {Watts}\ \emph {et~al.}(1989)\citenamefont {Watts},
  \citenamefont {Trucks},\ and\ \citenamefont {Bartlett}}]{Watts:1989ve}%
  \BibitemOpen
  \bibfield  {author} {\bibinfo {author} {\bibfnamefont {J.~D.}\ \bibnamefont
  {Watts}}, \bibinfo {author} {\bibfnamefont {G.~W.}\ \bibnamefont {Trucks}},\
  and\ \bibinfo {author} {\bibfnamefont {R.~J.}\ \bibnamefont {Bartlett}},\
  }\bibfield  {title} {\bibinfo {title} {{The unitary coupled-cluster approach
  and molecular properties. Applications of the UCC(4) method}},\ }\href@noop
  {} {\bibfield  {journal} {\bibinfo  {journal} {Chem. Phys. Lett.}\ }\textbf
  {\bibinfo {volume} {157}},\ \bibinfo {pages} {359} (\bibinfo {year}
  {1989})}\BibitemShut {NoStop}%
\bibitem [{\citenamefont {Musial}\ and\ \citenamefont
  {Bartlett}(2008{\natexlab{a}})}]{Musial:2008kn}%
  \BibitemOpen
  \bibfield  {author} {\bibinfo {author} {\bibfnamefont {M.}~\bibnamefont
  {Musial}}\ and\ \bibinfo {author} {\bibfnamefont {R.~J.}\ \bibnamefont
  {Bartlett}},\ }\bibfield  {title} {\bibinfo {title} {{Intermediate
  Hamiltonian Fock-space multireference coupled-cluster method with full
  triples for calculation of excitation energies}},\ }\href@noop {} {\bibfield
  {journal} {\bibinfo  {journal} {J. Chem. Phys.}\ }\textbf {\bibinfo {volume}
  {129}},\ \bibinfo {pages} {044101} (\bibinfo {year}
  {2008}{\natexlab{a}})}\BibitemShut {NoStop}%
\bibitem [{\citenamefont {Musial}\ and\ \citenamefont
  {Bartlett}(2008{\natexlab{b}})}]{Musial:2008ea}%
  \BibitemOpen
  \bibfield  {author} {\bibinfo {author} {\bibfnamefont {M.}~\bibnamefont
  {Musial}}\ and\ \bibinfo {author} {\bibfnamefont {R.~J.}\ \bibnamefont
  {Bartlett}},\ }\bibfield  {title} {\bibinfo {title} {{Multireference
  Fock-space coupled-cluster and equation-of-motion coupled-cluster theories:
  The detailed interconnections}},\ }\href@noop {} {\bibfield  {journal}
  {\bibinfo  {journal} {J. Chem. Phys.}\ }\textbf {\bibinfo {volume} {129}},\
  \bibinfo {pages} {134105} (\bibinfo {year} {2008}{\natexlab{b}})}\BibitemShut
  {NoStop}%
\bibitem [{\citenamefont {Crawford}\ and\ \citenamefont
  {Schaefer}(2000)}]{Crawford:2000ub}%
  \BibitemOpen
  \bibfield  {author} {\bibinfo {author} {\bibfnamefont {T.~D.}\ \bibnamefont
  {Crawford}}\ and\ \bibinfo {author} {\bibfnamefont {H.~F.}\ \bibnamefont
  {Schaefer}},\ }\bibfield  {title} {\bibinfo {title} {{An introduction to
  coupled cluster theory for computational chemists}},\ }\href@noop {}
  {\bibfield  {journal} {\bibinfo  {journal} {Rev. Comp. Chem.}\ }\textbf
  {\bibinfo {volume} {14}},\ \bibinfo {pages} {33} (\bibinfo {year}
  {2000})}\BibitemShut {NoStop}%
\bibitem [{\citenamefont {Mukherjee}(1997)}]{Mukherjee:1997tk}%
  \BibitemOpen
  \bibfield  {author} {\bibinfo {author} {\bibfnamefont {D.}~\bibnamefont
  {Mukherjee}},\ }\bibfield  {title} {\bibinfo {title} {{Normal ordering and a
  Wick-like reduction theorem for fermions with respect to a
  multi-determinantal reference state}},\ }\href@noop {} {\bibfield  {journal}
  {\bibinfo  {journal} {Chem. Phys. Lett.}\ }\textbf {\bibinfo {volume}
  {274}},\ \bibinfo {pages} {561} (\bibinfo {year} {1997})}\BibitemShut
  {NoStop}%
\bibitem [{\citenamefont {Kutzelnigg}\ and\ \citenamefont
  {Mukherjee}(1997)}]{Kutzelnigg:1997ut}%
  \BibitemOpen
  \bibfield  {author} {\bibinfo {author} {\bibfnamefont {W.}~\bibnamefont
  {Kutzelnigg}}\ and\ \bibinfo {author} {\bibfnamefont {D.}~\bibnamefont
  {Mukherjee}},\ }\bibfield  {title} {\bibinfo {title} {{Normal order and
  extended Wick theorem for a multiconfiguration reference wave function}},\
  }\href@noop {} {\bibfield  {journal} {\bibinfo  {journal} {J. Chem. Phys.}\
  }\textbf {\bibinfo {volume} {107}},\ \bibinfo {pages} {432} (\bibinfo {year}
  {1997})}\BibitemShut {NoStop}%
\bibitem [{\citenamefont {Andersson}\ \emph {et~al.}(1992)\citenamefont
  {Andersson}, \citenamefont {Malmqvist},\ and\ \citenamefont
  {Roos}}]{andersson1992second}%
  \BibitemOpen
  \bibfield  {author} {\bibinfo {author} {\bibfnamefont {K.}~\bibnamefont
  {Andersson}}, \bibinfo {author} {\bibfnamefont {P.-{\AA}.}\ \bibnamefont
  {Malmqvist}},\ and\ \bibinfo {author} {\bibfnamefont {B.~O.}\ \bibnamefont
  {Roos}},\ }\bibfield  {title} {\bibinfo {title} {Second-order perturbation
  theory with a complete active space self-consistent field reference
  function},\ }\href@noop {} {\bibfield  {journal} {\bibinfo  {journal} {J.
  Chem. Phys.}\ }\textbf {\bibinfo {volume} {96}},\ \bibinfo {pages} {1218}
  (\bibinfo {year} {1992})}\BibitemShut {NoStop}%
\bibitem [{\citenamefont {Mazziotti}(1998)}]{Mazziotti_1998}%
  \BibitemOpen
  \bibfield  {author} {\bibinfo {author} {\bibfnamefont {D.~A.}\ \bibnamefont
  {Mazziotti}},\ }\bibfield  {title} {\bibinfo {title} {Contracted schrödinger
  equation: Determining quantum energies and two-particle density matrices
  without wave functions},\ }\href@noop {} {\bibfield  {journal} {\bibinfo
  {journal} {Phys. Rev. A}\ }\textbf {\bibinfo {volume} {57}},\ \bibinfo
  {pages} {4219–4234} (\bibinfo {year} {1998})}\BibitemShut {NoStop}%
\bibitem [{\citenamefont {Kutzelnigg}\ and\ \citenamefont
  {Mukherjee}(1999)}]{Kutzelnigg_Mukherjee_1999}%
  \BibitemOpen
  \bibfield  {author} {\bibinfo {author} {\bibfnamefont {W.}~\bibnamefont
  {Kutzelnigg}}\ and\ \bibinfo {author} {\bibfnamefont {D.}~\bibnamefont
  {Mukherjee}},\ }\bibfield  {title} {\bibinfo {title} {Cumulant expansion of
  the reduced density matrices},\ }\href@noop {} {\bibfield  {journal}
  {\bibinfo  {journal} {The Journal of Chemical Physics}\ }\textbf {\bibinfo
  {volume} {110}},\ \bibinfo {pages} {2800–2809} (\bibinfo {year}
  {1999})}\BibitemShut {NoStop}%
\bibitem [{\citenamefont {Wang}\ \emph {et~al.}(2021)\citenamefont {Wang},
  \citenamefont {Li},\ and\ \citenamefont
  {Evangelista}}]{Wang.2021.10.1021/acs.jctc.1c00980}%
  \BibitemOpen
  \bibfield  {author} {\bibinfo {author} {\bibfnamefont {S.}~\bibnamefont
  {Wang}}, \bibinfo {author} {\bibfnamefont {C.}~\bibnamefont {Li}},\ and\
  \bibinfo {author} {\bibfnamefont {F.~A.}\ \bibnamefont {Evangelista}},\
  }\bibfield  {title} {\bibinfo {title} {{Analytic Energy Gradients for the
  Driven Similarity Renormalization Group Multireference Second-Order
  Perturbation Theory}},\ }\href {https://doi.org/10.1021/acs.jctc.1c00980}
  {\bibfield  {journal} {\bibinfo  {journal} {J. Chem. Theory Comput.}\
  }\textbf {\bibinfo {volume} {17}},\ \bibinfo {pages} {7666} (\bibinfo {year}
  {2021})}\BibitemShut {NoStop}%
\bibitem [{\citenamefont {He}\ \emph {et~al.}(2022)\citenamefont {He},
  \citenamefont {Li},\ and\ \citenamefont
  {Evangelista}}]{He.2022.10.1021/acs.jctc.1c01099}%
  \BibitemOpen
  \bibfield  {author} {\bibinfo {author} {\bibfnamefont {N.}~\bibnamefont
  {He}}, \bibinfo {author} {\bibfnamefont {C.}~\bibnamefont {Li}},\ and\
  \bibinfo {author} {\bibfnamefont {F.~A.}\ \bibnamefont {Evangelista}},\
  }\bibfield  {title} {\bibinfo {title} {{Second-Order Active-Space Embedding
  Theory}},\ }\href {https://doi.org/10.1021/acs.jctc.1c01099} {\bibfield
  {journal} {\bibinfo  {journal} {J. Chem. Theory Comput.}\ }\textbf {\bibinfo
  {volume} {18}},\ \bibinfo {pages} {1527} (\bibinfo {year}
  {2022})}\BibitemShut {NoStop}%
\bibitem [{\citenamefont {Evangelista}(2021)}]{Evangelista2021Forte}%
  \BibitemOpen
  \bibfield  {author} {\bibinfo {author} {\bibfnamefont {F.~A.}\ \bibnamefont
  {Evangelista}},\ }\href {https://github.com/evangelistalab/forte} {\bibinfo
  {title} {Forte: a suite of quantum chemistry methods for strongly correlated
  electrons}} (\bibinfo {year} {2021})\BibitemShut {NoStop}%
\bibitem [{\citenamefont {Smith}\ \emph {et~al.}(2020)\citenamefont {Smith},
  \citenamefont {Burns}, \citenamefont {Simmonett}, \citenamefont {Parrish},
  \citenamefont {Schieber}, \citenamefont {Galvelis}, \citenamefont {Kraus},
  \citenamefont {Kruse}, \citenamefont {Di~Remigio}, \citenamefont {Alenaizan},
  \citenamefont {James}, \citenamefont {Lehtola}, \citenamefont {Misiewicz},
  \citenamefont {Scheurer}, \citenamefont {Shaw}, \citenamefont {Schriber},
  \citenamefont {Xie}, \citenamefont {Glick}, \citenamefont {Sirianni},
  \citenamefont {O'Brien}, \citenamefont {Waldrop}, \citenamefont {Kumar},
  \citenamefont {Hohenstein}, \citenamefont {Pritchard}, \citenamefont
  {Brooks}, \citenamefont {Schaefer}, \citenamefont {Sokolov}, \citenamefont
  {Patkowski}, \citenamefont {DePrince}, \citenamefont {Bozkaya}, \citenamefont
  {King}, \citenamefont {Evangelista}, \citenamefont {Turney}, \citenamefont
  {Crawford},\ and\ \citenamefont {Sherrill}}]{Smith:2020ci}%
  \BibitemOpen
  \bibfield  {author} {\bibinfo {author} {\bibfnamefont {D.~G.~A.}\
  \bibnamefont {Smith}}, \bibinfo {author} {\bibfnamefont {L.~A.}\ \bibnamefont
  {Burns}}, \bibinfo {author} {\bibfnamefont {A.~C.}\ \bibnamefont
  {Simmonett}}, \bibinfo {author} {\bibfnamefont {R.~M.}\ \bibnamefont
  {Parrish}}, \bibinfo {author} {\bibfnamefont {M.~C.}\ \bibnamefont
  {Schieber}}, \bibinfo {author} {\bibfnamefont {R.}~\bibnamefont {Galvelis}},
  \bibinfo {author} {\bibfnamefont {P.}~\bibnamefont {Kraus}}, \bibinfo
  {author} {\bibfnamefont {H.}~\bibnamefont {Kruse}}, \bibinfo {author}
  {\bibfnamefont {R.}~\bibnamefont {Di~Remigio}}, \bibinfo {author}
  {\bibfnamefont {A.}~\bibnamefont {Alenaizan}}, \bibinfo {author}
  {\bibfnamefont {A.~M.}\ \bibnamefont {James}}, \bibinfo {author}
  {\bibfnamefont {S.}~\bibnamefont {Lehtola}}, \bibinfo {author} {\bibfnamefont
  {J.~P.}\ \bibnamefont {Misiewicz}}, \bibinfo {author} {\bibfnamefont
  {M.}~\bibnamefont {Scheurer}}, \bibinfo {author} {\bibfnamefont {R.~A.}\
  \bibnamefont {Shaw}}, \bibinfo {author} {\bibfnamefont {J.~B.}\ \bibnamefont
  {Schriber}}, \bibinfo {author} {\bibfnamefont {Y.}~\bibnamefont {Xie}},
  \bibinfo {author} {\bibfnamefont {Z.~L.}\ \bibnamefont {Glick}}, \bibinfo
  {author} {\bibfnamefont {D.~A.}\ \bibnamefont {Sirianni}}, \bibinfo {author}
  {\bibfnamefont {J.~S.}\ \bibnamefont {O'Brien}}, \bibinfo {author}
  {\bibfnamefont {J.~M.}\ \bibnamefont {Waldrop}}, \bibinfo {author}
  {\bibfnamefont {A.}~\bibnamefont {Kumar}}, \bibinfo {author} {\bibfnamefont
  {E.~G.}\ \bibnamefont {Hohenstein}}, \bibinfo {author} {\bibfnamefont
  {B.~P.}\ \bibnamefont {Pritchard}}, \bibinfo {author} {\bibfnamefont {B.~R.}\
  \bibnamefont {Brooks}}, \bibinfo {author} {\bibfnamefont {H.~F.~I.}\
  \bibnamefont {Schaefer}}, \bibinfo {author} {\bibfnamefont {A.~Y.}\
  \bibnamefont {Sokolov}}, \bibinfo {author} {\bibfnamefont {K.}~\bibnamefont
  {Patkowski}}, \bibinfo {author} {\bibfnamefont {A.~E.~I.}\ \bibnamefont
  {DePrince}}, \bibinfo {author} {\bibfnamefont {U.}~\bibnamefont {Bozkaya}},
  \bibinfo {author} {\bibfnamefont {R.~A.}\ \bibnamefont {King}}, \bibinfo
  {author} {\bibfnamefont {F.~A.}\ \bibnamefont {Evangelista}}, \bibinfo
  {author} {\bibfnamefont {J.~M.}\ \bibnamefont {Turney}}, \bibinfo {author}
  {\bibfnamefont {T.~D.}\ \bibnamefont {Crawford}},\ and\ \bibinfo {author}
  {\bibfnamefont {C.~D.}\ \bibnamefont {Sherrill}},\ }\bibfield  {title}
  {\bibinfo {title} {{PSI4 1.4: Open-source software for high-throughput
  quantum chemistry}},\ }\href@noop {} {\bibfield  {journal} {\bibinfo
  {journal} {J. Chem. Phys.}\ }\textbf {\bibinfo {volume} {152}},\ \bibinfo
  {pages} {184108} (\bibinfo {year} {2020})}\BibitemShut {NoStop}%
\bibitem [{\citenamefont {Dunning}(1989)}]{cc-pvdz-dk-H-Ne}%
  \BibitemOpen
  \bibfield  {author} {\bibinfo {author} {\bibfnamefont {T.~H.}\ \bibnamefont
  {Dunning}},\ }\bibfield  {title} {\bibinfo {title} {{Gaussian basis sets for
  use in correlated molecular calculations. I. The atoms boron through neon and
  hydrogen}},\ }\href@noop {} {\bibfield  {journal} {\bibinfo  {journal} {J.
  Chem. Phys.}\ }\textbf {\bibinfo {volume} {90}},\ \bibinfo {pages} {1007}
  (\bibinfo {year} {1989})}\BibitemShut {NoStop}%
\bibitem [{\citenamefont {Colmenero}\ \emph {et~al.}(1993)\citenamefont
  {Colmenero}, \citenamefont {C},\ and\ \citenamefont
  {Valdemoro}}]{Colmenero_C_Valdemoro_1993}%
  \BibitemOpen
  \bibfield  {author} {\bibinfo {author} {\bibnamefont {Colmenero}}, \bibinfo
  {author} {\bibfnamefont {P.~d.~V.}\ \bibnamefont {C}},\ and\ \bibinfo
  {author} {\bibnamefont {Valdemoro}},\ }\bibfield  {title} {\bibinfo {title}
  {Approximating q-order reduced density matrices in terms of the lower-order
  ones. i. general relations.},\ }\href@noop {} {\bibfield  {journal} {\bibinfo
   {journal} {Phys. Rev. A}\ }\textbf {\bibinfo {volume} {47}},\ \bibinfo
  {pages} {971–978} (\bibinfo {year} {1993})}\BibitemShut {NoStop}%
\bibitem [{\citenamefont {DePrince}\ and\ \citenamefont
  {Mazziotti}(2007)}]{DePrince_Mazziotti_2007}%
  \BibitemOpen
  \bibfield  {author} {\bibinfo {author} {\bibfnamefont {A.~E.}\ \bibnamefont
  {DePrince}}\ and\ \bibinfo {author} {\bibfnamefont {D.~A.}\ \bibnamefont
  {Mazziotti}},\ }\bibfield  {title} {\bibinfo {title} {Cumulant reconstruction
  of the three-electron reduced density matrix in the anti-hermitian contracted
  schrödinger equation},\ }\href@noop {} {\bibfield  {journal} {\bibinfo
  {journal} {J. Chem. Phys.}\ }\textbf {\bibinfo {volume} {127}},\ \bibinfo
  {pages} {104104} (\bibinfo {year} {2007})}\BibitemShut {NoStop}%
\bibitem [{\citenamefont {Abrams}\ and\ \citenamefont
  {Sherrill}(2004)}]{Abrams_Sherrill_2004}%
  \BibitemOpen
  \bibfield  {author} {\bibinfo {author} {\bibfnamefont {M.~L.}\ \bibnamefont
  {Abrams}}\ and\ \bibinfo {author} {\bibfnamefont {C.~D.}\ \bibnamefont
  {Sherrill}},\ }\bibfield  {title} {\bibinfo {title} {Natural orbitals as
  substitutes for optimized orbitals in complete active space wavefunctions},\
  }\href@noop {} {\bibfield  {journal} {\bibinfo  {journal} {Chem. Phys.
  Lett.}\ }\textbf {\bibinfo {volume} {395}},\ \bibinfo {pages} {227–232}
  (\bibinfo {year} {2004})}\BibitemShut {NoStop}%
\bibitem [{\citenamefont {Romero}\ \emph {et~al.}(2019)\citenamefont {Romero},
  \citenamefont {Babbush}, \citenamefont {McClean}, \citenamefont {Hempel},
  \citenamefont {Love},\ and\ \citenamefont {Aspuru-Guzik}}]{Romero:2019hk}%
  \BibitemOpen
  \bibfield  {author} {\bibinfo {author} {\bibfnamefont {J.}~\bibnamefont
  {Romero}}, \bibinfo {author} {\bibfnamefont {R.}~\bibnamefont {Babbush}},
  \bibinfo {author} {\bibfnamefont {J.~R.}\ \bibnamefont {McClean}}, \bibinfo
  {author} {\bibfnamefont {C.}~\bibnamefont {Hempel}}, \bibinfo {author}
  {\bibfnamefont {P.~J.}\ \bibnamefont {Love}},\ and\ \bibinfo {author}
  {\bibfnamefont {A.}~\bibnamefont {Aspuru-Guzik}},\ }\bibfield  {title}
  {\bibinfo {title} {{Strategies for quantum computing molecular energies using
  the unitary coupled cluster ansatz}},\ }\href@noop {} {\bibfield  {journal}
  {\bibinfo  {journal} {Quantum Sci. Technol.}\ }\textbf {\bibinfo {volume}
  {4}},\ \bibinfo {pages} {014008} (\bibinfo {year} {2019})}\BibitemShut
  {NoStop}%
\bibitem [{\citenamefont {Garrod}\ and\ \citenamefont
  {Percus}(1964)}]{Percus64_1756}%
  \BibitemOpen
  \bibfield  {author} {\bibinfo {author} {\bibfnamefont {C.}~\bibnamefont
  {Garrod}}\ and\ \bibinfo {author} {\bibfnamefont {J.~K.}\ \bibnamefont
  {Percus}},\ }\bibfield  {title} {\bibinfo {title} {Reduction of the
  $n$-particle variational problem},\ }\href@noop {} {\bibfield  {journal}
  {\bibinfo  {journal} {J. Math. Phys.}\ }\textbf {\bibinfo {volume} {5}},\
  \bibinfo {pages} {1756} (\bibinfo {year} {1964})}\BibitemShut {NoStop}%
\bibitem [{\citenamefont {Erdahl}(1978)}]{Erdahl78_697}%
  \BibitemOpen
  \bibfield  {author} {\bibinfo {author} {\bibfnamefont {R.~M.}\ \bibnamefont
  {Erdahl}},\ }\bibfield  {title} {\bibinfo {title} {Representability},\
  }\href@noop {} {\bibfield  {journal} {\bibinfo  {journal} {Int. J. Quantum
  Chem.}\ }\textbf {\bibinfo {volume} {13}},\ \bibinfo {pages} {697} (\bibinfo
  {year} {1978})}\BibitemShut {NoStop}%
\bibitem [{\citenamefont {Zhao}\ \emph {et~al.}(2004)\citenamefont {Zhao},
  \citenamefont {Braams}, \citenamefont {Fukuda}, \citenamefont {Overton},\
  and\ \citenamefont {Percus}}]{Percus04_2095}%
  \BibitemOpen
  \bibfield  {author} {\bibinfo {author} {\bibfnamefont {Z.}~\bibnamefont
  {Zhao}}, \bibinfo {author} {\bibfnamefont {B.~J.}\ \bibnamefont {Braams}},
  \bibinfo {author} {\bibfnamefont {M.}~\bibnamefont {Fukuda}}, \bibinfo
  {author} {\bibfnamefont {M.~L.}\ \bibnamefont {Overton}},\ and\ \bibinfo
  {author} {\bibfnamefont {J.~K.}\ \bibnamefont {Percus}},\ }\bibfield  {title}
  {\bibinfo {title} {The reduced density matrix method for electronic structure
  calculations and the role of three-index representability conditions},\
  }\href@noop {} {\bibfield  {journal} {\bibinfo  {journal} {J. Comp. Phys.}\
  }\textbf {\bibinfo {volume} {120}},\ \bibinfo {pages} {2095} (\bibinfo {year}
  {2004})}\BibitemShut {NoStop}%
\bibitem [{\citenamefont {Mazziotti}(2012)}]{Mazziotti12_263002}%
  \BibitemOpen
  \bibfield  {author} {\bibinfo {author} {\bibfnamefont {D.~A.}\ \bibnamefont
  {Mazziotti}},\ }\bibfield  {title} {\bibinfo {title} {Structure of fermionic
  density matrices: Complete $n$-representability conditions},\ }\href@noop {}
  {\bibfield  {journal} {\bibinfo  {journal} {Phys. Rev. Lett.}\ }\textbf
  {\bibinfo {volume} {108}},\ \bibinfo {pages} {263002} (\bibinfo {year}
  {2012})}\BibitemShut {NoStop}%
\bibitem [{\citenamefont {Rubin}\ \emph {et~al.}(2018)\citenamefont {Rubin},
  \citenamefont {Babbush},\ and\ \citenamefont
  {McClean}}]{rubin2018application}%
  \BibitemOpen
  \bibfield  {author} {\bibinfo {author} {\bibfnamefont {N.~C.}\ \bibnamefont
  {Rubin}}, \bibinfo {author} {\bibfnamefont {R.}~\bibnamefont {Babbush}},\
  and\ \bibinfo {author} {\bibfnamefont {J.}~\bibnamefont {McClean}},\
  }\bibfield  {title} {\bibinfo {title} {Application of fermionic marginal
  constraints to hybrid quantum algorithms},\ }\href@noop {} {\bibfield
  {journal} {\bibinfo  {journal} {New J. Phys.}\ }\textbf {\bibinfo {volume}
  {20}},\ \bibinfo {pages} {053020} (\bibinfo {year} {2018})}\BibitemShut
  {NoStop}%
\bibitem [{\citenamefont {{Google AI Quantum and
  Collaborators}}(2020)}]{GoogleAIQuantumandCollaborators:2020bs}%
  \BibitemOpen
  \bibfield  {author} {\bibinfo {author} {\bibnamefont {{Google AI Quantum and
  Collaborators}}},\ }\bibfield  {title} {\bibinfo {title} {{Hartree-Fock on a
  superconducting qubit quantum computer.}},\ }\href@noop {} {\bibfield
  {journal} {\bibinfo  {journal} {Science}\ }\textbf {\bibinfo {volume}
  {369}},\ \bibinfo {pages} {1084} (\bibinfo {year} {2020})}\BibitemShut
  {NoStop}%
\bibitem [{\citenamefont {Shavitt}\ and\ \citenamefont
  {Bartlett}(2009)}]{shavitt2009many}%
  \BibitemOpen
  \bibfield  {author} {\bibinfo {author} {\bibfnamefont {I.}~\bibnamefont
  {Shavitt}}\ and\ \bibinfo {author} {\bibfnamefont {R.~J.}\ \bibnamefont
  {Bartlett}},\ }\href@noop {} {\emph {\bibinfo {title} {Many-body methods in
  chemistry and physics: MBPT and coupled-cluster theory}}}\ (\bibinfo
  {publisher} {Cambridge university press},\ \bibinfo {year}
  {2009})\BibitemShut {NoStop}%
\bibitem [{\citenamefont {Raghavachari}\ \emph {et~al.}(1989)\citenamefont
  {Raghavachari}, \citenamefont {Trucks}, \citenamefont {Pople},\ and\
  \citenamefont {Head-Gordon}}]{Raghavachari:1989vx}%
  \BibitemOpen
  \bibfield  {author} {\bibinfo {author} {\bibfnamefont {K.}~\bibnamefont
  {Raghavachari}}, \bibinfo {author} {\bibfnamefont {G.~W.}\ \bibnamefont
  {Trucks}}, \bibinfo {author} {\bibfnamefont {J.~A.}\ \bibnamefont {Pople}},\
  and\ \bibinfo {author} {\bibfnamefont {M.}~\bibnamefont {Head-Gordon}},\
  }\bibfield  {title} {\bibinfo {title} {{A fifth-order perturbation comparison
  of electron correlation theories}},\ }\href@noop {} {\bibfield  {journal}
  {\bibinfo  {journal} {Chem. Phys. Lett.}\ }\textbf {\bibinfo {volume}
  {157}},\ \bibinfo {pages} {479} (\bibinfo {year} {1989})}\BibitemShut
  {NoStop}%
\bibitem [{\citenamefont {Evangelista}\ \emph {et~al.}(2012)\citenamefont
  {Evangelista}, \citenamefont {Hanauer}, \citenamefont {Koehn},\ and\
  \citenamefont {Gauss}}]{Evangelista:2012fo}%
  \BibitemOpen
  \bibfield  {author} {\bibinfo {author} {\bibfnamefont {F.~A.}\ \bibnamefont
  {Evangelista}}, \bibinfo {author} {\bibfnamefont {M.}~\bibnamefont
  {Hanauer}}, \bibinfo {author} {\bibfnamefont {A.}~\bibnamefont {Koehn}},\
  and\ \bibinfo {author} {\bibfnamefont {J.}~\bibnamefont {Gauss}},\ }\bibfield
   {title} {\bibinfo {title} {{A sequential transformation approach to the
  internally contracted multireference coupled cluster method}},\ }\href@noop
  {} {\bibfield  {journal} {\bibinfo  {journal} {J. Chem. Phys.}\ }\textbf
  {\bibinfo {volume} {136}},\ \bibinfo {pages} {204108} (\bibinfo {year}
  {2012})}\BibitemShut {NoStop}%
\bibitem [{\citenamefont {Leopold}\ \emph {et~al.}(1986)\citenamefont
  {Leopold}, \citenamefont {Miller},\ and\ \citenamefont
  {Lineberger}}]{Leopold:1986vo}%
  \BibitemOpen
  \bibfield  {author} {\bibinfo {author} {\bibfnamefont {D.}~\bibnamefont
  {Leopold}}, \bibinfo {author} {\bibfnamefont {A.}~\bibnamefont {Miller}},\
  and\ \bibinfo {author} {\bibfnamefont {W.}~\bibnamefont {Lineberger}},\
  }\bibfield  {title} {\bibinfo {title} {{Determination of the Singlet Triplet
  Splitting and Electron-Affinity of Ortho-Benzyne by Negative-Ion
  Photoelectron-Spectroscopy}},\ }\href@noop {} {\bibfield  {journal} {\bibinfo
   {journal} {J. Am. Chem. Soc.}\ }\textbf {\bibinfo {volume} {108}},\ \bibinfo
  {pages} {1379} (\bibinfo {year} {1986})}\BibitemShut {NoStop}%
\bibitem [{\citenamefont {Evangelista}\ \emph {et~al.}(2007)\citenamefont
  {Evangelista}, \citenamefont {Allen},\ and\ \citenamefont
  {Schaefer}}]{Evangelista:2007hz}%
  \BibitemOpen
  \bibfield  {author} {\bibinfo {author} {\bibfnamefont {F.~A.}\ \bibnamefont
  {Evangelista}}, \bibinfo {author} {\bibfnamefont {W.~D.}\ \bibnamefont
  {Allen}},\ and\ \bibinfo {author} {\bibfnamefont {H.~F.}\ \bibnamefont
  {Schaefer}},\ }\bibfield  {title} {\bibinfo {title} {{Coupling term
  derivation and general implementation of state-specific multireference
  coupled cluster theories}},\ }\href@noop {} {\bibfield  {journal} {\bibinfo
  {journal} {J. Chem. Phys.}\ }\textbf {\bibinfo {volume} {127}},\ \bibinfo
  {pages} {024102} (\bibinfo {year} {2007})}\BibitemShut {NoStop}%
\bibitem [{\citenamefont {Hanauer}\ and\ \citenamefont
  {K{\"o}hn}(2012)}]{Hanauer:2012gf}%
  \BibitemOpen
  \bibfield  {author} {\bibinfo {author} {\bibfnamefont {M.}~\bibnamefont
  {Hanauer}}\ and\ \bibinfo {author} {\bibfnamefont {A.}~\bibnamefont
  {K{\"o}hn}},\ }\bibfield  {title} {\bibinfo {title} {{Perturbative treatment
  of triple excitations in internally contracted multireference coupled cluster
  theory}},\ }\href@noop {} {\bibfield  {journal} {\bibinfo  {journal} {J.
  Chem. Phys.}\ }\textbf {\bibinfo {volume} {136}},\ \bibinfo {pages} {204107}
  (\bibinfo {year} {2012})}\BibitemShut {NoStop}%
\bibitem [{\citenamefont {Li}\ and\ \citenamefont {Paldus}(2008)}]{Li:2008tk}%
  \BibitemOpen
  \bibfield  {author} {\bibinfo {author} {\bibfnamefont {X.}~\bibnamefont
  {Li}}\ and\ \bibinfo {author} {\bibfnamefont {J.}~\bibnamefont {Paldus}},\
  }\bibfield  {title} {\bibinfo {title} {{Electronic structure of organic
  diradicals: Evaluation of the performance of coupled-cluster methods}},\
  }\href@noop {} {\bibfield  {journal} {\bibinfo  {journal} {J. Chem. Phys.}\
  }\textbf {\bibinfo {volume} {129}},\ \bibinfo {pages} {174101} (\bibinfo
  {year} {2008})}\BibitemShut {NoStop}%
\bibitem [{\citenamefont {Cramer}\ \emph {et~al.}(1997)\citenamefont {Cramer},
  \citenamefont {Nash},\ and\ \citenamefont {Squires}}]{Cramer:1997uh}%
  \BibitemOpen
  \bibfield  {author} {\bibinfo {author} {\bibfnamefont {C.~J.}\ \bibnamefont
  {Cramer}}, \bibinfo {author} {\bibfnamefont {J.}~\bibnamefont {Nash}},\ and\
  \bibinfo {author} {\bibfnamefont {R.}~\bibnamefont {Squires}},\ }\bibfield
  {title} {\bibinfo {title} {{A reinvestigation of singlet benzyne
  thermochemistry predicted by CASPT2, coupled-cluster and density functional
  calculations}},\ }\href@noop {} {\bibfield  {journal} {\bibinfo  {journal}
  {Chem. Phys. Lett.}\ }\textbf {\bibinfo {volume} {277}},\ \bibinfo {pages}
  {311} (\bibinfo {year} {1997})}\BibitemShut {NoStop}%
\bibitem [{\citenamefont {Lindh}\ \emph {et~al.}(1999)\citenamefont {Lindh},
  \citenamefont {Bernhardsson},\ and\ \citenamefont
  {Sch{\"u}tz}}]{Lindh:1999um}%
  \BibitemOpen
  \bibfield  {author} {\bibinfo {author} {\bibfnamefont {R.}~\bibnamefont
  {Lindh}}, \bibinfo {author} {\bibfnamefont {A.}~\bibnamefont
  {Bernhardsson}},\ and\ \bibinfo {author} {\bibfnamefont {M.}~\bibnamefont
  {Sch{\"u}tz}},\ }\bibfield  {title} {\bibinfo {title} {{Benzyne
  thermochemistry: A benchmark ab initio study}},\ }\href@noop {} {\bibfield
  {journal} {\bibinfo  {journal} {J. Phys. Chem. A}\ }\textbf {\bibinfo
  {volume} {103}},\ \bibinfo {pages} {9913} (\bibinfo {year}
  {1999})}\BibitemShut {NoStop}%
\bibitem [{\citenamefont {Slipchenko}\ and\ \citenamefont
  {Krylov}(2002)}]{Slipchenko:2002un}%
  \BibitemOpen
  \bibfield  {author} {\bibinfo {author} {\bibfnamefont {L.}~\bibnamefont
  {Slipchenko}}\ and\ \bibinfo {author} {\bibfnamefont {A.~I.}\ \bibnamefont
  {Krylov}},\ }\bibfield  {title} {\bibinfo {title} {{Singlet-triplet gaps in
  diradicals by the spin-flip approach: A benchmark study}},\ }\href@noop {}
  {\bibfield  {journal} {\bibinfo  {journal} {J. Chem. Phys.}\ }\textbf
  {\bibinfo {volume} {117}},\ \bibinfo {pages} {4694} (\bibinfo {year}
  {2002})}\BibitemShut {NoStop}%
\bibitem [{\citenamefont {Li}\ \emph {et~al.}(2007)\citenamefont {Li},
  \citenamefont {Yu}, \citenamefont {Huang},\ and\ \citenamefont
  {Wang}}]{li2007s1}%
  \BibitemOpen
  \bibfield  {author} {\bibinfo {author} {\bibfnamefont {H.}~\bibnamefont
  {Li}}, \bibinfo {author} {\bibfnamefont {S.-Y.}\ \bibnamefont {Yu}}, \bibinfo
  {author} {\bibfnamefont {M.-B.}\ \bibnamefont {Huang}},\ and\ \bibinfo
  {author} {\bibfnamefont {Z.-X.}\ \bibnamefont {Wang}},\ }\bibfield  {title}
  {\bibinfo {title} {The s1 states of o-, m-, and p-benzyne studied using
  multiconfiguration second-order perturbation theory},\ }\href@noop {}
  {\bibfield  {journal} {\bibinfo  {journal} {Chem. Phys. Lett.}\ }\textbf
  {\bibinfo {volume} {450}},\ \bibinfo {pages} {12} (\bibinfo {year}
  {2007})}\BibitemShut {NoStop}%
\bibitem [{\citenamefont {Wang}\ \emph {et~al.}(2008)\citenamefont {Wang},
  \citenamefont {Parish},\ and\ \citenamefont {Lischka}}]{wang2008extended}%
  \BibitemOpen
  \bibfield  {author} {\bibinfo {author} {\bibfnamefont {E.~B.}\ \bibnamefont
  {Wang}}, \bibinfo {author} {\bibfnamefont {C.~A.}\ \bibnamefont {Parish}},\
  and\ \bibinfo {author} {\bibfnamefont {H.}~\bibnamefont {Lischka}},\
  }\bibfield  {title} {\bibinfo {title} {An extended multireference study of
  the electronic states of para-benzyne},\ }\href@noop {} {\bibfield  {journal}
  {\bibinfo  {journal} {J. Chem. Phys.}\ }\textbf {\bibinfo {volume} {129}},\
  \bibinfo {pages} {044306} (\bibinfo {year} {2008})}\BibitemShut {NoStop}%
\bibitem [{\citenamefont {Li}\ and\ \citenamefont
  {Evangelista}(2016{\natexlab{b}})}]{Li:2016hb}%
  \BibitemOpen
  \bibfield  {author} {\bibinfo {author} {\bibfnamefont {C.}~\bibnamefont
  {Li}}\ and\ \bibinfo {author} {\bibfnamefont {F.~A.}\ \bibnamefont
  {Evangelista}},\ }\bibfield  {title} {\bibinfo {title} {{Towards numerically
  robust multireference theories: The driven similarity renormalization group
  truncated to one- and two-body operators}},\ }\href@noop {} {\bibfield
  {journal} {\bibinfo  {journal} {J. Chem. Phys.}\ }\textbf {\bibinfo {volume}
  {144}},\ \bibinfo {pages} {164114} (\bibinfo {year}
  {2016}{\natexlab{b}})}\BibitemShut {NoStop}%
\bibitem [{\citenamefont {Wang}\ \emph {et~al.}(2019)\citenamefont {Wang},
  \citenamefont {Li},\ and\ \citenamefont {Evangelista}}]{Wang:2019kf}%
  \BibitemOpen
  \bibfield  {author} {\bibinfo {author} {\bibfnamefont {S.}~\bibnamefont
  {Wang}}, \bibinfo {author} {\bibfnamefont {C.}~\bibnamefont {Li}},\ and\
  \bibinfo {author} {\bibfnamefont {F.~A.}\ \bibnamefont {Evangelista}},\
  }\bibfield  {title} {\bibinfo {title} {{Analytic gradients for the
  single-reference driven similarity renormalization group second-order
  perturbation theory}},\ }\href@noop {} {\bibfield  {journal} {\bibinfo
  {journal} {J. Chem. Phys.}\ }\textbf {\bibinfo {volume} {151}},\ \bibinfo
  {pages} {044118} (\bibinfo {year} {2019})}\BibitemShut {NoStop}%
\bibitem [{\citenamefont {ANIS}\ \emph {et~al.}(2021)\citenamefont {ANIS},
  \citenamefont {Abraham}, \citenamefont {AduOffei}, \citenamefont {Agarwal},
  \citenamefont {Agliardi}, \citenamefont {Aharoni}, \citenamefont {Akhalwaya},
  \citenamefont {Aleksandrowicz}, \citenamefont {Alexander}, \citenamefont
  {Amy}, \citenamefont {Anagolum}, \citenamefont {Arbel}, \citenamefont
  {Asfaw}, \citenamefont {Athalye}, \citenamefont {Avkhadiev}, \citenamefont
  {Azaustre}, \citenamefont {BHOLE}, \citenamefont {Banerjee}, \citenamefont
  {Banerjee}, \citenamefont {Bang}, \citenamefont {Bansal}, \citenamefont
  {Barkoutsos}, \citenamefont {Barnawal}, \citenamefont {Barron}, \citenamefont
  {Barron}, \citenamefont {Bello}, \citenamefont {Ben-Haim}, \citenamefont
  {Bennett}, \citenamefont {Bevenius} \emph {et~al.}}]{Qiskit}%
  \BibitemOpen
  \bibfield  {author} {\bibinfo {author} {\bibfnamefont {M.~S.}\ \bibnamefont
  {ANIS}}, \bibinfo {author} {\bibfnamefont {H.}~\bibnamefont {Abraham}},
  \bibinfo {author} {\bibnamefont {AduOffei}}, \bibinfo {author} {\bibfnamefont
  {R.}~\bibnamefont {Agarwal}}, \bibinfo {author} {\bibfnamefont
  {G.}~\bibnamefont {Agliardi}}, \bibinfo {author} {\bibfnamefont
  {M.}~\bibnamefont {Aharoni}}, \bibinfo {author} {\bibfnamefont {I.~Y.}\
  \bibnamefont {Akhalwaya}}, \bibinfo {author} {\bibfnamefont {G.}~\bibnamefont
  {Aleksandrowicz}}, \bibinfo {author} {\bibfnamefont {T.}~\bibnamefont
  {Alexander}}, \bibinfo {author} {\bibfnamefont {M.}~\bibnamefont {Amy}},
  \bibinfo {author} {\bibfnamefont {S.}~\bibnamefont {Anagolum}}, \bibinfo
  {author} {\bibfnamefont {E.}~\bibnamefont {Arbel}}, \bibinfo {author}
  {\bibfnamefont {A.}~\bibnamefont {Asfaw}}, \bibinfo {author} {\bibfnamefont
  {A.}~\bibnamefont {Athalye}}, \bibinfo {author} {\bibfnamefont
  {A.}~\bibnamefont {Avkhadiev}}, \bibinfo {author} {\bibfnamefont
  {C.}~\bibnamefont {Azaustre}}, \bibinfo {author} {\bibfnamefont
  {P.}~\bibnamefont {BHOLE}}, \bibinfo {author} {\bibfnamefont
  {A.}~\bibnamefont {Banerjee}}, \bibinfo {author} {\bibfnamefont
  {S.}~\bibnamefont {Banerjee}}, \bibinfo {author} {\bibfnamefont
  {W.}~\bibnamefont {Bang}}, \bibinfo {author} {\bibfnamefont {A.}~\bibnamefont
  {Bansal}}, \bibinfo {author} {\bibfnamefont {P.}~\bibnamefont {Barkoutsos}},
  \bibinfo {author} {\bibfnamefont {A.}~\bibnamefont {Barnawal}}, \bibinfo
  {author} {\bibfnamefont {G.}~\bibnamefont {Barron}}, \bibinfo {author}
  {\bibfnamefont {G.~S.}\ \bibnamefont {Barron}}, \bibinfo {author}
  {\bibfnamefont {L.}~\bibnamefont {Bello}}, \bibinfo {author} {\bibfnamefont
  {Y.}~\bibnamefont {Ben-Haim}}, \bibinfo {author} {\bibfnamefont {M.~C.}\
  \bibnamefont {Bennett}}, \bibinfo {author} {\bibfnamefont {D.}~\bibnamefont
  {Bevenius}}, \emph {et~al.},\ }\href {https://doi.org/10.5281/zenodo.2573505}
  {\bibinfo {title} {Qiskit: An open-source framework for quantum computing}}
  (\bibinfo {year} {2021})\BibitemShut {NoStop}%
\bibitem [{\citenamefont {Kinal}\ and\ \citenamefont
  {Piecuch}(2007)}]{kinal2007computational}%
  \BibitemOpen
  \bibfield  {author} {\bibinfo {author} {\bibfnamefont {A.}~\bibnamefont
  {Kinal}}\ and\ \bibinfo {author} {\bibfnamefont {P.}~\bibnamefont
  {Piecuch}},\ }\bibfield  {title} {\bibinfo {title} {Computational
  investigation of the conrotatory and disrotatory isomerization channels of
  bicyclo [1.1. 0] butane to buta-1, 3-diene: a completely renormalized
  coupled-cluster study},\ }\href@noop {} {\bibfield  {journal} {\bibinfo
  {journal} {J. Phys. Chem. A}\ }\textbf {\bibinfo {volume} {111}},\ \bibinfo
  {pages} {734} (\bibinfo {year} {2007})}\BibitemShut {NoStop}%
\bibitem [{\citenamefont {Berner}\ and\ \citenamefont
  {L{\"u}chow}(2010)}]{berner2010isomerization}%
  \BibitemOpen
  \bibfield  {author} {\bibinfo {author} {\bibfnamefont {R.}~\bibnamefont
  {Berner}}\ and\ \bibinfo {author} {\bibfnamefont {A.}~\bibnamefont
  {L{\"u}chow}},\ }\bibfield  {title} {\bibinfo {title} {Isomerization of
  bicyclo [1.1. 0] butane by means of the diffusion quantum monte carlo
  method},\ }\href@noop {} {\bibfield  {journal} {\bibinfo  {journal} {J. Phys.
  Chem. A}\ }\textbf {\bibinfo {volume} {114}},\ \bibinfo {pages} {13222}
  (\bibinfo {year} {2010})}\BibitemShut {NoStop}%
\bibitem [{\citenamefont {Shen}\ and\ \citenamefont
  {Piecuch}(2012)}]{shen2012combining}%
  \BibitemOpen
  \bibfield  {author} {\bibinfo {author} {\bibfnamefont {J.}~\bibnamefont
  {Shen}}\ and\ \bibinfo {author} {\bibfnamefont {P.}~\bibnamefont {Piecuch}},\
  }\bibfield  {title} {\bibinfo {title} {Combining active-space coupled-cluster
  methods with moment energy corrections via the cc (p; q) methodology, with
  benchmark calculations for biradical transition states},\ }\href@noop {}
  {\bibfield  {journal} {\bibinfo  {journal} {J. Chem. Phys.}\ }\textbf
  {\bibinfo {volume} {136}},\ \bibinfo {pages} {144104} (\bibinfo {year}
  {2012})}\BibitemShut {NoStop}%
\bibitem [{\citenamefont {Srinivasan}\ \emph {et~al.}(1965)\citenamefont
  {Srinivasan}, \citenamefont {Levi},\ and\ \citenamefont
  {Haller}}]{srinivasan1965thermal}%
  \BibitemOpen
  \bibfield  {author} {\bibinfo {author} {\bibfnamefont {R.}~\bibnamefont
  {Srinivasan}}, \bibinfo {author} {\bibfnamefont {A.}~\bibnamefont {Levi}},\
  and\ \bibinfo {author} {\bibfnamefont {I.}~\bibnamefont {Haller}},\
  }\bibfield  {title} {\bibinfo {title} {The thermal decomposition of bicyclo
  [1.1. 0] butane},\ }\href@noop {} {\bibfield  {journal} {\bibinfo  {journal}
  {J. Phys. Chem.}\ }\textbf {\bibinfo {volume} {69}},\ \bibinfo {pages} {1775}
  (\bibinfo {year} {1965})}\BibitemShut {NoStop}%
\bibitem [{\citenamefont {Wiberg}\ and\ \citenamefont
  {Fenoglio}(1968)}]{wiberg1968heats}%
  \BibitemOpen
  \bibfield  {author} {\bibinfo {author} {\bibfnamefont {K.~B.}\ \bibnamefont
  {Wiberg}}\ and\ \bibinfo {author} {\bibfnamefont {R.~A.}\ \bibnamefont
  {Fenoglio}},\ }\bibfield  {title} {\bibinfo {title} {Heats of formation of
  c4h6 hydrocarbons},\ }\href@noop {} {\bibfield  {journal} {\bibinfo
  {journal} {J. Am. Chem. Soc.}\ }\textbf {\bibinfo {volume} {90}},\ \bibinfo
  {pages} {3395} (\bibinfo {year} {1968})}\BibitemShut {NoStop}%
\bibitem [{\citenamefont {Blanchard~Jr}\ and\ \citenamefont
  {Cairncross}(1966)}]{blanchard1966bicyclo}%
  \BibitemOpen
  \bibfield  {author} {\bibinfo {author} {\bibfnamefont {E.}~\bibnamefont
  {Blanchard~Jr}}\ and\ \bibinfo {author} {\bibfnamefont {A.}~\bibnamefont
  {Cairncross}},\ }\bibfield  {title} {\bibinfo {title} {Bicyclo [1.1. 0]
  butane chemistry. i. the synthesis and reactions of 3-methylbicyclo [1.1. 0]
  butanecarbonitriles},\ }\href@noop {} {\bibfield  {journal} {\bibinfo
  {journal} {J. Am. Chem. Soc.}\ }\textbf {\bibinfo {volume} {88}},\ \bibinfo
  {pages} {487} (\bibinfo {year} {1966})}\BibitemShut {NoStop}%
\bibitem [{\citenamefont {Frey}\ and\ \citenamefont
  {Stevens}(1965)}]{frey1965thermal}%
  \BibitemOpen
  \bibfield  {author} {\bibinfo {author} {\bibfnamefont {H.}~\bibnamefont
  {Frey}}\ and\ \bibinfo {author} {\bibfnamefont {I.}~\bibnamefont {Stevens}},\
  }\bibfield  {title} {\bibinfo {title} {Thermal unimolecular isomerization of
  bicyclobutane},\ }\href@noop {} {\bibfield  {journal} {\bibinfo  {journal}
  {Trans. Faraday Soc.}\ }\textbf {\bibinfo {volume} {61}},\ \bibinfo {pages}
  {90} (\bibinfo {year} {1965})}\BibitemShut {NoStop}%
\bibitem [{\citenamefont {Wiberg}\ and\ \citenamefont
  {Lavanish}(1966)}]{wiberg1966formation}%
  \BibitemOpen
  \bibfield  {author} {\bibinfo {author} {\bibfnamefont {K.~B.}\ \bibnamefont
  {Wiberg}}\ and\ \bibinfo {author} {\bibfnamefont {J.~M.}\ \bibnamefont
  {Lavanish}},\ }\bibfield  {title} {\bibinfo {title} {Formation and thermal
  decomposition of bicyclo [1.1. 0] butane-2-exo-d11},\ }\href@noop {}
  {\bibfield  {journal} {\bibinfo  {journal} {J. Am. Chem. Soc.}\ }\textbf
  {\bibinfo {volume} {88}},\ \bibinfo {pages} {5272} (\bibinfo {year}
  {1966})}\BibitemShut {NoStop}%
\bibitem [{\citenamefont {Closs}\ and\ \citenamefont
  {Pfeffer}(1968)}]{closs1968steric}%
  \BibitemOpen
  \bibfield  {author} {\bibinfo {author} {\bibfnamefont {G.}~\bibnamefont
  {Closs}}\ and\ \bibinfo {author} {\bibfnamefont {P.}~\bibnamefont
  {Pfeffer}},\ }\bibfield  {title} {\bibinfo {title} {The steric course of the
  thermal rearrangements of methylbicyclobutanes},\ }\href@noop {} {\bibfield
  {journal} {\bibinfo  {journal} {J. Am. Chem. Soc.}\ }\textbf {\bibinfo
  {volume} {90}},\ \bibinfo {pages} {2452} (\bibinfo {year}
  {1968})}\BibitemShut {NoStop}%
\bibitem [{\citenamefont {Mazziotti}(2008)}]{mazziotti2008energy}%
  \BibitemOpen
  \bibfield  {author} {\bibinfo {author} {\bibfnamefont {D.~A.}\ \bibnamefont
  {Mazziotti}},\ }\bibfield  {title} {\bibinfo {title} {Energy barriers in the
  conversion of bicyclobutane to gauche-1, 3-butadiene from the anti-hermitian
  contracted schrodinger equation},\ }\href@noop {} {\bibfield  {journal}
  {\bibinfo  {journal} {J. Phys. Chem. A}\ }\textbf {\bibinfo {volume} {112}},\
  \bibinfo {pages} {13684} (\bibinfo {year} {2008})}\BibitemShut {NoStop}%
\bibitem [{\citenamefont {Nguyen}\ and\ \citenamefont
  {Gordon}(1995)}]{nguyen1995isomerization}%
  \BibitemOpen
  \bibfield  {author} {\bibinfo {author} {\bibfnamefont {K.~A.}\ \bibnamefont
  {Nguyen}}\ and\ \bibinfo {author} {\bibfnamefont {M.~S.}\ \bibnamefont
  {Gordon}},\ }\bibfield  {title} {\bibinfo {title} {Isomerization of bicyclo
  [1.1. 0] butane to butadiene},\ }\href@noop {} {\bibfield  {journal}
  {\bibinfo  {journal} {J. Am. Chem. Soc.}\ }\textbf {\bibinfo {volume}
  {117}},\ \bibinfo {pages} {3835} (\bibinfo {year} {1995})}\BibitemShut
  {NoStop}%
\bibitem [{\citenamefont {Lutz}\ and\ \citenamefont
  {Piecuch}(2008)}]{lutz2008extrapolating}%
  \BibitemOpen
  \bibfield  {author} {\bibinfo {author} {\bibfnamefont {J.~J.}\ \bibnamefont
  {Lutz}}\ and\ \bibinfo {author} {\bibfnamefont {P.}~\bibnamefont {Piecuch}},\
  }\bibfield  {title} {\bibinfo {title} {Extrapolating potential energy
  surfaces by scaling electron correlation: Isomerization of bicyclobutane to
  butadiene},\ }\href@noop {} {\bibfield  {journal} {\bibinfo  {journal} {J.
  Chem. Phys.}\ }\textbf {\bibinfo {volume} {128}},\ \bibinfo {pages} {154116}
  (\bibinfo {year} {2008})}\BibitemShut {NoStop}%
\bibitem [{\citenamefont {Boyn}\ and\ \citenamefont
  {Mazziotti}(2022)}]{boyn2022elucidating}%
  \BibitemOpen
  \bibfield  {author} {\bibinfo {author} {\bibfnamefont {J.-N.}\ \bibnamefont
  {Boyn}}\ and\ \bibinfo {author} {\bibfnamefont {D.~A.}\ \bibnamefont
  {Mazziotti}},\ }\bibfield  {title} {\bibinfo {title} {Elucidating the
  molecular orbital dependence of the total electronic energy in multireference
  problems},\ }\href@noop {} {\bibfield  {journal} {\bibinfo  {journal} {J.
  Chem. Phys.}\ }\textbf {\bibinfo {volume} {156}},\ \bibinfo {pages} {194104}
  (\bibinfo {year} {2022})}\BibitemShut {NoStop}%
\bibitem [{\citenamefont {Allen}\ and\ \citenamefont
  {Schaefer}(1986)}]{Allen:1986tf}%
  \BibitemOpen
  \bibfield  {author} {\bibinfo {author} {\bibfnamefont {W.~D.}\ \bibnamefont
  {Allen}}\ and\ \bibinfo {author} {\bibfnamefont {H.~F.}\ \bibnamefont
  {Schaefer}},\ }\bibfield  {title} {\bibinfo {title} {{Ab initio Studies of
  the Low-Lying Electronic States of Ketene}},\ }\href@noop {} {\bibfield
  {journal} {\bibinfo  {journal} {J. Chem. Phys.}\ }\textbf {\bibinfo {volume}
  {84}},\ \bibinfo {pages} {2212} (\bibinfo {year} {1986})}\BibitemShut
  {NoStop}%
\bibitem [{\citenamefont {Allen}\ and\ \citenamefont
  {Schaefer}(1987)}]{Allen:1987uu}%
  \BibitemOpen
  \bibfield  {author} {\bibinfo {author} {\bibfnamefont {W.~D.}\ \bibnamefont
  {Allen}}\ and\ \bibinfo {author} {\bibfnamefont {H.~F.}\ \bibnamefont
  {Schaefer}},\ }\bibfield  {title} {\bibinfo {title} {An examination of the 2
  1a1 states of formaldehyde and ketene including analytic configuration
  interaction energy first derivatives for singlet excited electronic states of
  the same symmetry as the ground state},\ }\href@noop {} {\bibfield  {journal}
  {\bibinfo  {journal} {J. Chem. Phys.}\ }\textbf {\bibinfo {volume} {87}},\
  \bibinfo {pages} {7076} (\bibinfo {year} {1987})}\BibitemShut {NoStop}%
\bibitem [{\citenamefont {Parrish}\ \emph {et~al.}(2019)\citenamefont
  {Parrish}, \citenamefont {Iosue}, \citenamefont {Ozaeta},\ and\ \citenamefont
  {McMahon}}]{parrish2019jacobi}%
  \BibitemOpen
  \bibfield  {author} {\bibinfo {author} {\bibfnamefont {R.~M.}\ \bibnamefont
  {Parrish}}, \bibinfo {author} {\bibfnamefont {J.~T.}\ \bibnamefont {Iosue}},
  \bibinfo {author} {\bibfnamefont {A.}~\bibnamefont {Ozaeta}},\ and\ \bibinfo
  {author} {\bibfnamefont {P.~L.}\ \bibnamefont {McMahon}},\ }\bibfield
  {title} {\bibinfo {title} {A jacobi diagonalization and anderson acceleration
  algorithm for variational quantum algorithm parameter optimization},\
  }\href@noop {} {\bibfield  {journal} {\bibinfo  {journal} {arXiv preprint
  arXiv:1904.03206}\ } (\bibinfo {year} {2019})}\BibitemShut {NoStop}%
\bibitem [{\citenamefont {Nakanishi}\ \emph {et~al.}(2020)\citenamefont
  {Nakanishi}, \citenamefont {Fujii},\ and\ \citenamefont
  {Todo}}]{Nakanishi:2020in}%
  \BibitemOpen
  \bibfield  {author} {\bibinfo {author} {\bibfnamefont {K.~M.}\ \bibnamefont
  {Nakanishi}}, \bibinfo {author} {\bibfnamefont {K.}~\bibnamefont {Fujii}},\
  and\ \bibinfo {author} {\bibfnamefont {S.}~\bibnamefont {Todo}},\ }\bibfield
  {title} {\bibinfo {title} {{Sequential minimal optimization for
  quantum-classical hybrid algorithms}},\ }\href@noop {} {\bibfield  {journal}
  {\bibinfo  {journal} {Phys. Rev. Research}\ }\textbf {\bibinfo {volume}
  {2}},\ \bibinfo {pages} {043158} (\bibinfo {year} {2020})}\BibitemShut
  {NoStop}%
\end{thebibliography}%

\end{document}